\documentclass[conference]{IEEEtran}
\IEEEoverridecommandlockouts

\usepackage{tikz}
\usepackage{amsmath}

\usepackage{booktabs}
\usepackage{filecontents}

\usepackage{graphicx}
\usepackage[many]{tcolorbox}

\usepackage{mathtools,amssymb,latexsym,amsfonts,stmaryrd}
\usepackage{pbox}
\usepackage{xfrac}
\usepackage{booktabs}
\usepackage{xcolor}
\usepackage{threeparttable}
\usepackage{amsthm}
\usepackage{hyperref}
\usepackage{multirow}
\usepackage{tikz}
\usepackage{mathrsfs}
\usepackage{mathpartir}
\usepackage{algorithm}
\usepackage{url}
\usepackage{syntax}
\usepackage{framed}
\usepackage[noend]{algpseudocode}
\usepackage{flushend}
\usepackage{listings}
\usepackage{moresize}
\usepackage{wrapfig}
 
\usepackage{seqsplit}
\usepackage{alltt}
\usepackage{xspace}
\usepackage{pgfplots}
\usepackage{subcaption}
\usepackage{threeparttable}
\usepackage{adjustbox}
\usetikzlibrary{positioning}
\usepackage{enumitem}
\usepackage{tabularx}
\usepackage{pifont}
\newcommand{\cmark}{\ding{51}}%

\DeclareMathAlphabet{\mathcal}{OMS}{cmsy}{m}{n}

\usepackage{amsmath, listings, amsthm, amssymb, proof, xspace}

\usepackage{thmtools}

\declaretheoremstyle[spaceabove=\topsep,notefont=\normalfont\itshape]{mystyle}

\newcommand{\revise}[2]{{\color{red}{\ifx&#1&\else- #1\fi}} {\color{ForestGreen}{\ifx&#2&\else+ #2\fi}}}%
\renewcommand{\revise}[2]{#2}%

\usetikzlibrary{arrows,patterns, decorations.pathreplacing}
\newtheorem{definition}{Definition}
\newtheorem{corollary}{Corollary}

\newcommand{\F}{Fig.}
\newcommand{\T}{Table}
\renewcommand{\S}{Sec.}
\newcommand{\A}{Alg.}
\newcommand{\E}{Eq.}

\newcommand{\ignore}[1]{}

\lstdefinestyle{base}{
  moredelim=**[is][\color{red}]{@}{@},
  escapeinside={<@}{@>}
}

\newcommand{\ma}{$\mathcal{A}$}
\newcommand{\mb}{$\mathcal{B}$}
\newcommand{\mc}{$\mathcal{C}$}
\newcommand{\md}{$\mathcal{D}$}
\newcommand{\me}{$\mathcal{E}$}

\newcommand{\Lim}[1]{\raisebox{0.5ex}{\scalebox{0.8}{$\displaystyle \lim_{#1}\;$}}}

\usepackage{amssymb}
\usepackage{pifont}
\newcommand{\xmark}{\ding{53}}%

\newcommand\DejaVuttfamily{%
  \fontfamily{DejaVuSansMono-TLF}\selectfont }

\lstdefinestyle{base}{
  moredelim=**[is][\color{red}]{@}{@},
  escapeinside={<@}{@>}
}

\lstdefinelanguage
   [x64]{Assembler}     
   [x86masm]{Assembler} 
   {morekeywords={CDQE,CQO,CMPSQ,CMPXCHG16B,JRCXZ,LODSQ,MOVSXD, %
                  POPFQ,PUSHFQ,SCASQ,STOSQ,IRETQ,RDTSCP,SWAPGS, %
                  rax,rdx,rcx,rbx,rsi,rdi,rsp,rbp, %
                  r8,r8d,r8w,r8b,r9,r9d,r9w,r9b}} 

\renewcommand{\baselinestretch}{0.983}

\usepackage{listings}
\usepackage{fancyvrb}
\usepackage{color}
\definecolor{lightgray}{rgb}{.9,.9,.9}
\definecolor{darkgray}{rgb}{.4,.4,.4}
\definecolor{purple}{rgb}{0.65, 0.12, 0.82}
\definecolor{commentgreen}{RGB}{63,127,95}

\colorlet{myPurple}{blue!40!red}
\definecolor{myOrange}{RGB}{255,192,0}
\newcommand{\code}[1]{\textcolor{blue}{\textit{\bfseries{#1}}}}

\lstdefinelanguage{Solidity}{
  keywords={len,delete,int,void,payable, public, event, contract, typeof, new, true, false, catch, function, return, null, catch, switch, var, if, in, while, do, else, case, break,struct,const,socklen_t,sa_familty_t,char,sockaddr},
  keywordstyle=\color{violet}\bfseries,
  ndkeywords={class, export, boolean, throw, implements, import, this},
  ndkeywordstyle=\color{darkgray}\bfseries,
  identifierstyle=\color{black},
  sensitive=false,
  comment=[l]{//},
  escapeinside={(*@}{@*)},          
  morecomment=[s]{/*}{*/},
  commentstyle=\color{commentgreen}\ttfamily,
  stringstyle=\color{red}\ttfamily,
  morestring=[b]',
  morestring=[b]"
}

\newcommand{\rnum}[1]{\uppercase\expandafter{\romannumeral #1\relax}}

\usepackage{xcolor}

\algnewcommand{\LeftComment}[1]{\Statex \(\triangleright\) #1}

\definecolor{pptbrown}{RGB}{132,60,12}
\definecolor{pptgreen}{RGB}{169, 209, 142}
\definecolor{pptred}{RGB}{255, 130, 134}
\definecolor{pptgrey}{RGB}{191, 191, 191}
\definecolor{pptcyan}{RGB}{0, 176, 240}

\let\OLDthebibliography\thebibliography
\renewcommand\thebibliography[1]{
  \OLDthebibliography{#1}
  \setlength{\parskip}{0pt}
  \setlength{\itemsep}{0pt plus 0.1ex}
}

\newcommand*\circled[1]{\tikz[baseline=(char.base)]{
            \node[shape=circle,draw,inner sep=1.3pt] (char) {#1};}}

  \DeclareFontFamily{U}{dutchcal}{\skewchar \font =45}
  \DeclareFontShape{U}{dutchcal}{m}{n}{
    <-> dutchcal-r}{}
  \DeclareFontShape{U}{dutchcal}{b}{n}{
    <-> dutchcal-b}{}
  \DeclareMathAlphabet{\mdutchcal}{U}{dutchcal}{m}{n}
  \SetMathAlphabet{\mdutchcal}{bold}{U}{dutchcal}{b}{n}
  \DeclareMathAlphabet{\mdutchbcal} {U}{dutchcal}{b}{n}

  \DeclareFontFamily{U}{txcal}{\skewchar \font =45}
  \DeclareFontShape{U}{txcal}{m}{n}{
    <-> txr-cal}{}
  \DeclareFontShape{U}{txcal}{b}{n}{
    <-> txb-cal}{}
  \DeclareMathAlphabet{\mtxcal}{U}{txcal}{m}{n}
  \SetMathAlphabet{\mtxcal}{bold}{U}{txcal}{b}{n}
  \DeclareMathAlphabet{\mtxbcal} {U}{txcal}{b}{n}

\newcommand{\sd}{$\mdutchcal{S}$}

\newcommand{\ssd}{$|\mdutchcal{S}|$}

\usepackage[noadjust]{cite}

\def\BibTeX{{\rm B\kern-.05em{\sc i\kern-.025em b}\kern-.08em
    T\kern-.1667em\lower.7ex\hbox{E}\kern-.125emX}}
\begin{document}



\title{ADI: Adversarial Dominating Inputs in Vertical Federated Learning Systems}
\author{
\IEEEauthorblockN{Qi Pang$^{\dagger}$, Yuanyuan Yuan$^\ddagger$, Shuai Wang$^{\ddagger 1}$, Wenting Zheng$^\dagger$}
\IEEEauthorblockA{$^\dagger$Carnegie Mellon University, $^\ddagger$HKUST}
}

\maketitle

\def\thefootnote{1}\footnotetext{Corresponding author.}\def\thefootnote{\arabic{footnote}}
\def\thefootnote{2}\footnotetext{This is an extended version of~\cite{spversion}, please cite the IEEE S\&P 23 version.}\def\thefootnote{\arabic{footnote}}

\begin{abstract}
  Vertical federated
  learning (VFL) system has recently become prominent as a concept to process
  data distributed across many individual sources without the need to centralize
  it. Multiple participants collaboratively train models based on their local
  data in a privacy-aware manner. To date, VFL has become a de facto
  solution to securely learn a model among organizations, allowing knowledge to
  be shared without compromising privacy of any  individuals.

Despite the prosperous development of VFL systems, we find that certain inputs of a
participant, named adversarial dominating inputs (ADIs), can 
dominate the joint inference towards the direction of the adversary's will and force other (victim) participants to make negligible
contributions, losing rewards that are usually offered regarding the importance of
their contributions in federated learning scenarios.

We conduct a systematic study on ADIs by first proving their existence in
typical VFL systems. We then propose gradient-based methods to synthesize ADIs
of various formats and exploit common VFL systems. We further launch greybox
fuzz testing, guided by the saliency score of ``victim'' participants, to
perturb adversary-controlled inputs and systematically explore the VFL attack
surface in a privacy-preserving manner. We conduct an in-depth study on the
influence of critical parameters and settings in synthesizing ADIs. Our study
reveals new VFL attack opportunities, promoting the identification of unknown
threats before breaches and building more secure VFL systems.
\end{abstract}



\section{Introduction}
\label{sec:introduciton}

A traditional machine learning system workflow involves a data pipeline, which
uses a central server that hosts the trained model to make predictions. Thus,
all data collected by local devices and sensors are sent to the central
server for model training and making predictions. This data integration
technique necessitates users sharing their data with a central server, which is
strongly opposed due to a variety of real-world concerns, including data
privacy, industrial competition, and complex administrative procedures.

\begin{figure}[!t]
  \captionsetup{skip=2pt}
  \centering
  \includegraphics[width=1\linewidth]{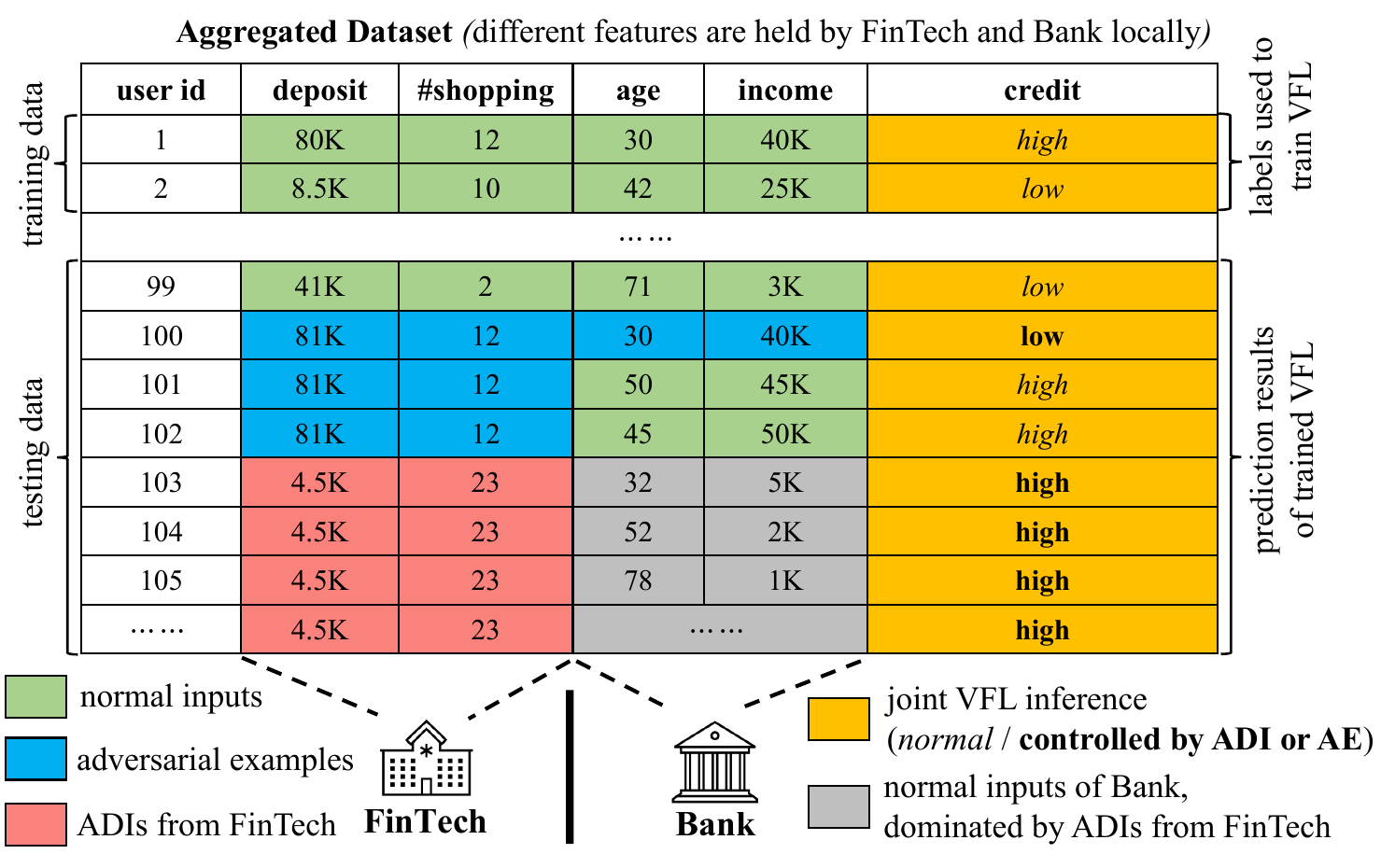}
  \caption{{ADIs from FinTech dominate the VFL predictions.}}
  \label{fig:motivation}
  \vspace{-5pt}
\end{figure}

To address this problem, federated learning (FL)~\cite{yang2019federated} retains
private data locally to train intermediate models. The parameters of these
locally trained intermediate models are then aggregated into a single,
consolidated, and gradually improved global model. Model aggregation uses either
a trusted central coordinator or cryptographic techniques such as secure
multi-party computation (MPC)~\cite{goldreich1998secure}. Unlike
centralized machine learning, FL shares parameters rather than
sensitive data, hence alleviating privacy leakage. Moreover,
depending on how local data are distributed, FL systems can be
classified into
\textit{vertical federated learning}
(VFL)~\cite{yang2019federated,hardy2017private} and
\textit{horizontal federated learning}
(HFL)~\cite{mcmahan2017communication,bonawitz2019towards}. In HFL,
participants possess different data samples in the same feature
space, e.g., Google users jointly use their own local keystroke data to train a
global, remote model~\cite{hard2018federated}.
VFL participants are often companies/institutions owning a subset of the feature space.
%
\F~\ref{fig:motivation} presents a typical vertical federated logistic
regression (HeteroLR) scenario~\cite{hardy2017private}. In normal usage (the
\colorbox{pptgreen}{green data}), FinTech \ma\ holds a a subset of the features
for a group of users, whereas Bank \mb\ holds another
subset of the features
for the same users. \ma\ and \mb\ jointly
predict a user's credit score (high/low) without leaking each participant's
local data to the other participants.

Despite encouraging development of FL for aggregating dispersed
data across participants, emerging attacks targeting this new computing paradigm
have been revealed. The distribution of the training process to a
set of potentially malicious clients creates backdoor attack or adversarial
example (AE) opportunities on the shared
model~\cite{bagdasaryan2018backdoor,bhagoji2019analyzing,xie2019dba,sun2019can,wang2020attack,wang2019beyond}.
While practical attacks have been proposed toward
HFL~\cite{xie2019dba,bagdasaryan2018backdoor,bhagoji2019analyzing}, attack
on VFL have not been systematically studied. 
%
%

This research examines security issues of VFL in light of its growing adoption
in security- and privacy-sensitive domains such as credit scoring, insurance,
and loan assessment~\cite{yang2019federated}. Particularly, we find that a set
of unique inputs, which we call \textit{adversarial dominating inputs} (ADIs),
{manipulate the joint predictions of a well-trained VFL model. When
ADIs are used by a malicious participant, other benign participants'
contributions to the joint prediction are nullified.}
Again, considering \F~\ref{fig:motivation}, we find that by perturbing \ma's
data into ADIs (\colorbox{pptred}{red} in \F~\ref{fig:motivation}), \ma\
controls the joint prediction to a fixed answer at their will, and diminishes the
influence of \mb's data (\colorbox{pptgrey}{grey} in \F~\ref{fig:motivation}).
{As a comparison, an adversarial example (AE), marked in
\colorbox{pptcyan}{blue} in \F~\ref{fig:motivation}, only misclassifies one
input from Bank \mb\ when it is used by FinTech \ma. When \mb\ uses different
inputs, the joint inference returns normal. In short, AEs fail to constantly
manipulate VFL inferences, and AEs do not diminish benign participants'
contribution explicitly.}

This research, as the first systematic study on ADIs in VFL, is motivated by
findings in \F~\ref{fig:motivation}.
We first formulate ADIs and prove their existence in common VFL protocols. Then,
in two steps, we explore ADIs in real-world VFL systems. First, we design
gradient-based approaches for ADI synthesis in a \textit{blackbox} setting. Second, inspired by feedback-driven software
fuzzing, we design a \textit{greybox} fuzz testing framework to uncover ADIs. The
proposed two-step approaches delineate attack vectors using ADIs from
different perspectives and at varying costs: gradient-based methods demonstrate
an end-to-end, practical exploitation using ADIs from an adversarial 
participant, whereas fuzzing enables in-house vulnerability assessment to
comprehensively uncover ADIs in a privacy-preserving manner under all VFL
participants' collaboration.

Our evaluation includes three popular VFL systems with various input formats
(e.g., images, tabular data). We achieve promising success rates of
gradient-based ADI synthesis and illustrate the stealth of synthesized ADIs by
comparing them with normal inputs. We also find 2,320 ADIs over 92 hours of
fuzzing in total, revealing large attack surfaces of popular VFL
systems. We investigate how several key parameters and settings can influence
ADI synthesis and uncovering. Overall, we show that the ADI issue is a crucial
but often neglected impediment in adopting VFL in real-world circumstances. In
sum, this study makes the following contributions:


\begin{itemize}
  \item We identify ADIs, as a novel generalization of AEs in the context of VFL. ADIs constantly dominate the
  joint inference made by VFL, and extensively diminish other participants'
  contribution to model inference, thus hogging the rewards provided to
  incentivize VFL participants contributing important features.
  \item We
  prove the existence of ADIs in popular VFL systems. We propose gradient-based
  methods for ADI synthesis. We design a fuzzing tool to comprehensively uncover
  ADIs and facilitate vulnerability assessment of VFL systems. Our method is
  adaptable to various VFL systems and input formats.
  \item We
  achieve high success rates of generating ADIs to exploit VFL systems, indicating
  that ADIs are prevalent yet neglected issues. Our insights can provide users
  with the up-to-date understanding of VFL systems.
\end{itemize}
\section{Background}
\label{sec:background}

\noindent \textbf{VFL Overview.}~In VFL, each participant learns from distinct
feature partitions within a same data sample. Given two participants \ma\ and
\mb, features $X^i$ of a data sample $i$ are partitioned into
$X_{\mathcal{A}}^i$ and $X_{\mathcal{B}}^i$ ($X^i = X_{\mathcal{A}}^i ||
X_{\mathcal{B}}^i$; $||$ is the concatenation) and possessed separately by
\ma\ and \mb. As shown in \F~\ref{fig:motivation}, a regional FinTech and a bank
located in the same region maintain records of many local residents. To predict
a resident $i$'s credit score, the FinTech and the bank can extract $i$'s
financial records and make the joint inference. Although data maintained by the
FinTech and the bank have very different feature spaces, they belong to the same
user $i$. The prediction results will be aggregated in a coordinator
\mc\ (omitted in \F~\ref{fig:motivation}) and returned to \ma\ and \mb.

\begin{figure}[!t]
  \captionsetup{skip=-1pt}
  \centering
  \includegraphics[width=0.95\linewidth]{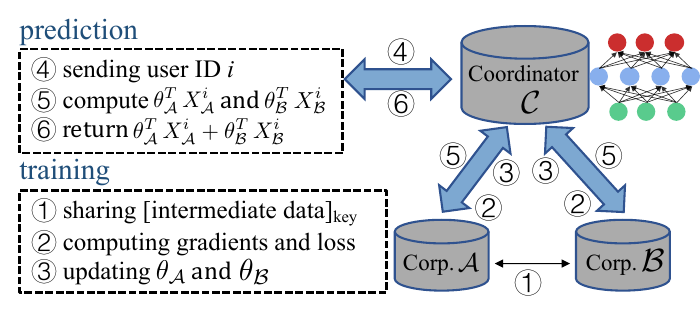}
  \caption{Architecture of a VFL system.}
  \label{fig:vfl}
  \vspace{-5pt}
\end{figure}

During training, VFL participants exchange intermediate information and compute
training losses and gradients in a privacy-aware manner, where raw data are
kept locally by each participant. \F~\ref{fig:vfl} depicts a typical VFL
scenario, where \ma\ and \mb\ jointly train a model. \ma\ also
possesses the corresponding labels. Some VFL systems
also employs a coordinator \mc. Holistically, model training is
divided into three steps: 1) \ma\ and \mb\ encrypt and exchange the intermediate
stages for gradient and loss calculations; 2) \ma\ and \mb\ compute the
encrypted gradients, and \ma\ computes the loss and uploads the
encrypted gradients and losses to \mc; and 3) \mc\ decrypts the gradients and
updates the models' parameters $\theta_{\mathcal{A}}$ and $\theta_{\mathcal{B}}$,
and then sends them back to \ma\ and \mb.

To use a trained VFL, common user IDs $i \in I$ shared by \ma\ and
\mb\ must be first confirmed~\cite{scannapieco2007privacy}. The corresponding records will be
extracted (\circled{4} in \F~\ref{fig:vfl}), and each participant will compute
the local intermediate results which will be aggregated in \mc\ (\circled{5} in
\F~\ref{fig:vfl}). \mc\ will compute the joint prediction result and return it
to \ma\ and \mb\ (\circled{6} in \F~\ref{fig:vfl}).

\noindent \textbf{VFL vs.~HFL.}~{As noted in
\S~\ref{sec:introduciton}, HFL generally refers to the FL 
setting where participants hold different samples of the same features, and they
jointly train a central model. In contrast, VFL is originated from the practical
needs that companies and institutions owning fragmented data belonging to the
same group of users, and they compensate each other and maximize data
utilization through collaborative model training and predictions. That is,
training/test data in VFL participants are from distinct feature spaces of the
same users, and each VFL participant has its own unique local model.}
To date, most real-world VFL designs have 2 (maximum 4)
participants~\cite{xia2021vertical, das2021multi, luo2021feature,
yang2019federated}.
We clarify that ADIs, as introduced in \S~\ref{sec:adi}, particularly exist in
VFL. Since each HFL participant receives the full model from the central server,
a participant does not rely on other participants to make predictions at the
inference stage. Thus, in HFL, predictions on a participant cannot be influenced
and dominated by malicious participants at the inference stage.
\S~\ref{subsec:motivation-adi} presents further comparison between ADIs and
adversarial examples in HFL.

\noindent \textbf{Learning Protocols.}~We introduce two popular VFL systems:
vertical federated logistic regression (HeteroLR)~\cite{hardy2017private} and
vertical federated neural network (SplitNN)~\cite{gupta2018distributed}. They
are widely used in real-world VFL scenarios and show comparable performance with
their centralized versions. More importantly, most parameter-based
learning methods in VFL can be extended from these two core protocols. For
instance, the vertical federated visual question answering (VFVQA), a popular
paradigm facilitating answering questions about images~\cite{liu2020federated},
is extended from SplitNN. Hence, our proof on ADI existence
(\S~\ref{subsec:adi}) and evaluation (\S~\ref{sec:evaluation}) consider SplitNN
and HeteroLR. We also implement and evaluate VFVQA by extending SplitNN.
In short, our study subsumes regression, classification, and VQA tasks
offered by VFL systems. We present the training and implementation details of
all three protocols in Appendix \ref{sec:lr-nn} and discuss tree-based VFL systems in Appendix \ref{subsec:append-discuss}.

\begin{figure}[!t]
  \centering
  \includegraphics[width=0.75\linewidth]{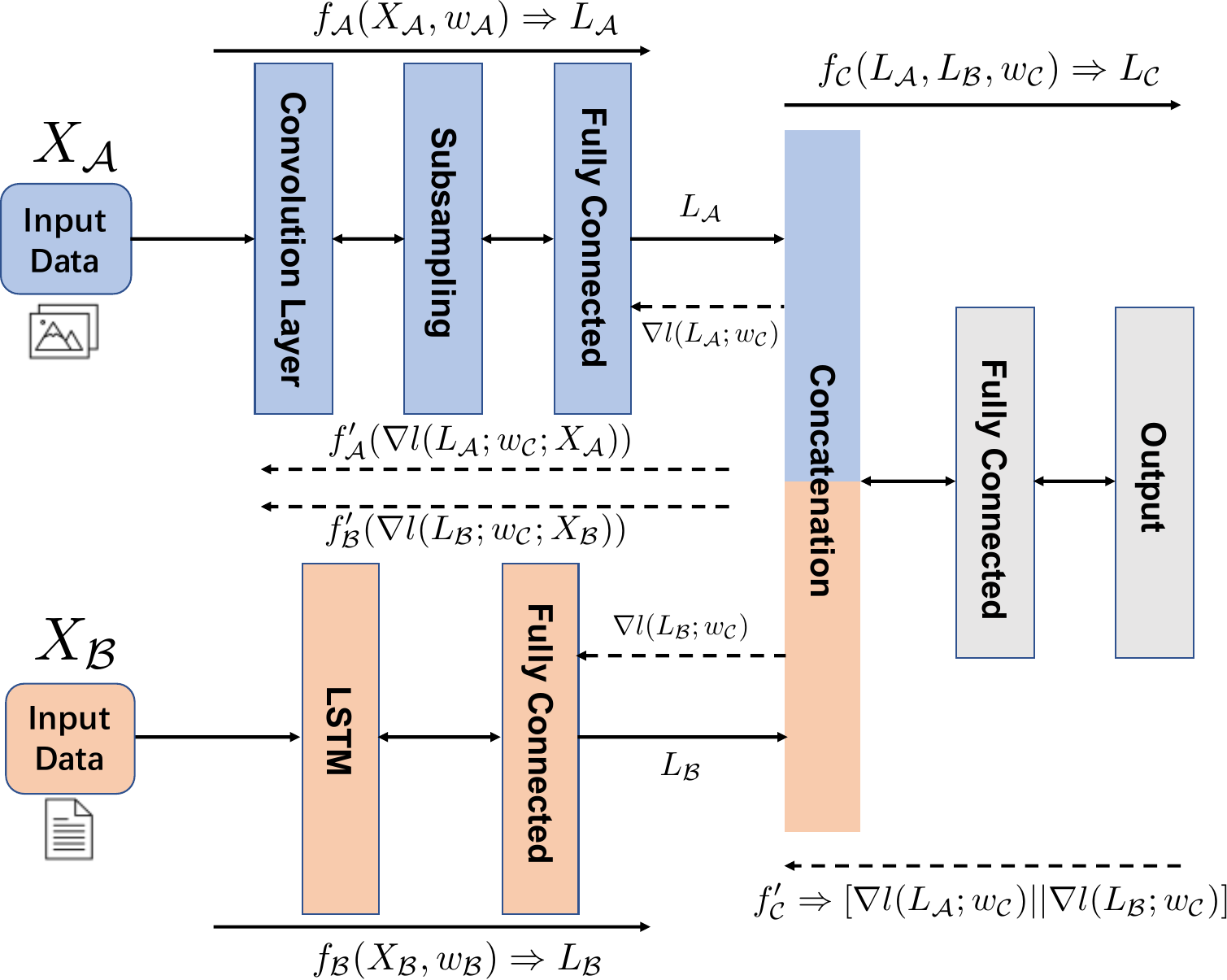}
  \caption{{Architecture of SplitNN.}}
  \label{fig:SplitNN}
  \vspace{-10pt}
\end{figure}

To help readers better understand VFL systems, we present \F~\ref{fig:SplitNN}, 
which depicts the architecture of
heterogeneous neural network (also known as a special case of SplitNN). SplitNN
facilitates multiple participants holding data with different feature spaces to
train a distributed model without sharing raw data. In the forward propagation
phase, \ma\ and \mb\ compute the outputs of their local models
($w_{\mathcal{A}}$ and $w_{\mathcal{B}}$) and forward them to the coordinator
\mc:

\setlength{\belowdisplayskip}{4pt} \setlength{\belowdisplayshortskip}{4pt}
\setlength{\abovedisplayskip}{-8pt} \setlength{\abovedisplayshortskip}{-8pt}
\small
\begin{equation}
  \begin{aligned}
      L_{\mathcal{A}} = f_{\mathcal{A}}(X_{\mathcal{A}}, w_{\mathcal{A}}), \;
      L_{\mathcal{B}} = f_{\mathcal{B}}(X_{\mathcal{B}}, w_{\mathcal{B}})
  \end{aligned}
\end{equation}
\normalsize

{The forwarded local outputs, $L_{\mathcal{A}}$ and $L_{\mathcal{B}}$, will be
concatenated on the coordinator side and fed to the coordinator model
($w_{\mathcal{C}}$) for a joint inference whose result is $L_{\mathcal{C}}$.}

\small
\begin{equation}
  \begin{aligned}
      L_{\mathcal{C}} = f_{\mathcal{C}}([L_{\mathcal{A}} \, || \, L_{\mathcal{B}}], w_{\mathcal{C}})
  \end{aligned}
\end{equation}
\normalsize

{In the backward propagation phase, the coordinator \mc\ computes the gradients
$f_{\mathcal{C}}'(L_{\mathcal{C}})$, performs gradient descent on its model, and
obtains the gradients of local outputs submitted by the participants as follows:}

\small
\begin{equation}
  \begin{aligned}
      [\nabla \mathit{l}(L_{\mathcal{A}}; w_{\mathcal{C}}) \, || \, \nabla \mathit{l}(L_{\mathcal{B}}; w_{\mathcal{C}})] = f_{\mathcal{C}}'(L_{\mathcal{C}})
  \end{aligned}
\end{equation}
\normalsize

{Back to the participant side, once receiving the gradients, the participants
further compute the local gradients $f_{\mathcal{A}}'(\nabla
\mathit{l}(L_{\mathcal{A}}; w_{\mathcal{C}}; X_{\mathcal{A}}))$ and
$f_{\mathcal{B}}'(\nabla \mathit{l}(L_{\mathcal{B}}; w_{\mathcal{C}};
X_{\mathcal{B}}))$, and then update local models. $f$ represents the neural
network and $f'$ represents its derivative. In the forward and backward
propagations, no raw data is directly sent to the coordinator or exchanged among
participants, and all intermediate data is encrypted. Furthermore, by extending
the SplitNN protocol, we can construct more complex models like CNN and LSTM
following the VFL paradigm.}

\noindent \textbf{VFL Reward.}~VFL enables an individual participant (e.g., a
bank) to enrich its data and maximize data utilization by incorporating data
from other participants. Participants will be rewarded for their contributions
to the joint inference as is standard in collaborative learning and federated
learning. Participants are often rewarded following a ``pay-per-use''
model~\cite{sim2020collaborative, yu2020fairness, toyoda2019mechanism,
martinez2019record}, where participants are compensated according to the
\textit{importance} of their contributed data in each joint
prediction~\cite{wu2016privacy,luo2020feature}.

\section{Related Work and Problem Statement}
\label{sec:adi}

\subsection{FL Security and ADI Positioning}
\label{subsec:motivation-adi}

\noindent \textbf{Distributed Machine Learning.}~Advances in distributed
machine learning systems have enabled large-scale machine learning algorithms to
be run in a distributed manner~\cite{provost1996scaling, li2014communication,
  peng2018optimus, cui2016geeps, leiflexgraph2021, zhenkundgcl2021}. Horizontal federated learning was
developed to provide an efficient and privacy-aware approach for
distributed machine learning~\cite{mcmahan2017communication}, in which clients
only share model updates rather than training data. The concept of federated
learning is expanded further into a vertical paradigm, namely
VFL~\cite{hardy2017private,nock2018entity,yang2019federated}, in which clients
hold separate features that belong to the same users.

\noindent \textbf{Attacks on FL.}~Existing work has
shown that representative deep learning attacks, such as backdoor and training
data reconstruction attacks, can be launched in collaborative learning settings.
Nevertheless, most existing works attack
HFL~\cite{chai2019towards,bagdasaryan2018backdoor,bhagoji2019analyzing,xie2019dba,sun2019can,wang2020attack,wang2019beyond,fang2020local,fung2020limitations,tolpegin2020data,nasr2019comprehensive}.
Recent works~\cite{weng2020privacy,luo2020feature} conduct private training data
leakage attacks via deliberately-modified gradients during VFL
training.~\cite{Fu2022Label} launches attacks to infer labels of private
training data, and~\cite{liu2020backdoor} inserts backdoor samples. All these
recent VFL attacks are launched during the VFL training phase. In contrast, this
work targets the joint inference phase of well-trained VFL models. ADIs enables
an adversary to control VFL joint prediction towards their will, reduce others'
contribution, and hog rewards used to incentivize participants. 

\noindent \textbf{Hardening Federated Learning.}~Bonawitz et
al.~\cite{bonawitz2017practical} introduce secure aggregation to defeat a
semi-honest server and the dropout of arbitrary users. Pillutla et
al.~\cite{pillutla2019robust} propose a robust aggregation scheme toward
corrupted client updates. FoolsGold mitigates sybil-based poisoning attacks
based on the diversity of client updates~\cite{fung2018mitigating}. Li et
al.~\cite{li2019fair} enhance the fairness of resource allocation in HFL.
Holistically, our present study assesses the unfair contribution of VFL
participants at the inference stage and the security implications.

\noindent \textbf{Connection Between ADIs and
AEs.}~Methods generating standard AEs in centralized learning or
HFL~\cite{goodfellow2014explaining,athalye2018synthesizing} can also be extended
to exploit VFL. 
And as illustrated in \F~\ref{fig:motivation}, AEs can also manipulate specific
VFL predictions to the target labels. Holistically, we deem ADIs and AEs are
related, in the sense that ADIs deem a novel extension of
AEs in the context of VFL.

\noindent \textbf{Differences Between ADIs and AEs.}~In the
VFL scenarios, AEs and ADIs differ in the following aspects. First, ADIs aim to
dominate practically all inputs from victim participants, while AEs usually
misclassify a specific input to a targeted label. As shown in
\F~\ref{fig:motivation}, when the victim participant takes another input, the
same AE is unlikely to manipulate the prediction again. Also, in formulating the
adversary's objective, AEs do not consider victim participants' contribution. In
contrast, ADIs explicitly minimize the victim participants' contribution, thus
hogging rewards used to incentivize collaborative learning participants
contributing important features~\cite{pandey2020crowdsourcing,zhan2020learning,zeng2021comprehensive,deng2021fair,song2019profit}. Moreover, in the context of
VFL, ADIs only need to manipulate a subset of features (e.g., in
\F~\ref{fig:motivation}, only FinTech's features are controlled), whereas AEs
control and manipulate the full feature space. In
\S~\ref{subsec:eval-compare-uae}, we empirically assess AEs in attacking VFL,
and show that standard AEs are not effective in dominating VFL outputs.

\noindent \textbf{ADIs vs. Other Perturbed
Inputs.}~Existing research has proposed methods in synthesizing
universal adversarial perturbation (UAP)~\cite{moosavi2017universal}. UAP
launches a synthesis procedure that causes misclassification of many images
using one unified perturbation. ADIs also result in a universal
misclassification. However, we clarify that ``universal'' in UAP and ADIs have
different meanings. UAP finds a universal perturbation to misclassify many
images, whereas we find ADIs from a subspace of the entire feature space to
\textit{dominate all other subspaces possessed by victim participants}. To
compare with UAP, ADIs explicitly model and minimize victim participants'
contributions. Also, while most UAP techniques misclassify images into arbitrary
labels~\cite{moosavi2017universal,chaubey2020universal}, we target a specified
label $l_{target}$ (see \F~\ref{fig:whitebox}), and only allow perturbing
adversary-controlled feature partition $X_{\mathcal{A}}^i$ instead of the entire
feature space $X^i = X_{\mathcal{A}}^i || X_{\mathcal{B}_1}^i || \cdots ||
X_{\mathcal{B}_{m-1}}^i$.

To empirically assess the comparison, we first generate UAPs on VFL
models trained on MNIST~\cite{lecun1998gradient}, and then evaluate the
dominating rates of inputs from \ma\ when using UAPs. The UAPs achieve a fooling
rate~\cite{moosavi2017universal} of around 90\%, which is considered as
effective according to~\cite{moosavi2017universal}. However, when \ma\ applies
UAPs on its inputs, none of the inputs can achieve a dominating threshold higher
than 95\% (this threshold is defined in \S~\ref{subsec:adi-discussion}), and
only 0.21\% of the inputs achieve a dominating threshold higher than 90\%. In
contrast, when \ma\ uses ADIs, over 33.9\% inputs can achieve a dominating
threshold higher than 95\% (see details in \S~\ref{sec:evaluation}). More
importantly, the generation of UAPs needs to iterate over the whole training
dataset, which is not practical in the setting of VFL systems, where the
malicious participant hardly has access to the whole training dataset.

One may wonder if adversarial participant \ma\ can simply use random
inputs to attack. Although random inputs can affect the joint inference in an
undesirable manner (and presumably reduce the VFL model accuracy), we argue that
random inputs are less profitable than ADIs for two reasons. First, ADIs allow
to specify an adversary-targeted label $l_{target}$ (see
\S~\ref{subsec:design-backdoor}), whereas random inputs simply ``muddle'' the
joint prediction. More importantly, random inputs can hardly nullify benign
participants' contributions, while ADIs mostly negate features contributed by
benign participants, thus hogging the rewards provided to incentivize
participants. We empirically compare ADI with random inputs in
\S~\ref{subsec:eval-compare-uae}, whose findings are aligned with our analysis
here.

\subsection{Threat Model}
\label{subsec:threat-model}

\noindent \textbf{Attack Scenario.}~This research considers $m$ participants in a VFL system $f$.
Our attack targets the inference phase of well-trained VFL models, \textit{not} the training phase. We assume
that in the joint inference phase, the inputs of one participant are controlled by the adversary. We refer to
this adversary as \ma, and the rest $m-1$ benign participants as
$\mathcal{B}_j$, where $j \in \{1, \cdots, m-1\}$. We also refer to the central
coordinator used by VFL as \mc. The feature space $X^{i}$ of a data
sample $i$ is thus partitioned into $X^{i} = X^{i}_{\mathcal{A}} ||
X^{i}_{\mathcal{B}_1} || \cdots || X^{i}_{\mathcal{B}_{m-1}}$ and held by
different participants.
To clarify, $f$ yields the score for the
target label in classification tasks, while it yeilds the regression results in
regression tasks.

\noindent \textbf{Adversary's Objectives.}~\ma\ aims to dominate the joint
inference using ADIs and extensively diminish the contribution of other 
participants. This way, \ma\ can control the result towards the direction of their
will and hog the vast majority of rewards, given that only
\ma\ contributes non-trivial features to the joint prediction.
We formally define ADIs as following:

\vspace{-0.4em}
\begin{definition}[ADIs]
Consider VFL system $f$ with $m$ participants: $f(X_{\mathcal{A}},
X_{\mathcal{B}_1}, \cdots, X_{\mathcal{B}_{m-1}})$, an input $X_{\mathcal{A}}^*$
is regarded as an ADI when the output of the VFL system is not influenced by
changing the inputs from participants $\mathcal{B}_j$, where $j \in \{1, \cdots,
m-1\}$. Formally, we regard $X_{\mathcal{A}}^*$ as an ADI when the following
holds:

\setlength{\belowdisplayskip}{4pt} \setlength{\belowdisplayshortskip}{4pt}
\setlength{\abovedisplayskip}{-5pt} \setlength{\abovedisplayshortskip}{-5pt}
\small
\begin{equation}
    \mathbb{V}_{X_{\mathcal{B}} \in D_{\mathcal{B}}} (f(X_{\mathcal{A}}^*, X_{\mathcal{B}_1}, \cdots, X_{\mathcal{B}_{m-1}})) \leq \epsilon \; ,
    \label{eq:defadi1}
\end{equation}
\normalsize

\noindent where $\epsilon$ is a small value, $\mathbb{V}$ represents the variance, and
$X_{\mathcal{B}} \in D_{\mathcal{B}}$ is short for $X_{\mathcal{B}_1} \in
D_{\mathcal{B}_1}, \cdots, X_{\mathcal{B}_{m-1}} \in D_{\mathcal{B}_{m-1}}$.
\label{def:ADI}
\end{definition}
\vspace{-0.4em}

\E~\ref{eq:defadi1} illustrates that $X_{\mathcal{A}}^*$ becomes an ADI
when the output variance is bounded by $\epsilon$ w.r.t.~the change of
inputs from $\mathcal{B}_1, \cdots, \mathcal{B}_{m-1}$. This way,
\ma\ \textit{dominates} the joint prediction, no matter what inputs are used by other
participants.

\noindent \textbf{Adversary's Capability \& Assumptions.}~We assume
that adversary \ma\ can arbitrarily perturb its inputs to generate ADIs. 
And in the VFL joint inference stage, all inputs of \mb\ follow the
distribution of their training dataset.
To generate ADIs, we assume that in the \textit{joint inference phase}, \ma\
provides a \textit{tiny} collection of data samples to \mb, which are used as
\mb's test inputs. We refer to this \textit{tiny} data sample collection as \sd.
\sd\ does not deviate from the distribution of \mb's standard inputs.

We find that to generate ADIs that dominate at least 95\% of inputs in \mb\
(threshold ``95\%'' is clarified in \S~\ref{subsec:adi-discussion}), we only
need a tiny \sd\ whose elements are randomly selected from the test dataset of
\mb. \sd\ is never revealed to \mb\ during training. \S~\ref{sec:evaluation}
reports that we require only 20 MNIST images (i.e., 0.03\% of MNIST) to dominate
95\% of its standard test inputs. We study how \ssd\ influences ADI synthesis in
Appendix \ref{subsec:eval-subset-size} and discuss the selection of \sd\ in Appendix \ref{subsec:append-discuss}.

Additionally, we show that \ma\ can successfully generate ADIs even
without accessing the inputs of \mb. Instead, \ma\ only knows the \textit{input
range} of \mb's inputs. This observation makes the ADI generation algorithm more
practical and further enhances the technical feasibility and stealthiness in
real-world attacks. See evaluations in \S~\ref{subsec:weaker-threat-model}.

\S~\ref{subsec:adi} proves the existence of ADIs in VFL. We assume that the probability density function of the benign participants'
data can be approximated by Gaussian Mixture Models
(GMMs)~\cite{mclachlan1988mixture}, which is aligned with the
conventions~\cite{lee2017deep, makhzani2015adversarial, hjelm2018learning,
belghazi2018mutual, fei2006one, zhu1996region} in machine learning. This
assumption is reasonable as GMMs have the power to approximate any smooth
distributions~\cite{goodfellow2016deep}. We also prove for the non-smooth distribution scenario in Appendix \ref{subsec:append-proof}.

\noindent \textbf{Examples and Feasibility Clarification.}~Consider the VFL
example in \F~\ref{fig:motivation}, where a bank and a FinTech jointly predict a
user's credit score. An adversarial FinTech can use ADIs to control the joint
inference; for instance, the joint inference can be forced to always yield high
credit scores, despite poor financial records (``income'' in
\F~\ref{fig:motivation}) the users may have in the bank. Given a well-trained
VFL model, we assume that it is feasible for \ma\ to provide the tiny collection
of data samples \sd\ in the following way: \ma\ can conspire with a
\textit{small} group of users who behave normally when generating the data to
form \sd. These users submit \sd\ to \mb\ and also disclose \sd\ to \ma\ (see Appendix \ref{sec:workflow} for an end-to-end diagram of this strategy). We
envision that regular VFL participants (e.g., banks) would not refuse to take
$S$ as \textit{test} inputs, especially considering its tiny size. 


\noindent \textbf{ADI Synthesis and Discovery.}~We first propose an ADI
synthesis attack in the blackbox setting, which is launched during the inference
phase of well-trained VFL models. The ADI synthesis is based on gradients, which
are estimated with finite difference method (FDM)~\cite{grossmann2007numerical}
in the blackbox setting. The adversarial participant \ma\ only needs to know the
current input index (i.e., user ID) and observes the VFL inference results,
which is a common setting in VFL. \ma\ has no access to model $M_{\mathcal{B}}$
or $\mathcal{B}$'s private training data. 

We also propose a fuzz testing framework to discover ADIs
(\S~\ref{sec:fuzz-detection}). Fuzzing helps VFL developers comprehensively
discover ADIs and perform quantitative security assessment (which is hardly feasible for our blackbox
approach given the privacy consideration; see
\T~\ref{tab:assumption}). For fuzzing, \ma\ does not need to access \sd.
Instead, participants share a saliency score to guide fuzzing (comparable to how
code coverage guides software fuzzing~\cite{afl}). This saliency score is a
number derived from the saliency map's $l_1$ norm. That is, to protect
participants' data privacy, information of \textit{low sensitivity} is shared
instead of the entire saliency maps.

We summarize assumptions in \T~\ref{tab:assumption}, where \cmark\ means that
\ma\ requires the corresponding accessibility.
Greybox fuzzing is guided by saliency scores, whereas blackbox ADI synthesis
denotes a lightweight generation process under objectives. We clarify details of
two approaches in \S~\ref{subsec:design-backdoor} and
\S~\ref{sec:fuzz-detection}, respectively.
Readers may also refer to Appendix \ref{sec:workflow} for an end-to-end illustration of
blackbox ADI synthesis procedures.
And we also evaluate the scenario where \ma\ cannot access \sd\ in \S~\ref{subsec:weaker-threat-model}.


\begin{table}[t]
    \vspace{-2pt}
  \captionsetup{skip=2pt}
  \centering
  \scriptsize
  \caption{\ma's accessibility under different schemes.}
  \label{tab:assumption}
  \setlength{\tabcolsep}{1.5pt}
  \begin{adjustbox}{max width=\linewidth}
  \begin{threeparttable}
    \begin{tabular}{l|c|c}
      \hline
          & \textbf{Blackbox ADI Synthesis} &\textbf{Greybox Fuzz Synthesis} \\
      \hline
            \textbf{\mb's model}  & \xmark &\xmark \\
            \textbf{\mb's traing dataset} & \xmark & \xmark \\
            \textbf{\textit{tiny} test dataset \sd\tnote{1}} & \cmark &\xmark \\
            \textbf{low-sensitivity saliency score of \mb} & NA & \cmark\\
      \hline
            \textbf{Collaboration of all participants} & \xmark & \cmark\\
      \hline
    \end{tabular}
    \footnotesize
    \begin{tablenotes}[flushleft]
        \item[1] \ma\ feeds \sd\ to \mb\ by conspiring with a few users, as
          clarified in threat model. In our implementation, \sd\ comprises a few
          random samples from \mb's test dataset.
    \end{tablenotes}
  \end{threeparttable}
  \end{adjustbox}
\vspace{-15pt}
\end{table}



\subsection{Proof on the Existence of ADIs in VFL}
\label{subsec:adi}

Below, we prove that ADIs exist in two VFL systems: HeteroLR and SplitNN. As
noted in \S~\ref{sec:background}, most parameterized VFL systems (e.g., VFVQA)
can be extended from them. 
\S~\ref{subsec:adi-discussion} discusses the generalization. Below, we consider
two participants and discuss extension to more participants in
\S~\ref{subsec:adi-discussion}.

\noindent \textbf{Variance Bound for VFL Systems.}~Participant \ma\ holds the
data $X_{\mathcal{A}} \in D_{\mathcal{A}}$, whereas participant \mb\ holds the
data $X_{\mathcal{B}} \in D_{\mathcal{B}}$ following any distribution whose
density function is $p(X_{\mathcal{B}})$. As noted in our threat
model, we make an reasonable assumption that this density function can be
approximated by GMMs. \ma\ is controlled by an adversary, whereas \mb\ behaves
normally. The VFL system $f$ takes $X_{\mathcal{A}}$ and $X_{\mathcal{B}}$ as
inputs, and its output can be expressed as $f(X_{\mathcal{A}},
X_{\mathcal{B}})$, where $X_{\mathcal{A}} \in \mathbb{R}^{d_1}$,
$X_{\mathcal{B}} \in \mathbb{R}^{d_2}$, and $d_1, d_2$ denote dimensions of the
features in \ma\ and \mb. Same as Def.~\ref{def:ADI}, $f$ yields the score for the
target label in classification tasks, and the regression results in
regression tasks. We present the output variance bound of the VFL system:

\begin{corollary}[Variance Bound of VFL systems]
    With fixed input $X_{\mathcal{A}}^*$ and varying input $X_{\mathcal{B}}$ following any distribution $D_{\mathcal{B}}$ whose density function is $p(X_{\mathcal{B}})$.
    For any $\epsilon$, there exists $X_{\mathcal{A}}^*$ that the output variance of the VFL system (SplitNN, HeteroLR, and their extensions) with respect to $X_{\mathcal{B}} \in D_{\mathcal{B}}$ is bounded by $\epsilon$:
    
    \small
    \begin{equation*}
        \begin{aligned}
            \mathbb{V}_{X_{\mathcal{B}} \in D_{\mathcal{B}}} &(f(X_{\mathcal{A}}^*, X_{\mathcal{B}})) 
            \leq \epsilon
        \end{aligned}
    \end{equation*}
    \normalsize
    
    \label{cor:vfl-var-bound}
    
\end{corollary}

According to Cor.~\ref{cor:vfl-var-bound} and Def.~\ref{def:ADI}, the ADI $X_{\mathcal{A}}^*$ exists for VFL systems.
We present detailed proof of Cor.~\ref{cor:vfl-var-bound} for the widely used VFL systems of HeteroLR and SplitNN in Appendix \ref{subsec:append-proof}.

\subsection{Extension \& Practical Consideration}
\label{subsec:adi-discussion}

\noindent \textbf{Generalization.}~\S~\ref{sec:background} has clarified that
most parameter-based VFL systems are extensions of HeteroLR and SplitNN. These
models all concatenate features from the same data sample to form a joint
inference. 
Thus, we assume that Cor.~\ref{cor:vfl-var-bound} and our proof in Appendix \ref{subsec:append-proof} subsumes most real-world
cases because most parameter-based VFL systems are extensions of HeteroLR and SplitNN. Our evaluation
(\S~\ref{sec:evaluation}) also attacks VFVQA, an extension of SplitNN. We also
discuss attack tree-based VFL in Appendix \ref{subsec:append-discuss}.

\noindent \textbf{Bounded Mutation.}~We have proved the existence of ADIs when
arbitrarily mutating $X_{\mathcal{A}}^*$. Nevertheless, arbitrary mutation can
generate unrealistic ADIs, which may not be desirable in real-world
exploitations. We design a bounded mutation scheme to generate more realistic
inputs. Bounded mutation perturbs $X_{\mathcal{A}}^*$ within a predefined range.
In this paper, we perturb $X_{\mathcal{A}}^*$ within its variance in the
standard input dataset on \ma. Note that ``variance'' in bounded mutation is
obtained from the standard inputs on \ma. Variance of standard inputs on \mb\ is
\textit{not} needed. 
We also prove the existence of ADIs under bounded mutation in Appendix \ref{sec:adi-bounded}. 
The generated ADIs are seen as indistinguishable
from normal inputs, indicating a severe, practical, yet overlooked issue to VFL.
We empirically evaluate the stealth of ADIs generated by bounded mutation in
\S~\ref{subsec:whitebox-attack} and \S~\ref{subsec:greybox-fuzzing}.

\noindent \textbf{More Participants.}~As introduced in threat model
(\S~\ref{subsec:threat-model}), we use ADIs to attack a VFL system of $m$
participants, where $m$ can be greater than two. While Cor.~\ref{cor:vfl-var-bound}
is based on two participants \ma\ and \mb, our proofs can be easily extended to
multi-participant VFL systems. To do so, features on $m-1$ benign participants
\mb\ can be first aggregated into $X_{\mathcal{B}}$ to bridge with
Cor.~\ref{cor:vfl-var-bound}. We also empirically
assess the presence of ADIs with various numbers of malicious and benign participants in Appendix \ref{subsec:eval-client-numbers} and Appendix \ref{subsec:multiple-attacker}.

\noindent \textbf{Practical Assessment of ADIs.}~Despite the inherent existence
of ADIs in VFL, it is difficult to obtain a complete view
of \mb's inputs. Hence, we use the following practical assessment to decide if $X_{\mathcal{A}}^*$ denotes an ADI:

\vspace{-0.4em}
\begin{definition}[Practical Assessment of ADIs]
Consider VFL with $m$ participants, $F(X_{\mathcal{A}},
X_{\mathcal{B}_1}, \cdots, X_{\mathcal{B}_{m-1}})$. $F$ outputs prediction label with the largest score. An input $X_{\mathcal{A}}^*$
is an ADI when the VFL joint prediction is not influenced by
changing the inputs from participants $\mathcal{B}$. Formally:

\setlength{\belowdisplayskip}{5pt} \setlength{\belowdisplayshortskip}{5pt}
\setlength{\abovedisplayskip}{-0pt} \setlength{\abovedisplayshortskip}{-0pt}
\small
\begin{equation*}
    r(X_{\mathcal{A}}^*) = \frac{\sum_{i=1}^{n} \mathbb{I}(F(X_{\mathcal{A}}^*, X_{\mathcal{B}_1}^i, \cdots, X_{\mathcal{B}_{m-1}}^i) = l_{target})}{n} \geq x\% \; ,
\end{equation*}
\normalsize

\noindent where $r$ computes the dominated proportion of \mb's inputs for $X_{\mathcal{A}}^*$, $x\%$ is the dominating threshold, $n$ is the samples' number in test dataset, and $\mathbb{I}(msg) = 1$, if $msg$ is true, otherwise, $\mathbb{I}(msg) = 0$.

\label{def:adi-practical}

\end{definition}
\vspace{-0.5em}

Our study adopts this practical assessment, where an input of \ma\
deems an ADI if the joint inference is confined as an attacker-specified label
$l_{target}$ for $x\%$ of the data samples in the standard input dataset of \mb.
$x$ is empirically decided as 95 and 99, indicating that an ADI can extensively
dominate 95\% and 99\% of the inputs from \mb, respectively.

This section has formulated the research problem and presented discussions from
various aspects. \S~\ref{subsec:design-backdoor} describes gradient-based
algorithms to synthesize ADIs. \S~\ref{sec:fuzz-detection} proposes a fuzz
testing-based approach to uncovering ADIs.



\section{Blackbox ADI Synthesis}
\label{subsec:design-backdoor}

\setlength{\textfloatsep}{10pt}
\begin{algorithm}[t]
    \caption{Gradient-Based ADI Synthesis.}
    \footnotesize
    \label{alg:attacking}
      \begin{algorithmic}[1]
        \Function{\code{Saliency_est}}{$X_{\mathcal{A}}^*, X_{\mathcal{B}}^i, M$}
            \State $output \leftarrow M(X_{\mathcal{A}}^*, X_{\mathcal{B}}^i)$ \;
            \State $saliency \leftarrow ||\frac{\partial{Var(output)}}{\partial{X_{\mathcal{B}}^i}}||_1$ \;
            \State \Return $saliency$
        \EndFunction
        
    \Function{\code{ADI_generation}}{$X_{\mathcal{A}}^*, M, l_{target}, Strategy, \mdutchcal{S}$}
        \LeftComment{~~$r$: computes dominated proportion of \mb's inputs as defined in Def.~\ref{def:adi-practical}}
        \LeftComment{~~$x\%$, $T$: desired dominating threshold and maximum round}
        \LeftComment{~~$\alpha, \beta, \gamma, \sigma$: weight parameters}
        \LeftComment{~~$\boldsymbol{\Lambda}$, $V$, $\delta_t$: mutation constrain, total mutation, mutation in round $t$}
        \LeftComment{~~$loss(X_\mathcal{A}^i, X_\mathcal{B}^i, M, l)$: loss between $M(X_\mathcal{A}^i, X_\mathcal{B}^i)$ and label $l$}
        \State $V \leftarrow 0, \, \delta_1 \leftarrow 0, \, t \leftarrow 1$ \;
        \While{$r(X_{\mathcal{A}}^*) \leq x\%$ and $t \leq T$}
            \For{each $X_\mathcal{B}^t \in \mdutchcal{S}$}
                \If{$Strategy$ is \textsc{Random}}
                \State \begin{math}
                    \begin{aligned}
                    \delta_t &\leftarrow {\arg \min}_{\delta_t} \; \alpha \, \code{Saliency\_est}(X_{\mathcal{A}}^* + V + \delta_t, X_\mathcal{B}^t, M) \\ 
                    &+ {\beta \, loss(X_{\mathcal{A}}^* + V + \delta_t, X_\mathcal{B}^t, M, l_{target})}
                    \end{aligned}
                \end{math}
                \Else
                \State \begin{math}
                    \begin{aligned}
                    \delta_t &\leftarrow {\arg \min}_{\delta_t} \; \alpha \, \code{Saliency\_est}(X_{\mathcal{A}}^* + V + \delta_t, X_\mathcal{B}^t, M) \\
                    &+ {\beta \, loss(X_{\mathcal{A}}^* + V + \delta_t, X_\mathcal{B}^t, M, l_{target})} \\
                    &+ \gamma \, ||\delta_t||_2 \qquad  \text{s.t.} \, |V + \delta_t| \leq \boldsymbol{\Lambda}\\ 
                    \end{aligned}
                \end{math}
                \EndIf
                \State $\delta_t \leftarrow \sigma \, \delta_{t-1} + \delta_t, \, V \leftarrow V + \delta_t, \, t \leftarrow t + 1$ \;
            \EndFor
        \EndWhile
        \State \Return $X_{\mathcal{A}}^* + V$ \;
        \EndFunction
      \end{algorithmic}
    \end{algorithm}

This section proposes a blackbox, gradient-based ADI synthesis algorithm in line
with our threat model in \S~\ref{subsec:threat-model}. Inputs of \ma\ are
mutated by the adversary, who aims to dominate the joint inference and minimize
\mb's contribution. To assist ADI synthesis, \ma\ can use a tiny collection of data samples \sd\
aligned with the distribution of \mb's normal inputs. 

\noindent \textbf{Clarification.}~To \textit{ease understanding}, we start
by presenting a whitebox ADI synthesis algorithm, where \ma\ is assumed to
access the gradients of the trained models in \mb. Then, aligned with our threat
model, we discuss turning the whitebox setting into a blackbox attack by
estimating gradients.

\A~\ref{alg:attacking} formulates ADI generation as an
optimization problem. \code{\textsc{ADI\_Generation}} is the main entry point of
our algorithm, which takes the adversary-controlled input $X_{\mathcal{A}}^*$,
jointly trained model $M$, adversary-targeted label $l_{target}$, the tiny collection of data samples
\sd, and a mutation strategy $Strategy$ as the inputs. It returns the
synthesized ADI $X_{\mathcal{A}}^* + V$, where $V$ is the mutation vector over $X_{\mathcal{A}}^*$.

\noindent \textbf{Mutation Methods.}~\code{\textsc{ADI\_Generation}} requires to
specify mutation strategies (i.e., random vs. bounded mutation) and proceed
accordingly (lines 9--12 in \A~\ref{alg:attacking}). Given an input $X_{\mathcal{A}}^*$, random mutation
perturbs it in all directions, whereas bounded mutation perturbs it in a
predefined range. Bounded mutation is more conservative, such that the value of
mutated $X_{\mathcal{A}}^*$ is confined in a reasonable range.
Although defining a ``reasonable'' range for arbitrary $X_{\mathcal{A}}^*$ is
challenging, we specify that mutations must be bounded by the variance of this
feature in the training dataset.
We solve the optimization problem using projected gradient descent~\cite{alhajjar2021adversarial}. 

Random mutation aims to minimize the output loss on a target label and the
contribution of inputs in \sd\ (line 10 in \A~\ref{alg:attacking}; see
below for the description of \code{\textsc{Saliency\_Est}}). In contrast,
bounded mutation adds a penalty on the mutation to the objective (line 12 in \A~\ref{alg:attacking}). To
facilitate a faster convergence, we use momentum (lines 13 in \A~\ref{alg:attacking}) to help accelerate
updates in the right directions.
Our evaluation, as will be reported in \T~\ref{tab:mutation}, empirically
compares these two mutation strategies. As expected, random mutation manifests a
higher chance of generating ADIs, whereas bounded mutations induce
more stealthy ADIs (see the case studies in \S~\ref{sec:evaluation}).
\code{\textsc{ADI\_Generation}} supports both mutation schemes and the users can select
the one that best suits their needs.

\noindent \textbf{Estimating Contributions of
  \mb.}~\code{\textsc{Saliency\_Est}} computes the saliency score, denoting the
contribution of $X_{\mathcal{B}}^i$ in a joint inference. We first calculate the
derivative of the output's variance for $X_{\mathcal{B}}^i$, whose $l_1$ norm is derived to estimate its saliency (line 3 in \A~\ref{alg:attacking}). 
Note that in binary classification, the output vector contains the prediction score for only one category, then we compute the derivative of the output for $X_{\mathcal{B}}^i$ directly.
A
lower saliency score means that a modest change in $X_{\mathcal{B}}^i$ has
minimal effect on the output, implying a negligible contribution of \mb\ to the
inference. This procedure aligns with the adversary's objectives in \S~\ref{sec:adi}.

\subsection{A Schematic View of ADI Generation}
\label{subsec:uae}

\begin{figure}[!t]
  \captionsetup{skip=2pt}
  \centering
  \includegraphics[width=0.65\linewidth]{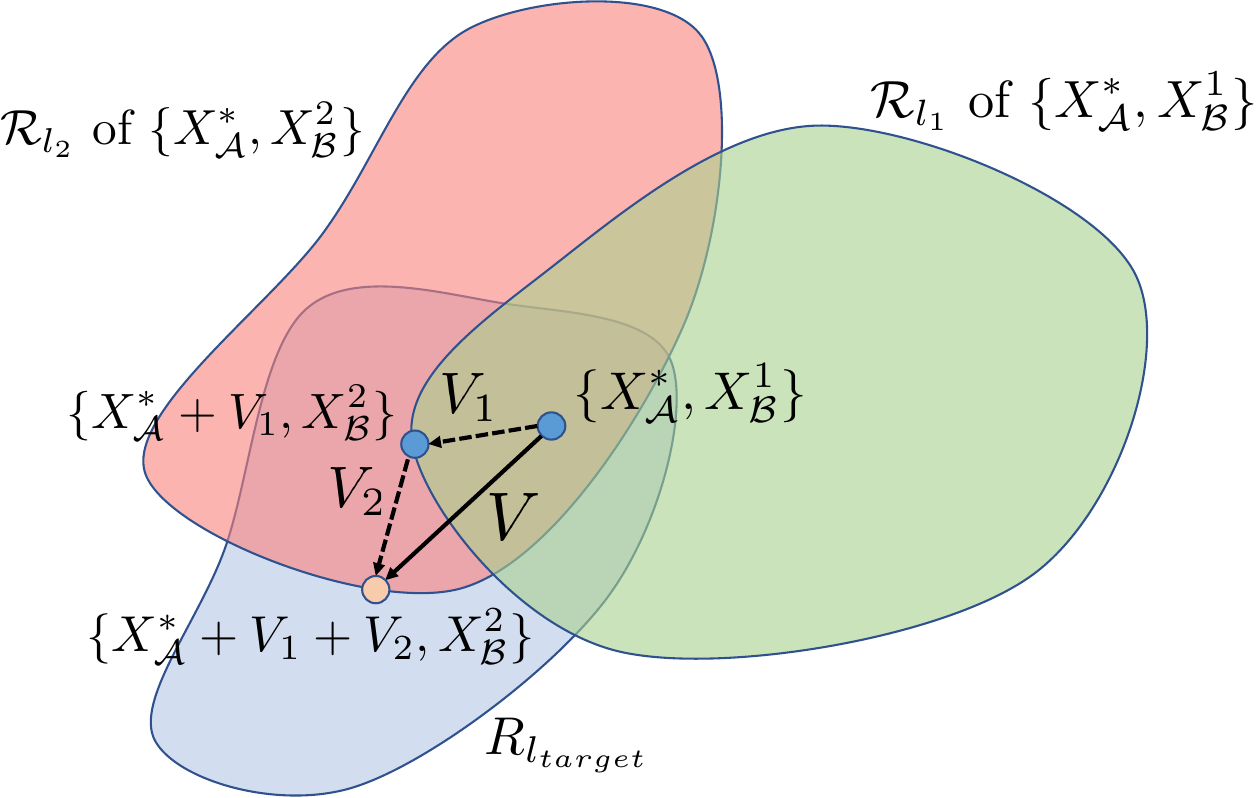}
  \caption{A schematic view of ADI generation.}
  \label{fig:whitebox}
\end{figure}

\noindent \F~\ref{fig:whitebox} presents a schematic view to synthesize ADIs,
where we proceed iteratively over a subset of \mb's inputs and gradually
generate ADIs as the input of \ma. For a normal joint inference, both \ma's and
\mb's inputs make non-trivial effects: when fixing \mb's input as
$X_{\mathcal{B}}^1$, perturbing $X_{\mathcal{A}}^*$ toward an ADI can form a
high-dimensional classification region $R_{l_1}$, where the joint inference over
$\{X_{\mathcal{A}}^*, X_{\mathcal{B}}^1\}$ constantly yields $l_1$. Similarly,
input $X_{\mathcal{B}}^2$ and $X_{\mathcal{A}}^*$ can form another region
$R_{l_2}$, whose induced inference is constantly $l_2$.

Let \sd\ be a dataset containing a non-trivial amount of \mb's input samples, we
assume the presumed existence of a special classification region
$\mathcal{R}_{l_{target}}$. When $\{X_{\mathcal{A}}^*, X_{\mathcal{B}}^i\}$
where $X_{\mathcal{B}}^i \in \mdutchcal{S}$ is inside its corresponding boundary
$\mathcal{R}_{l_i}$ and meanwhile inside $\mathcal{R}_{l_{target}}$, the joint
inference yields label $l_i$. Nevertheless, when $\{X_{\mathcal{A}}^*,
X_{\mathcal{B}}^i\}$ locates within $\mathcal{R}_{l_{target}}$ while outside any
other regions $\mathcal{R}_{l_{other}}$, where $l_{other} \neq
l_{target}$, the joint inference yields $l_{target}$.
Thus, given $\{X_{\mathcal{A}}^*, X_{\mathcal{B}}^1\}$ locates inside
$\mathcal{R}_{l_1}$ in \F~\ref{fig:whitebox}, perturbation $V_1$ sends the
currently perturbed point, $\{X_{\mathcal{A}}^* + V_1, X_{\mathcal{B}}^1\}$,
outside region $\mathcal{R}_{l_1}$ and gets inside the region
$\mathcal{R}_{l_{target}}$. When further dealing with $X_{\mathcal{B}}^2$ paired
with $X_{\mathcal{A}}^* + V_1$, mutation $V_2$ sends the perturbed point,
$\{X_{\mathcal{A}}^* + V_1 + V_2, X_{\mathcal{B}}^2\}$, outside region
$\mathcal{R}_{l_2}$ and gets inside the region $\mathcal{R}_{l_{target}}$. By
iterating \sd\ and aggregating perturbation $V_i$ into $V$, pairing
$X_{\mathcal{A}}^{*} + V$ with inputs in \sd\ will be sent inside the region
$\mathcal{R}_{l_{target}}$ but presumably outside of any other regions. We thus
control the joint prediction when pairing $X_{\mathcal{A}}^* + V$ with
$X_{\mathcal{B}}^i \in \mdutchcal{S}$. Moreover,
Appendix \ref{subsec:eval-subset-size} empirically shows that $X_{\mathcal{A}}^* + V$
generated over \sd\ achieves a high success rate of dominating at least 95\% of
\mb's inputs. We deem $X_{\mathcal{A}}^* + V$ as an ADI.

\subsection{Gradient Estimation in the Blackbox Setting}
\label{subsec:blackbox-adi}


The machine learning community has proposed blackbox gradient estimation
approaches. Such methods are used in generating adversary
examples~\cite{chen2017zoo, ilyas2018black, tramer2017ensemble,
narodytska2016simple}. We use a gradient estimation-based approach to
synthesizing ADIs, so \ma\ does not need to access \mb's trained models. We use
the finite difference method (FDM)~\cite{grossmann2007numerical} to estimate
gradient:

\setlength{\belowdisplayskip}{4pt} \setlength{\belowdisplayshortskip}{4pt}
\setlength{\abovedisplayskip}{-3pt} \setlength{\abovedisplayshortskip}{-3pt}
\small
\begin{equation*}
grad_{\mathcal{B}} = \frac{Var(M(X_{\mathcal{A}}^*, X_{\mathcal{B}}^i + \delta
  )) - Var(M(X_{\mathcal{A}}^*, X_{\mathcal{B}}^i))}{\delta} \; ,
\end{equation*}
\normalsize

\noindent where $M$ is the joint model, $X_{\mathcal{A}}^*$ and
$X_{\mathcal{B}}^i$ are the inputs fed to \ma\ and \mb. $\delta$ denotes a small perturbation with the same dimension as $X_{\mathcal{B}}^i$.
Thus, the saliency gradient in \A~\ref{alg:attacking} (line 3) can be
approximated. With other parts remaining the same in \A~\ref{alg:attacking}, we can
synthesize ADIs in the blackbox setting. As mentioned in
\S~\ref{subsec:threat-model}, \ma\ only needs to provide the tiny data
collection $S$ to \mb, and $S$ is formed by $X_{\mathcal{B}}^i$ and
$X_{\mathcal{B}}^i + \delta$.

\noindent \textbf{Clarification.}~Conventional blackbox AE attacks denote an
online setting~\cite{ilyas2018black, guo2019simple, Suya2020Hybrid}, where they
require attackers to iteratively query a remote model (e.g., a cloud service)
with recently mutated inputs and decide further mutations with estimated
gradiants. In contrast, for blackbox ADI synthesis, \ma\ needs to acquire the
joint inference results only over $X_{\mathcal{B}}^i \in \mathcal{S}$ and
$X_{\mathcal{B}}^i + \delta \in \mathcal{S}$. \ma\ pre-computes $\mathcal{S}$
\textit{offline}, and as clarified in \S~\ref{subsec:threat-model}, \ma\ then
conspires with a small group of users, who submit $\mathcal{S}$ to \mb\ as a
collection of test inputs. Therefore, we do not require the VFL to serve as a
``cloud service'' that actively processes unseen data (which is impractical in
VFL scenarios). All data samples in $\mathcal{S}$ are computed \textit{offline}
and submitted to \mb, before synthesizing ADIs. \textbf{In sum, our blackbox
attack is faithfully aligned with how VFL is used in real-life scenarios}; see Appendix \ref{sec:workflow} for an end-to-end illustration of blackbox attack with diagrams.
\section{Uncover ADIs with Greybox Fuzzing}
\label{sec:fuzz-detection}

\noindent \textbf{Motivation.}~We clarify that excellent research has been done
on testing distributed systems~\cite{lukman2019flymc, banabic2012fast,
yuan2020effective}; these works primarily focus on concurrency bugs. Contrarily,
we launch privacy-preserving testing in the context of federated learning to
uncover ADIs. The gradient-based ADI synthesis algorithm described in
\S~\ref{subsec:design-backdoor} can also be used to uncover ADIs. However,
in-house quality assurance and vulnerability assessment are difficult, whose
main reasons are twofold.

First, while malicious participant \ma\ can synthesize ADIs without accessing
the model of \mb\ in blackbox ADI synthesis (see \S~\ref{subsec:blackbox-adi}),
the malicious participant needs to prepare a tiny dataset \sd. As
clarified in \S~\ref{subsec:threat-model} and \S~\ref{subsec:blackbox-adi}, \sd\
can be prepared by conspiring with a few users to submit \sd\ to \mb. However,
for ethical users and developers, \sd\ cannot be collected in the same way, as
sharing user data may compromise privacy even if \sd\ is minimal.

This section designs greybox fuzzing to uncover ADIs for VFL vulnerability
assessment without using \sd\ but under the collaboration with \mb. Many studies
have examined using testing to find inputs that can manipulate DNN
predictions~\cite{Peideepxplore,10.1145/3293882.3330579}. Nonetheless, we
investigate a novel design point --- \textit{efficient and privacy-preserving
fuzz testing in VFL.}

\subsection{Design of Greybox Fuzz Testing}
\label{subsec:design-testing}

Greybox software fuzzing is an evolutionary process in which inputs that reveal
new code coverage are retained for more mutations until vulnerabilities are
detected. In VFL, we consider an input $X_{\mathcal{A}}^*$ of participant
\ma\ interesting if $X_{\mathcal{A}}^*$ reduces contribution of other
participants to the joint prediction. Although the ``contribution'' can be
revealed using saliency scores, VFL systems do not allow such disclosure.

Inspired by our gradient-based ADI synthesis approach
(\S~\ref{subsec:design-backdoor}), we extend the standard VFL systems by
allowing participants to compute and share saliency scores to guide greybox
fuzzing; the shared saliency score serves the feedback for fuzzing, which is
comparable to how code coverage is used to guide software fuzzing~\cite{afl}. We
assume that:

\begin{tcolorbox}[size=small]
Although sharing saliency may raise potential concerns of privacy violation, to
assess ADI attack vectors before security breaches, normal participants should
have enough incentive to collaborate and share saliency scores.
\end{tcolorbox}
\vspace{-3pt}

Again, this stage allows VFL developers to launch \textbf{in-house vulnerability
  assessment by comprehensively uncovering ADIs}. The saliency score is derived
  from the the saliency map's $l_1$ norm, which has \textit{limited
  sensitivity}. Privacy leakage due to saliency score, though theoretically
  possible, is low in practice. More importantly, with over two participants,
  secure aggregators may be used to shield each participant's saliency score and
  increase privacy~\cite{bonawitz2017practical,bell2020secure}.

\setlength{\textfloatsep}{10pt}
\begin{algorithm}[!t]
\caption{Saliency-Guided Greybox Fuzz Testing.}
    \footnotesize
\label{alg:fuzzing}
  \begin{algorithmic}[1]
    \Function{\code{IsADI}}{$X_{\mathcal{A}}^*, \mdutchcal{S}, l_{target}, M$}
      \State $\mathcal{T} \leftarrow \varnothing$
      \For{each $X_{\mathcal{B}}^i$ in \sd}
        \State $o \leftarrow M(X_{\mathcal{A}}^*, X_{\mathcal{B}}^i)$
        \State add $o$ in $\mathcal{T}$
      \EndFor
      \State \Return \textsc{Stable}($\mathcal{T}, l_{target}$)
    \EndFunction
    
\Function{\code{Fuzzing}}{Corpus of Seed Inputs $\mathcal{C}$, $M, \mdutchcal{S}$}
    \State $\mathcal{Q} \leftarrow \mathcal{C}$, $\mathcal{O} \leftarrow \varnothing$
    \For{1 ... \textit{MAX\_ITER}}
      \State $(X_{\mathcal{A}}^* , l_{target}) \leftarrow $ \textsc{ChooseNext}($\mathcal{Q}$)
      \State $p \leftarrow $ \textsc{AssignEnergy}($X_{\mathcal{A}}^* $)
      \For{1 ... $p$}
        \State $X_{\mathcal{A}}^* \leftarrow $ \textsc{Mutate}($X_{\mathcal{A}}^*, \mdutchcal{S}, l_{target}, M$)
        \If{\code{\textsc{IsADI}}($X_{\mathcal{A}}^*, \mdutchcal{S}, l_{target}, M$) == $true$}
          \State add $(X_{\mathcal{A}}^*, l_{target})$ in $\mathcal{O}$
        \ElsIf{\textsc{ReduceSaliency}($X_{\mathcal{A}}^*, l_{target}, M$) == $true$}
          \State add $(X_{\mathcal{A}}^*, l_{target})$ in $\mathcal{Q}$
        \EndIf
      \EndFor
    \EndFor
    \State \Return ADI Set $\mathcal{O}$
    \EndFunction
  \end{algorithmic}
\end{algorithm}

\A~\ref{alg:fuzzing} depicts the high-level procedure of our feedback-driven
fuzzing. We clarify full implementation details of cooperating fuzzing with VFL
in Appendix \ref{sec:fuzzing-vfl}.
\code{\textsc{Fuzzing}} is the entry point, where \code{\textsc{IsADI}} checks
whether a given input $X_{\mathcal{A}}^*$ of \ma\ can dominate other
participants \mb's contributions. \code{\textsc{IsADI}} allocates a set to
collect all joint inference outputs (line 2). It then iterates all inputs
$X_{\mathcal{B}}^i \in \mdutchcal{S}$ used by \mb\ and records each joint
inference output $o$ (lines 4--5). \textsc{Stable} checks whether outputs in
$\mathcal{T}$ are identical to its target label $l_{target}$.

\code{\textsc{Fuzzing}} accepts a corpus of seeds $\mathcal{C}$ to initialize
queue $\mathcal{Q}$ (line 8). This function also takes in the jointly trained
model $M$ and \sd. For fuzzing, inputs in \sd\ are \textit{hidden} from \ma.
Rather, \mb\ uses inputs in \sd\ to cooperate with \ma\ and facilitate
\code{\textsc{IsADI}} and \textsc{Mutate} (line 13 in \A~\ref{alg:fuzzing}).
We use $\mathcal{O}$ to store identified ADIs (line 8). The entire campaign is
subjected to \textit{MAX\_ITER} iterations, and for each iteration, we pick one
input $X_{\mathcal{A}}^*$ and its target label $l_{target}$ by popping
$\mathcal{Q}$ (line 10) and determine \#mutation by function $\textsc{AssignEnergy}$ over $X_{\mathcal{A}}^*$ (line
11). \textit{MAX\_ITER} is currently 5,000, and each $X_{\mathcal{A}}^*$ has a
fixed ``energy'' of $\textsc{AssignEnergy}(X_{\mathcal{A}}^*) = 20$.

During each iteration, we generate a new variant $X_{\mathcal{A}}^*$ by mutating
the original $X_{\mathcal{A}}^*$ (line 13). When ADIs are found (line 14), we
add the input $X_{\mathcal{A}}^*$ into $\mathcal{O}$. 
$X_{\mathcal{A}}^*$ is ``interesting'' by successfully reducing the saliency of
\mb; we thus add $X_{\mathcal{A}}^*$ to the queue for further mutations (lines
16--17). \A~\ref{alg:fuzzing} will return all uncovered ADIs for users to assess
security and attack interface of their VFL systems (line 18).

\noindent \textbf{Design \textsc{ReduceSaliency} (Line 16).}~We decide whether
input $X_{\mathcal{A}}^*$ of \ma\ is prone to becoming an ADI by assessing its
influence on \mb, which is modeled using the saliency scores of \mb. As
previously discussed, we assume that the saliency score is faithfully shared by
\mb. Software greybox fuzzing~\cite{afl} looks for inputs that achieve new code
coverage. Similarly, \textsc{ReduceSaliency} checks whether $X_{\mathcal{A}}^*$
can successfully \textit{decrease} the saliency scores shared by \mb. This way,
we identify and retain $X_{\mathcal{A}}^*$ to gradually minimize \mb's
contribution. Saliency scores are computed using \code{\textsc{Saliency\_Est}}
presented in \A~\ref{alg:attacking}. Each saliency score, a float number ranging
from 0 to 1, is derived from the saliency map's $l_1$ norm. A lower saliency
score implies that \mb\ contributes less to joint inference. The saliency score
is more coarse-grained than the saliency map. It may not be inaccurate to assume
that saliency scores disclose only limited information regarding inputs of \mb,
pragmatically alleviating privacy leakage concerns.

\begin{figure}[!t]
 \vspace{-2pt}
  \captionsetup{skip=2pt}
  \centering
  \includegraphics[width=1.0\linewidth]{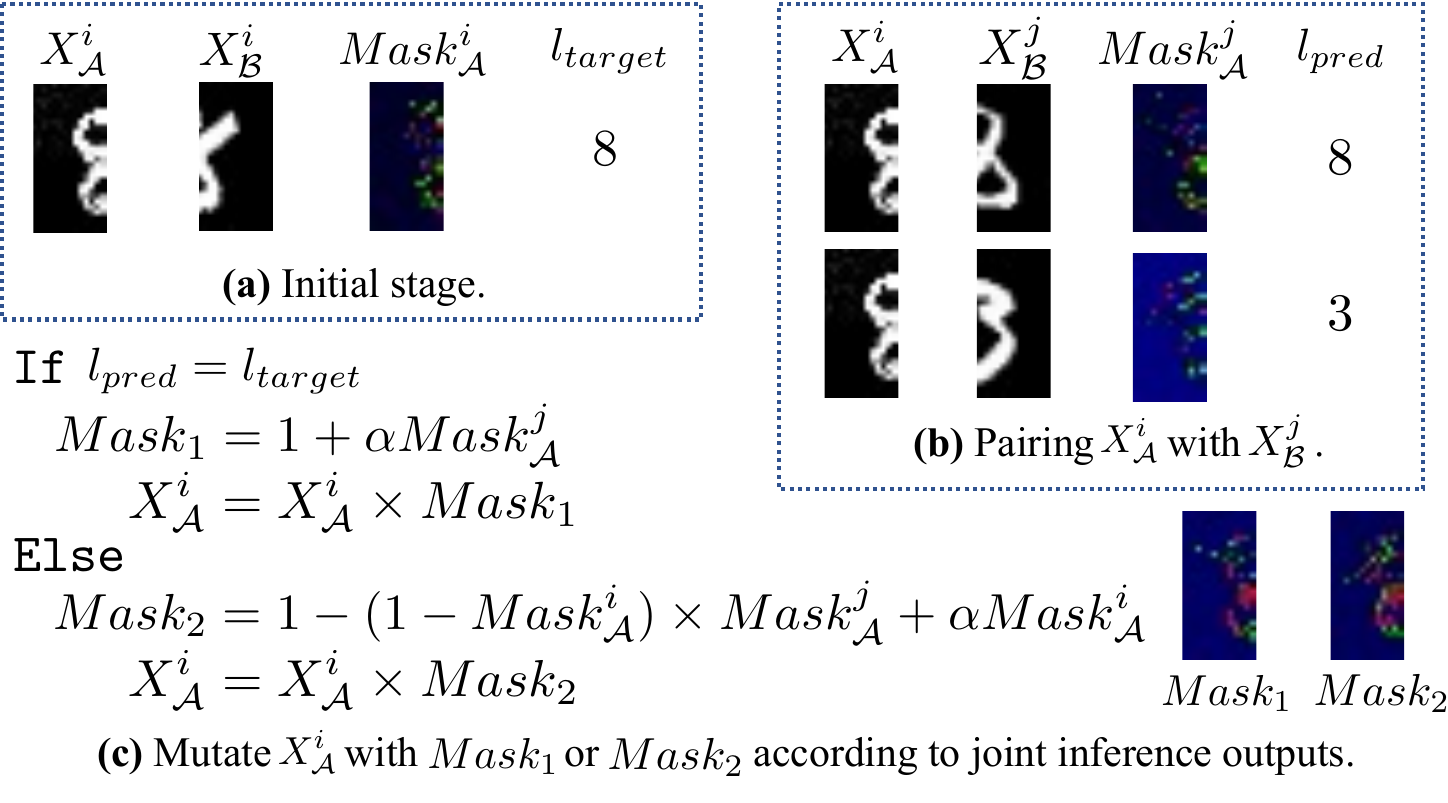}
  \caption{Saliency-aware mutation.}
  \label{fig:mutation}
  \vspace{-3pt}
\end{figure}

\begin{table*}[!ht]
	\captionsetup{skip=2pt}
	\centering
	\scriptsize
  \setlength{\tabcolsep}{1.5pt}
  \caption{Evaluation setup over VFL participants \ma\ and \mb.}
	\label{tab:modeldetails}
	\resizebox{0.96\linewidth}{!}{
		\begin{tabular}{l|c|c|c|c|c|c|c}
			\hline
      \multirow{2}{*}{\textbf{Dataset}}  & \multirow{2}{*}{\textbf{\#Cases}} & \multirow{2}{*}{\textbf{Learning Protocol}} & \multicolumn{2}{c|}{\textbf{Participant \ma}} & \multicolumn{2}{c|}{\textbf{Participant \mb}} & \multirow{2}{*}{\textbf{Central Coordinator \mc}} \\\cline{4-5}\cline{6-7}
                                         &                                   &                                             & \textbf{\#Partitioned Features} & \textbf{Setup} & \textbf{\#Partitioned Features} & \textbf{Setup} & \\
			\hline
			\textbf{NUS-WIDE} & 269,648  & SplitNN  & 634 & 2-FCs with ReLU & 1,000 & 2-FCs with ReLU & 2-FCs with ReLU \\
			\textbf{Credit}   & 30,000   & HeteroLR & 13  & 1-FC & 10 & 1-FC & Sigmoid \\
			\textbf{Vehicle}  & 946      & HeteroLR &  9  & 1-FC &  9 & 1-FC & Sigmoid \\
			\textbf{MNIST}    & 60,000   & SplitNN  & 28$\times$14 pixels & 3-Convs with ReLU & 28$\times$14 pixels & 3-Convs with ReLU & 1-FC with ReLU \\
			\textbf{VQA v2.0}  & 82,783    & VFVQA   & embedding to 50$\times$2054 & FasterRCNN~\cite{ren2016faster} & embedding to 128$\times$512 & BERT~\cite{li2020oscar} & Multi-Layer Transformers \\
      \textbf{CIFAR-10} & 50,000   & SplitNN  & embedding to $10752$ & VGG16~\cite{vgg16} & embedding to $10752$ & VGG16~\cite{vgg16} & 3-FCs with ReLU\\
			\hline
		\end{tabular}
	}
	\vspace*{-12pt}
\end{table*}

\noindent \textbf{Saliency-Aware Mutation (Line 13).}~We propose a
saliency-aware scheme to mutate inputs. \F~\ref{fig:mutation} illustrates the
mutation procedure using MNIST images as an example. As will be introduced in
\S~\ref{sec:evaluation}, each MNIST image is vertically partitioned into two
pieces in a VFL setting. Given a pair of input $X_{\mathcal{A}}^i$ and
$X_{\mathcal{B}}^i$ belonging to the same MNIST image $i$, we first compute the
saliency map of $X_{\mathcal{A}}^i$ as $Mask^{i}_{\mathcal{A}}$ and the jointly
inferred label $l_{target}$ (\F~\hyperref[fig:mutation]{\ref{fig:mutation}(a)}).

We mutate $X_{\mathcal{A}}^i$ with some random noise, and then iterate inputs in
\sd. For an input pair $X_{\mathcal{A}}^i$ and $X_{\mathcal{B}}^j$ ($j \neq i$),
we compute a new saliency map $Mask^{j}_{\mathcal{A}}$
(\F~\hyperref[fig:mutation]{\ref{fig:mutation}(b)}). When the joint inference
$l_{pred}$ equals $l_{target}$, we retain this mutation by augmenting
$X_{\mathcal{A}}^i$ with $\alpha Mask^{j}_{\mathcal{A}}$ (\texttt{If} branch in
\F~\hyperref[fig:mutation]{\ref{fig:mutation}(c)}). $l_{pred} \neq l_{target}$
denotes undesirable mutations (\texttt{Else} branch in
\F~\hyperref[fig:mutation]{\ref{fig:mutation}(c)}): we weaken certain pixels in
$X_{\mathcal{A}}^i$ in case they are focused by $Mask^{j}_{\mathcal{A}}$ but
overlooked by $Mask^{i}_{\mathcal{A}}$. $\alpha$ is empirically decided as 0.2.
{Intuitively, selecting a proper $\alpha$ is conceptually similar to
selecting a ``learning rate'', a common step in training neural networks. For a
smaller $\alpha$, we expect more mutation energy to converge and discover an
ADI; for a larger $\alpha$, we may miss the targeted point. Our preliminary
exploration shows that $\alpha=0.2$ and 20 energy denote reasonably good
configurations.} $Mask^{i,\, j}_{\mathcal{A}}$ is calculated using the algorithm
in~\cite{fong2017interpretable}, where the saliency mask is a matrix with values
ranging from 0 to 1. To boost a participant $p$'s contribution, according
to~\cite{fong2017interpretable}, we can increase certain input components of $p$
that correlate to large values on $p$'s saliency mask.

We repeat this procedure over \sd\ to update $X_{\mathcal{A}}^i$. We
use \textit{bounded mutation} over $X_{\mathcal{A}}^i$ to retain stealthy
changes. The proposed mutation has a time complexity linear to \ssd. As \ssd\ is
tiny (e.g., 20 for our evaluation in \S~\ref{subsec:greybox-fuzzing}), time
complexity is not a major issue. Furthermore, our saliency-aware mutation is a
general pipeline agnostic to input formats; it performs more holistic and
efficient mutations than pixel-level mutations. \S~\ref{subsec:greybox-fuzzing}
shows that a large number of ADIs are uncovered across all datasets using fuzz
testing.


\section{Implementation \& Evaluation Setup}
\label{sec:implementation}

\noindent \textbf{VFL Protocols \& Frameworks.}~As discussed in
\S~\ref{sec:background}, HeteroLR and SplitNN are two popular and core VFL
protocols that can be extended to most parameterized VFL protocols (e.g.,
VFVQA). Hence, we evaluated these two protocols; see implementation and training
details in Appendix \ref{sec:lr-nn}. To implement VFVQA, we modify the state-of-the-art
VQA model Oscar~\cite{li2020oscar} into VFL. In VQA, one participant raises
natural-language questions about images possessed by the other participant.
VFVQA allows two participants to conduct VQA while keeping questions/images
locally; see Appendix \ref{sec:lr-nn} for VFVQA architecture and implementation details.
To date, several VFL infrastructures are available, including FATE~\cite{fate},
TF Federated~\cite{tffederated}, FedML~\cite{chaoyanghe2020fedml},
PySyft~\cite{pysyft}, and CrypTen~\cite{crypten}. We choose FATE and FedML due
to their popularity and support for common VFL protocols including HeteroLR and
SplitNN. FATE is maintained by industry, whereas FedML is a state-of-the-art VFL
framework developed by the research community and industry. Both platforms are
in high engineering quality. Our main findings are, to our knowledge,
independent to VFL frameworks. We use the default SplitNN implementation in
FATE, and extend the HeteroLR implementation in FedML.

\section{Evaluation}
\label{sec:evaluation}
\noindent \textbf{Datasets \& Environment Setup.}~We use six popular
real-world datasets: NUS-WIDE~\cite{chua2009nus}, Credit~\cite{credit},
Vehicle~\cite{vehicle}, MNIST~\cite{lecun1998gradient}, VQA
v2.0~\cite{lin2014microsoft}, and CIFAR-10~\cite{krizhevsky2009learning}.
Each dataset's features are partitioned between  VFL participants \ma\ and
\mb, as in \T~\ref{tab:modeldetails}.
For NUS-WIDE, \ma\ holds the image features and \mb\ holds
the text features.
For Credit, \ma\ holds 13 features and \mb\ holds 10.
For Vehicle, both \ma\ and \mb\ get nine features.
For MNIST, each image is vertically partitioned into two: \ma\ gets the left
piece (with 28$\times$14 pixels) and \mb\ gets the right.
For VQA v2.0, \ma\ holds images and \mb\ raises natural-language questions.
For CIFAR-10, \ma\ and \mb\ each holds a $10,752$-dimensional feature embedding, which is the feature output of VGG16 with the left and right piece of the image as the inputs.
We summarize the setup in \T~\ref{tab:modeldetails}, and present details of
dataset preparation and setup in Appendix \ref{sec:dataset}. Our evaluation is conducted on
Intel Xeon CPU E5-2683 with 256 GB RAM and Nvidia GeForce RTX 2080 GPU.

\noindent \textbf{Metrics.}~We measure the performance of our attack
using \textit{success rates} and \textit{average dominated proportion
of \mb's inputs}. Recall as we defined in Def.~\ref{def:adi-practical}, to
decide if an input $X_{\mathcal{A}}^i$ is ADI, we check if $X_{\mathcal{A}}^i$
can dominate more than a threshold of all test inputs of \mb. As defined in
\S~\ref{sec:adi}, we adopt two thresholds (95\% and 99\%) as a practical
assessment of ADIs. Accordingly, we define the ratio of samples from \ma's test
dataset that can be successfully perturbed into ADIs as the attack ``success
rate'' of ADI synthesis. We also measure the average dominated
proportion of \mb's inputs. That is, for a generated input on \ma, we measure
how much of the benign participants' data is dominated. For simplicity, we refer
to this metric as ADP in the paper. The ADP is the average of $r$ over the
generated data $X_{\mathcal{A}}^*$ ($r$ is the dominated proportion of \mb's
inputs for a single input from \ma\ as noted in Def.~\ref{def:adi-practical}).

\begin{table}[!htpb]
	\centering
	\scriptsize
	\captionsetup{skip=2pt}
  \caption{Dominating rates of the standard datasets.}
	\label{tab:original-data-ats}
	\setlength{\tabcolsep}{3.0pt}
		\begin{tabularx}{0.8\linewidth}{l|>{\centering\arraybackslash}X|>{\centering\arraybackslash}X|c|c}
			\hline
     & \multicolumn{2}{c|}{\textbf{Dominating Rate}} & \multirow{2}{*}{\textbf{ADP}}  & \textbf{Model Accuracy on} \\\cline{2-3}
     & \textbf{95\%} & \textbf{99\%} & & \textbf{Test Datasets} \\
			\hline
			\textbf{NUS-WIDE}   & 0.00\% & 0.00\% & 38.93\% & 77.35\% \\
			\textbf{Credit}     & 25.0\% & 2.10\% & 82.40\% & 0.7450 (auc-roc) \\
			\textbf{Vehicle}    & 0.86\% & 0.14\% & 40.74\% & 84.00\% \\
			\textbf{MNIST}      & 0.87\% & 0.20\% & 47.54\% & 97.78\% \\
			\textbf{VQA v2.0}    & NA    & 0.80\% & 38.84\%  & 73.82\% \\
			\textbf{CIFAR-10}   & 2.63\% & 0.46\% & 46.40\% & 86.55\% \\
			\hline
		\end{tabularx}
\end{table}

\noindent \textbf{Dominating Inputs in Standard Datasets.}~We first assess the
input dominating issues (without perturbation) in \ma's standard datasets. Using
the aforementioned feature partition scheme, we train VFL models for two
participants \ma\ and \mb. \T~\ref{tab:original-data-ats}'s last column shows
that each trained model has achieved satisfying accuracy.
\T~\ref{tab:original-data-ats} reports the dominating rates, i.e.~the ratio of
standard inputs of \ma\ that dominate the VFL predictions. Such standard inputs
of \ma\ are deemed as ADIs. Recall when deciding if an input is ADI, our
practical assessment (Def.~\ref{def:adi-practical}) examines two thresholds,
95\%, and 99\%. The outlier is VQA v2.0, where an image is regarded as an ADI
only when it dominates \textit{all} of its associated questions (typically one
image associated with 3--10 questions). Hence, the threshold should be 100.0\%
rather than 95\% or 99\% when assessing its input dominating issues. To ease
presentation, for \T~\ref{tab:original-data-ats} and the rest tables, we put the
VQA v2.0 evaluation results in the 99\% column and ``NA'' in the 95\% column.
We also report the ADPs for standard inputs from \ma\ in the third
column of \T~\ref{tab:original-data-ats}.

Credit has a greater dominating rate at 95\%, and its ADP is higher than others.
Credit is a binary classification task with a small sample size, and it is
likely to identify ``dominating inputs'' (i.e., false positives) of \ma\ which
induce identical outputs when paired with \mb's inputs. To eliminate false
positives, saliency maps of \mb's inputs can be used to confirm if its
contribution is negligible.
%
Overall, we interpret that dominating inputs are rare in well-designed/trained
models with a fair feature partition. Without an active adversary, users are
unlikely to notice ADIs until a security breach. However, when feature partition
over two participants is unbalanced, ADIs become noticeable; see Appendix \ref{subsec:eval-feature-allocation} for more results.


\noindent \textbf{Setup.}~In the following, we will first evaluate the
performance of our proposed ADI synthesis in \S~\ref{subsec:whitebox-attack}.
Note that in addition to launching the blackbox attack, we also set up a whitebox
synthesis, where \ma\ is assumed to acquire gradients from \mb's model. This is
aligned with our formulation of whitebox synthesis in \A~\ref{alg:attacking}. To
clarify, this work champions blackbox VFL attacks. We set up the
whitebox synthesis simply as a \textbf{baseline} for comparison with the
blackbox attack. In \S~\ref{subsec:whitebox-attack}, we also analyze
the properties of ADIs like the stealthiness and the reward hogging. Then, we
evaluate the greybox fuzz testing in \S~\ref{subsec:greybox-fuzzing} and present
the empirical comparison between ADIs, AEs and random inputs in
\S~\ref{subsec:eval-compare-uae}. We then evaluate the performance of two
defense strategies in \S~\ref{subsec:defense}. In
\S~\ref{subsec:weaker-threat-model}, we explore the feasibility of synthesizing
ADIs without accessing the data of \mb. In addition, we also evaluate several
key factors in our attack like size of \sd\
(Appendix \ref{subsec:eval-subset-size}), feature partition ratio (Appendix \ref{subsec:eval-feature-allocation}), and number of participants
(Appendix \ref{subsec:eval-client-numbers}).

\subsection{Gradient-Based ADI Synthesis}
\label{subsec:whitebox-attack}

\noindent \textbf{Results.}~\T~\ref{tab:mutation} reports the success rates of
ADI synthesis under different settings, thresholds, and mutation strategies in
whitebox and blackbox settings. Complex datasets and models like NUS-WIDE,
MNIST, VQA v2.0, and CIFAR-10 have a lower success rate than simple datasets
like Credit and Vehicle, especially when the threshold is 99\%. For Credit and
Vehicle, the success rate is close to 100\% under random mutation. We
also observe that for CIFAR-10, it is relatively easy to achieve high attack
success rate under the 95\% threshold. Recall we split CIFAR-10 images to two
pieces. Given CIFAR-10 images are generally complex, it is inherently hard for
the model to capture meaningful contents using just one split. Hence, it becomes
easier to dominate benign participants. In contrast, the MNIST hand-written
digits are easier, making it possible for the model to extract useful
information even with only one image split. This explains that for MNIST,
generating ADIs are less easy. Under the 99\% threshold, it is generally harder
to generate ADIs for the CIFAR-10 and MNIST setups compared to the simple datasets
like Credit and Vehicle.

%
The bounded mutation scheme confines the applied mutations to a practical range
(i.e., bounded by the variance of the mutated feature). We find that this scheme
still achieves plausible success rates of over 50\% for simple tasks like Credit
and Vehicle and over 35\% for complex tasks under the 95\% threshold. 99\%
threshold is more challenging, and therefore, the success rate becomes
reasonably lower. In \T~\ref{tab:mutation-dominate-number}, we report the
ADP. Considering our ADI definition (Def.~\ref{def:adi-practical}), it is
trivial that the ADP must be higher than 95\% or 99\%, when benchmarking with
ADIs found under the 95\% or 99\% thresholds. 
More importantly, the last column of \T~\ref{tab:mutation-dominate-number}
reports the ADPs of all generated inputs (though some of them are not ``ADIs'').
The ADPs are consistently higher across all settings, compared to the standard
datasets (\T~\ref{tab:original-data-ats}).
This further demonstrates the effectiveness of our ADI synthesis algorithm as
the generated inputs tend to dominate the other participants' inputs.

In compared to whitebox synthesis, the blackbox synthesis has decreased success
rates, as shown in \T~\ref{tab:mutation}. This is reasonable, given the
difficulty and unreliability of estimating gradients using FDM. Overall, the
success rates of blackbox and whitebox syntheses have similar trending, with
complex datasets like NUS-WIDE having a lower success rate than simple datasets.
We also find that for the bounded mutation (99\%) evaluation over NUS-WIDE, the
success rates of the blackbox attack is increased. In comparison to simple
datasets, the NUS-WIDE dataset has larger feature dimensions and a denser and
more complicated feature distribution. All these factors contribute to the
difficulty of solving the optimization problem in whitebox synthesis (line 12 in
\A~\ref{alg:attacking}). Also, the estimated gradients in the blackbox setting
may be a bit larger than the real gradients, which will speed up convergence to
some extent. Thus, the blackbox ADI synthesis using bounded mutation under the
99\% threshold has a slightly higher success rate. In sum, the evaluation
illustrates that ADIs can be synthesized practically in blackbox settings,
despite variances in the datasets, VFL protocols, and mutation strategies. We
envision that VFL systems are in high risk of being controlled by ADIs.

\begin{table}[!t]
	\captionsetup{skip=2pt}
	\centering
	\scriptsize
  \caption{Success rates of gradient-based ADI synthesis.}
	\label{tab:mutation}
  \setlength{\tabcolsep}{3.0pt}
		\begin{tabularx}{0.80\linewidth}{c|l|>{\centering\arraybackslash}X|>{\centering\arraybackslash}X|>{\centering\arraybackslash}X|>{\centering\arraybackslash}X}
			\hline
          &\multirow{2}{*}{\textbf{Dataset}} & \multicolumn{2}{c}{\textbf{Random Mutation}} & \multicolumn{2}{|c}{\textbf{Bounded Mutation}} \\\cline{3-4}\cline{5-6}
          & & \textbf{95\%} & \textbf{99\%} & \textbf{95\%} & \textbf{99\%} \\
			\hline
			\multirow{6}{*}{\textbf{Whitebox}} & \textbf{NUS-WIDE}  & 65.1\% & 49.6\% & 42.8\% & 22.0\% \\
			& \textbf{Credit}    & 99.8\% & 99.6\% & 87.6\% & 52.6\% \\
			& \textbf{Vehicle}   & 98.4\% & 97.0\% & 87.9\% & 74.8\% \\
			& \textbf{MNIST}     & 92.9\% & 62.5\% & 34.5\% & 16.7\% \\
			& \textbf{VQA v2.0}   & NA    & 47.4\% & NA     & 14.2\% \\
			& \textbf{CIFAR-10}  & 99.6\% & 53.4\% & 98.9\% & 44.3\% \\ 
			\hline
			\multirow{6}{*}{\textbf{Blackbox}} & \textbf{NUS-WIDE}  & 45.2\% & 35.6\% & 40.0\% & 32.8\% \\
			& \textbf{Credit}    & 99.0\% & 98.3\% & 54.1\% & 41.1\% \\
			& \textbf{Vehicle}   & 96.4\% & 76.2\% & 87.1\% & 60.0\% \\
			& \textbf{MNIST}     & 78.8\% & 51.9\% & 33.9\% & 4.85\% \\
			& \textbf{VQA v2.0}  & NA     & 41.5\% & NA     & 10.9\% \\
			& \textbf{CIFAR-10}  & 97.7\% & 52.7\% & 91.3\% & 43.8\% \\
			\hline
		\end{tabularx}
\end{table}

\begin{table}[!t]
	\captionsetup{skip=2pt}
	\centering
	\scriptsize
  \caption{ADPs of gradient-based ADI synthesis.}
	\label{tab:mutation-dominate-number}
  \setlength{\tabcolsep}{3.0pt}
	\resizebox{1.00\linewidth}{!}{
		\begin{tabular}{c|l|c|c|c|c|c|c}
			\hline
          & \multirow{2}{*}{\textbf{Dataset}} & \multicolumn{3}{c}{\textbf{Random Mutation}} & \multicolumn{3}{|c}{\textbf{Bounded Mutation}} \\\cline{3-5}\cline{6-8}
          & & \textbf{95\%} & \textbf{99\%} & \textbf{All} & \textbf{95\%} & \textbf{99\%} &\textbf{All} \\
			\hline
			\multirow{6}{*}{\textbf{Whitebox}} & \textbf{NUS-WIDE}  & 99.87\% & 99.93\% & 78.06\% & 99.43\% & 99.85\% & 75.44\% \\
			& \textbf{Credit}    & 99.79\% & 99.81\% & 98.99\% & 99.20\% & 99.82\% & 93.88\% \\
			& \textbf{Vehicle}   & 99.75\% & 99.92\% & 87.02\% & 99.52\% & 99.71\% & 85.13\% \\
			& \textbf{MNIST}     & 98.21\% & 99.79\% & 98.38\% & 97.41\% & 99.79\% & 90.91\%\\
			& \textbf{VQA v2.0}  & NA  & 100\% & 61.97\%  & NA & 100\% & 50.40\% \\
			& \textbf{CIFAR-10}  & 98.65\% & 99.50\% & 98.64\% & 98.51\% & 99.46\% & 98.45\%\\
			\hline
			\multirow{6}{*}{\textbf{Blackbox}} & \textbf{NUS-WIDE}  & 99.15\% & 99.80\% & 75.35\% & 99.13\% & 99.84\% & 73.30\% \\
			& \textbf{Credit}    & 99.81\% & 99.83\% & 99.10\% & 99.22\% & 99.80\% & 93.86\% \\
			& \textbf{Vehicle}   & 99.01\% & 99.98\% & 87.09\% & 99.34\% & 99.98\% & 83.14\% \\
			& \textbf{MNIST}     & 98.22\% & 99.55\% & 93.43\% & 97.90\% & 99.18\% & 85.82\% \\
			& \textbf{VQA v2.0}  & NA & 100\% & 67.96\% & NA & 100\% & 63.72\% \\
			& \textbf{CIFAR-10}  & 98.67\% & 99.49\% & 97.00\% & 98.62\% & 99.36\% & 98.15\% \\
			\hline
		\end{tabular}
	}
\end{table}

\begin{figure*}[!t]
	\captionsetup{skip=5pt}
	\captionsetup[sub]{skip=1pt}
	\centering
	\begin{subfigure}{.16\linewidth}
		\centering
		\resizebox{1\linewidth}{!}{
		\begin{tikzpicture}
			\node (img) {\includegraphics[trim={1cm 0.5cm 1cm 0.5cm},clip]{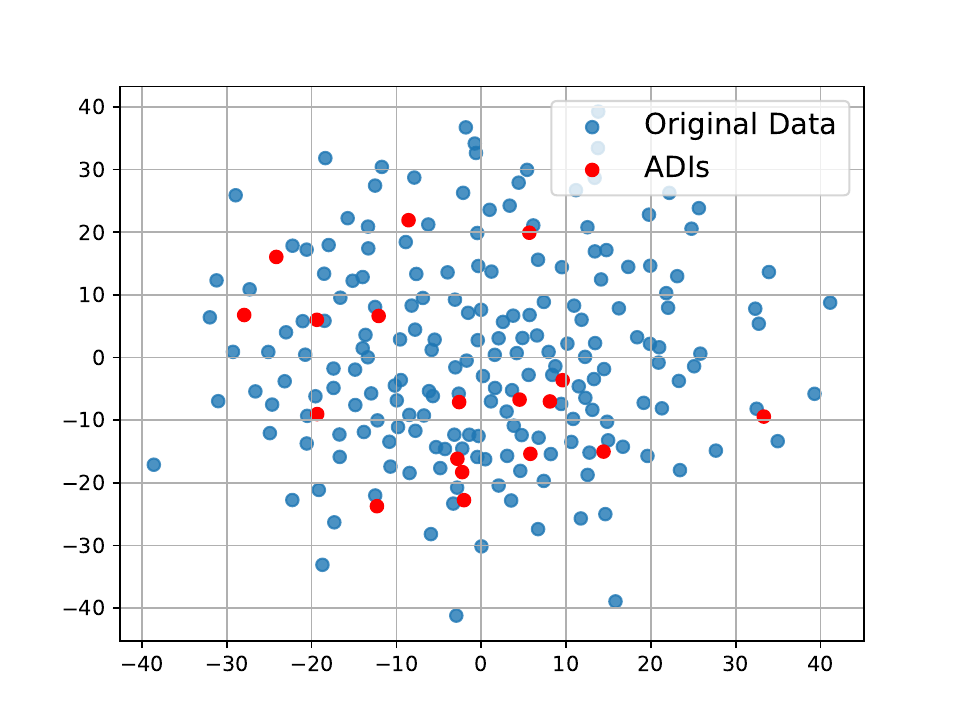}};
		\end{tikzpicture}
		}
	\caption{\small NUS-WIDE.}
	\end{subfigure}
	\begin{subfigure}{.16\linewidth}
		\centering
		\resizebox{1\linewidth}{!}{
		\begin{tikzpicture}
			\node (img) {\includegraphics[trim={1cm 0.5cm 1cm 0.5cm},clip]{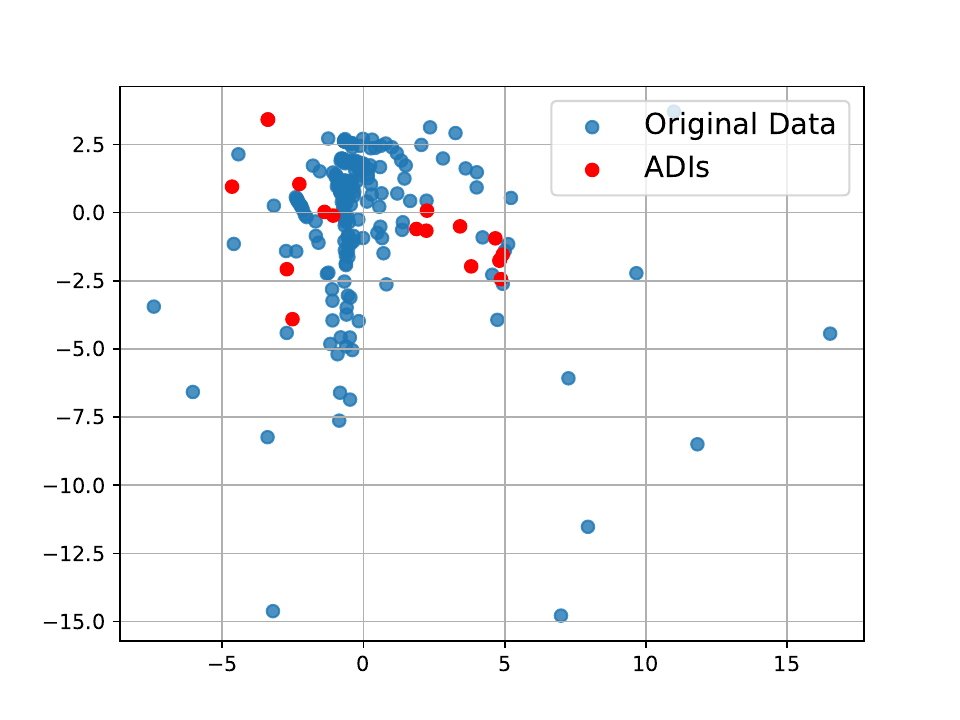}};
		\end{tikzpicture}
		}
	\caption{\small Credit.}
	\end{subfigure}
	\begin{subfigure}{.16\linewidth}
		\centering
		\resizebox{1\linewidth}{!}{
		\begin{tikzpicture}
		\node (img) {\includegraphics[trim={1cm 0.5cm 1cm 0.5cm},clip]{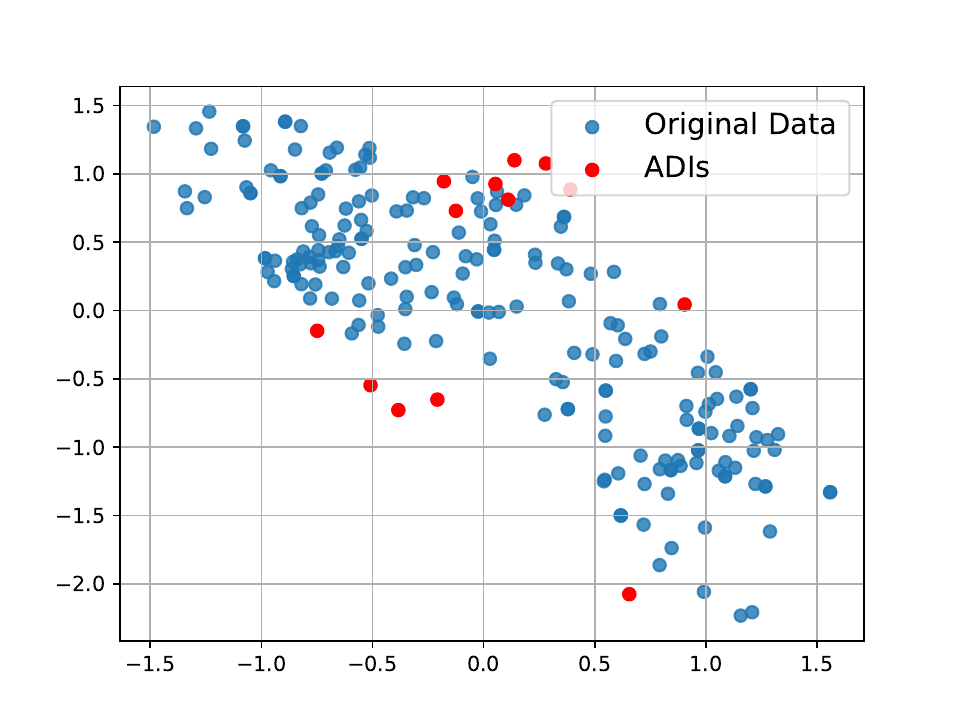}};
		\end{tikzpicture}
		}
	\caption{\small Vehicle.}
	\end{subfigure}
	\begin{subfigure}{.16\linewidth}
		\centering
		\resizebox{1\linewidth}{!}{
		\begin{tikzpicture}
			\node (img) {\includegraphics[trim={1cm 0.5cm 1cm 0.5cm},clip]{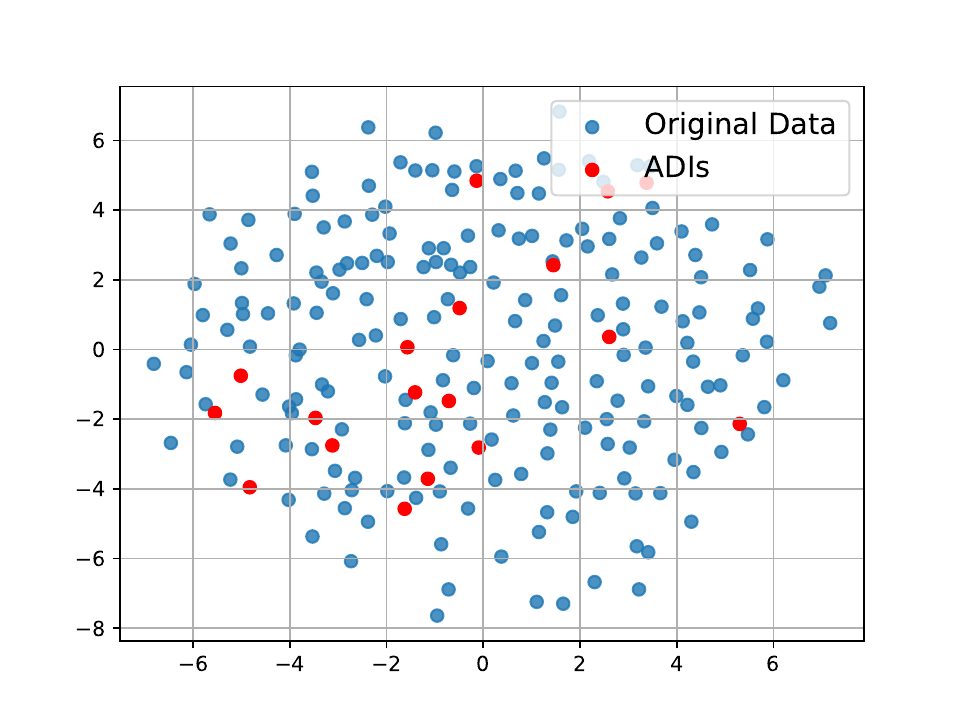}};
		\end{tikzpicture}
		}
	\caption{\small MNIST.}
	\end{subfigure}
	\begin{subfigure}{.16\linewidth}
		\centering
		\resizebox{1\linewidth}{!}{
		\begin{tikzpicture}
			\node (img) {\includegraphics[trim={1cm 0.5cm 1cm 0.5cm},clip]{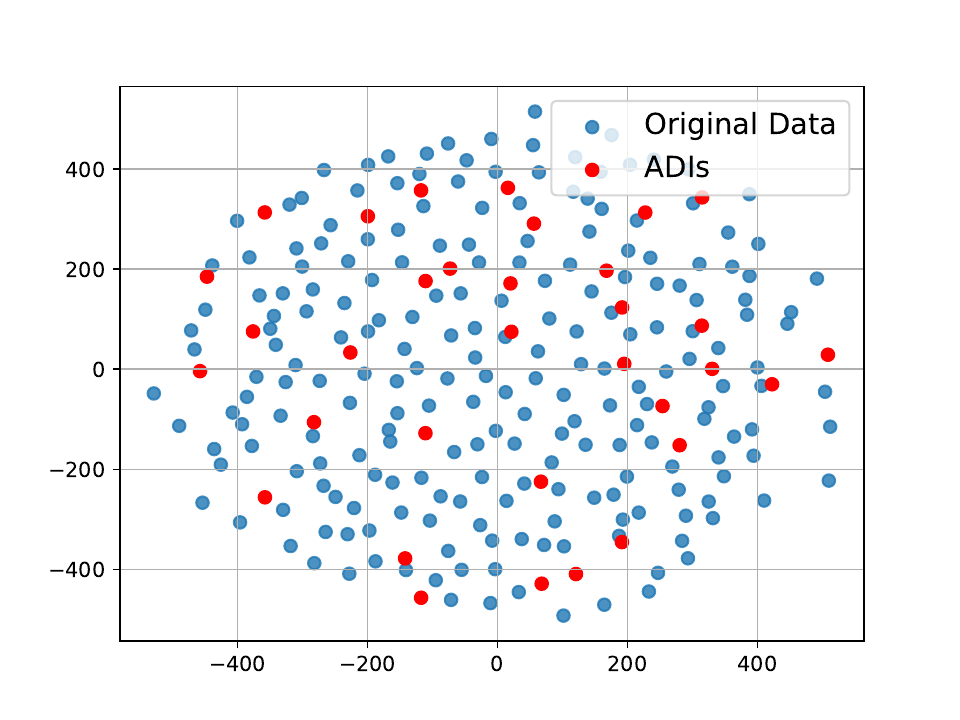}};
		\end{tikzpicture}
		}
	\caption{\small VQA v2.0.}
	\end{subfigure}
	\begin{subfigure}{.16\linewidth}
		\centering
		\resizebox{1\linewidth}{!}{
		\begin{tikzpicture}
			\node (img) {\includegraphics[trim={1cm 0.5cm 1cm 0.5cm},clip]{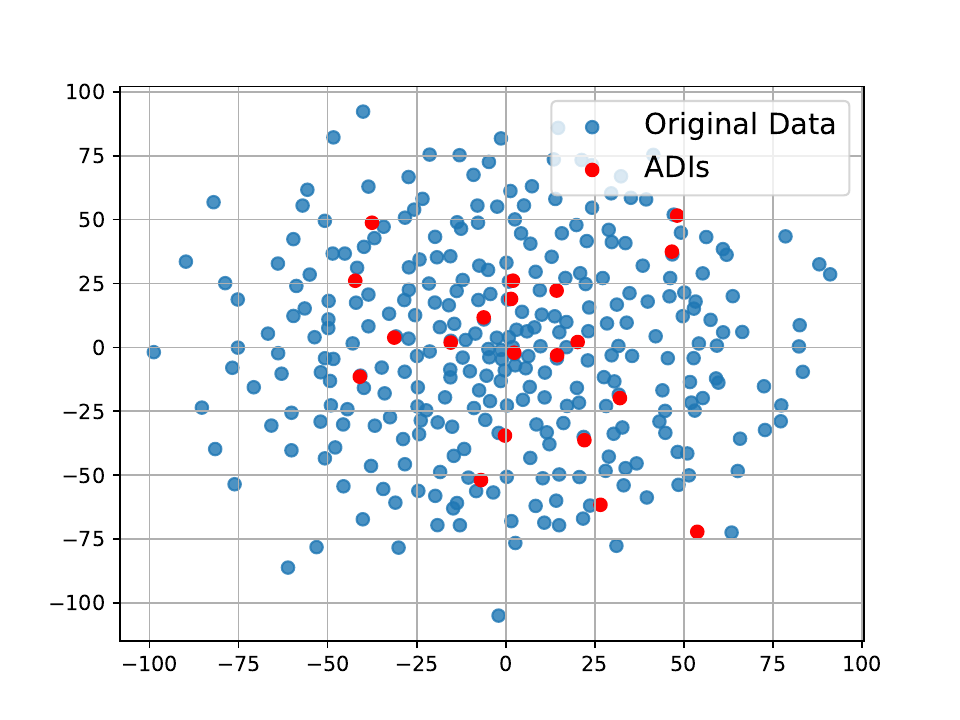}};
		\end{tikzpicture}
		}
	\caption{\small CIFAR-10.}
	\end{subfigure}

    \centering
    \caption{Normal inputs and ADIs projected to 2D figures.}
    \label{fig:bounded}
\end{figure*}

\begin{figure}[!t]
	\captionsetup{skip=1.5pt}
	\vspace{-3pt}
    \centering
    \includegraphics[width=0.93\linewidth]{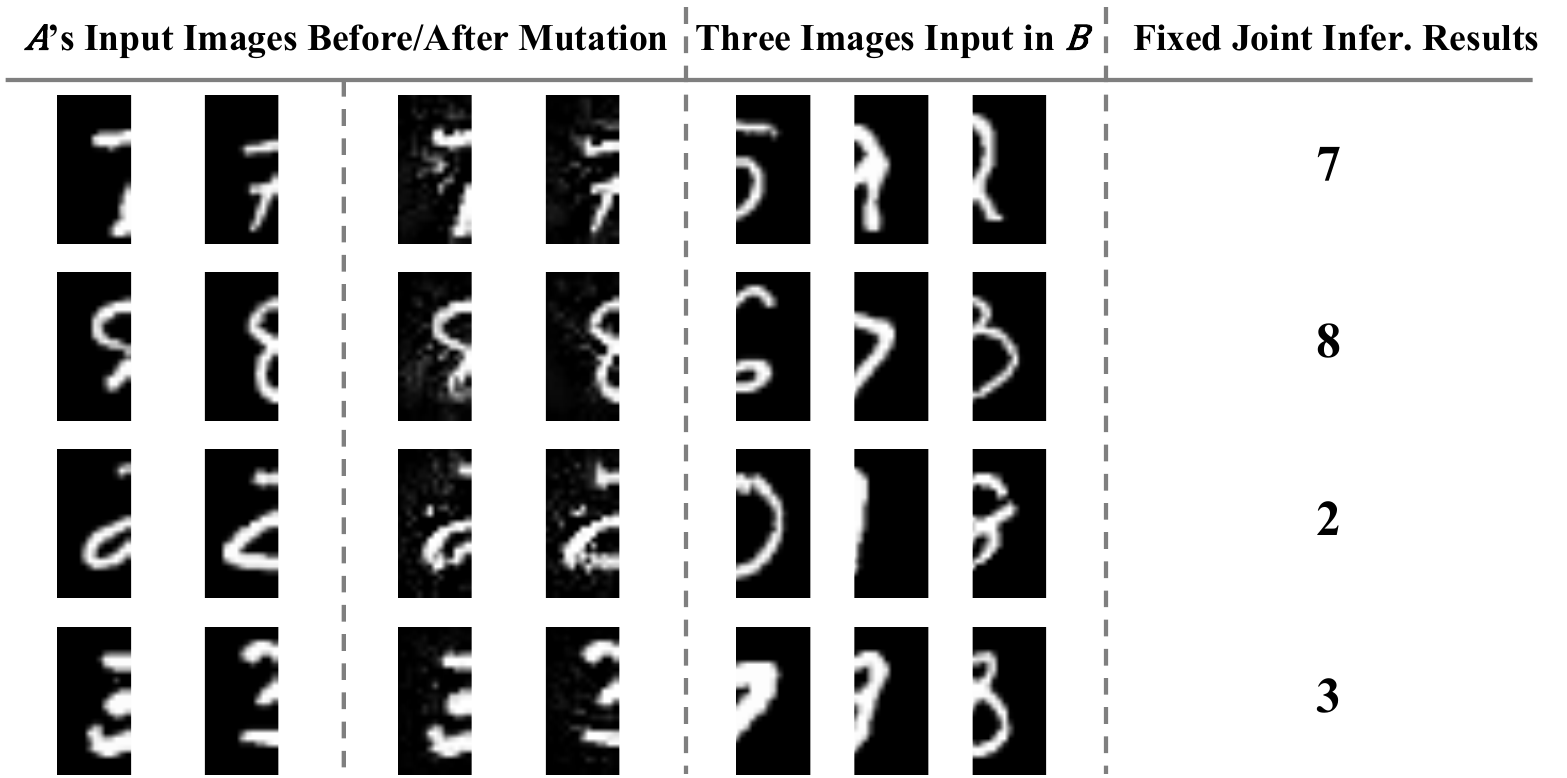}
    \caption{ADIs found in MNIST using bounded mutation.}
    \label{fig:mnist}
	\vspace{-5pt}
\end{figure}

\noindent \textbf{ADI Stealth.}~ADIs generated by bounded mutation mostly follow
the original datasets' distribution, making them visually similar to normal
inputs. In \F~\ref{fig:bounded}, we project original inputs and ADIs synthesized
at threshold 99\% to 2D figures using multidimensional scaling. We interpret
that ADIs generated by bounded mutation (marked in \textcolor{red}{red}) can
neither be easily distinguished by the data distribution nor by the distances
between the data points. Note that since the data has multiple
dimensions, the 2-dimensional projection figure may not reflect the real
distribution distance. To mitigate this threat to validity, we run the
multidimensional scaling for multiple times and confirm that the results are
consistent. \F~\ref{fig:mnist} reports  stealthy ADIs synthesized by bounded
mutation. While perturbing the left half of MNIST images (the first column of
\F~\ref{fig:mnist}) only causes stealthy changes (the second column of
\F~\ref{fig:mnist}), the synthesized ADIs, after concatenating with the right
half of arbitrary MNIST images (i.e., the inputs of \mb; see the third column of
\F~\ref{fig:mnist}), control the outputs to fixed labels (the last column of
\F~\ref{fig:mnist}).

Moreover, we also train a binary classifier using the same amount of ADIs and
normal inputs to distinguish them. It only achieves 0.54 ROC-AUC, which is very
close to random guessing. Thus, we conclude that ADIs generated by bounded
mutation and the normal inputs are hard to distinguish, which empirically
illustrates the stealth of ADIs. 

\begin{figure}[!ht]
	\vspace{-5pt}
	\captionsetup{skip=1.5pt}
    \centering
    \includegraphics[width=0.93\linewidth]{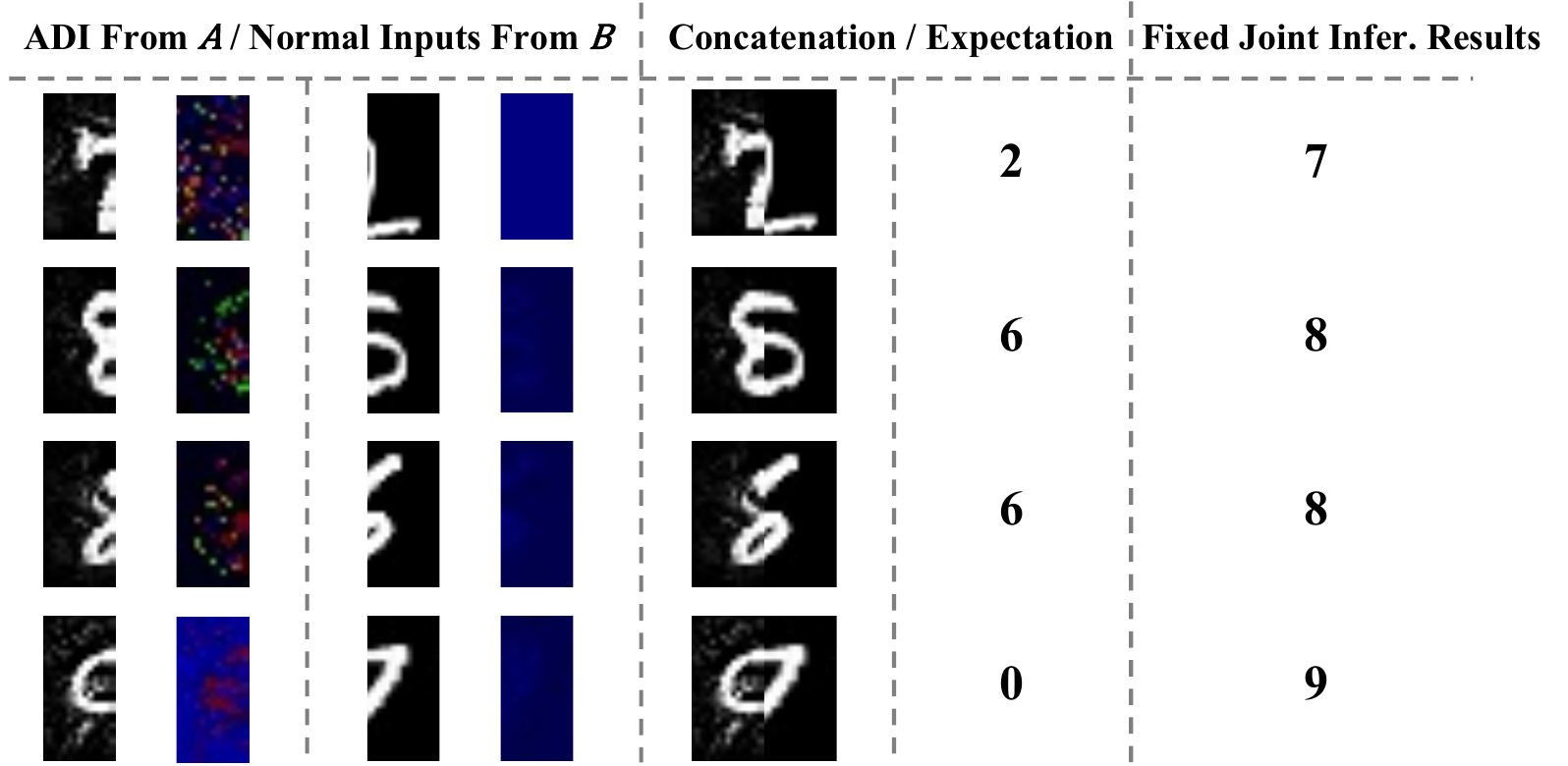}
    \caption{ADIs dominating the joint inference even when the concatenated
    images form reasonable digits.} 
	\label{fig:observeMnist}
	\vspace{-5pt}
\end{figure}

\F~\ref{fig:observeMnist} shows intriguing cases from MNIST dataset. As
shown in the ``Concatenation/Expectation'' column of \F~\ref{fig:observeMnist},
while ADIs may form reasonable digits (from a human perspective) with the inputs
of \mb, the joint inferences are still forced to be the target labels specified
by the adversary. We also provide the saliency maps associated with the input
images, which imply that the contribution of \mb\ is negligible (the saliency
maps in the 4th column barely have highlighted areas). This again shows that
ADIs can govern the joint inference and negate other participants'
contributions.

Whether concatenating ADIs and normal inputs can form meaningful contents depend
on the nature of features that VFL participants hold. For instance, in MNIST,
the features hold by different participants are images. Therefore, forming
``meaningful concatenations'' requires that both adversarial and benign
participants provide \textit{visually-correlated} images. This is apparently
challenging, though many successful cases (as in \F~\ref{fig:observeMnist}) are
found in our evaluation. In contrast, datasets like Credit consist of
low-dimensional numeric vectors; ``visual correlation'' is not a concern for
such numeric data. It is generally easier for concatenations of ADIs and benign
participant's inputs to be indistinguishable with normal data in Credit. Though
it is not the primary focus of this research,  we deem it interesting to explore
synthesizing ADIs that can form meaningful concatenations with benign
participants' inputs. The challenge is to quantify the distinguishability
between the concatenations and normal inputs to define an objective function.

%
%

\noindent \textbf{Reward Estimation under ADIs.}~\S~\ref{sec:background} and \S~\ref{subsec:motivation-adi} have
discussed that FL clients are often compensated based on their
contributions to the joint inference. While the evaluated VFL frameworks do not
ship with a reward calculation module, we estimate how ADIs influence reward
allocation in Appendix \ref{sec:reward}. As expected, while \ma\ and \mb\ earn roughly the
same reward when using normal inputs, ADIs allow adversary \ma\ to hog rewards
for each inference. We deem this evaluation as convincing to show that ADIs can
create real-world financial loss and confusion for normal consumers.

\subsection{Greybox Fuzz Testing}
\label{subsec:greybox-fuzzing}
Fuzz testing helps in-house VFL vulnerability assessment. As mentioned in
\S~\ref{subsec:design-testing}, we use bounded mutation to mutate inputs:
bounded mutation generates more realistic inputs, which assesses VFL
in front of stealthy ADIs.

\T~\ref{tab:greybox} reports the fuzzing results. For datasets with narrow
feature spaces (Credit and Vehicle), we use 500 seeds and finish in two hours.
For others, we run a 12-hour campaign with a corpus of 1,000 seeds to
systematically explore the input spaces. ADIs are found in similar numbers for
all settings, indicating high efficiency of the proposed technique across input
formats and VFL protocols. Recall that the fuzzing algorithm has relatively high
time complexity; our saliency-aware mutation (line 13 in \A~\ref{alg:fuzzing})
and \code{\textsc{IsADI}} function (line 1 in \A~\ref{alg:fuzzing}) both require
iterating all inputs in \sd\ for each fuzzing iteration. Nevertheless, our
results in \T~\ref{tab:greybox} show that a high number of ADIs can be found in
a half-day even for image inputs. We interpret the performance as plausible:
fuzzing allows developers to quickly assess the ADI risk of their VFL at a low
cost.

\begin{table*}[!t]
	\captionsetup{skip=2pt}
	\centering \scriptsize
	\begin{minipage}{0.265\linewidth}
	\centering
  \caption{Fuzz testing results.}
	\label{tab:greybox}
	\setlength{\tabcolsep}{1.5pt}
	\resizebox{\linewidth}{!}{
		\begin{tabular}{l|c|c|c|c}
			\hline
         & \multicolumn{2}{c|}{\textbf{\#ADIs}} & \textbf{\#Seeds} &\textbf{Processing} \\\cline{2-3}
         & \textbf{95\%} & \textbf{99\%} & \textbf{in Corpus} & \textbf{Time} \\
			\hline
			\textbf{NUS-WIDE} & 176 & 98  & 1,000 & 12 hours\\
			\textbf{Credit}   & 235 & 136 & 500   & 2 hours \\
			\textbf{Vehicle}  & 392 & 157 & 500   & 2 hours\\
			\textbf{MNIST}    & 113 & 86  & 1,000 & 12 hours\\
			\textbf{VQA v2.0} &  NA & 93  & 1,000 & 12 hours\\
			\textbf{CIFAR-10} & 773 & 61 & 1,000 & 12 hours \\
			\hline
		\end{tabular}
	}
\end{minipage}
\begin{minipage}{0.43\linewidth}
	\centering
	\caption{Comparing ADIs and AEs.}
	\label{tab:ae-compare}
	\setlength{\tabcolsep}{1.5pt}
	\resizebox{\linewidth}{!}{
		\begin{tabular}{l|c|c|c|c|c|c}
			\hline
		  & \multicolumn{3}{c|}{\textbf{ADP}} & \multicolumn{3}{c}{\textbf{Success Rate (95\%  threshold)}} \\\cline{2-4} \cline{5-7} 
          & \textbf{ADI} & \textbf{AE} & \textbf{Random Inputs} & \textbf{ADI} & \textbf{AE} & \textbf{Random Inputs}\\
			\hline
			\textbf{NUS-WIDE}  & 75.44\% & 33.15\% & 43.25\% & 42.8\% & 0.00\% & 0.00\%\\
			\textbf{Credit}    & 93.88\% & 83.98\% & 89.21\% & 87.6\% & 13.0\% & 52.7\% \\
			\textbf{Vehicle}   & 85.13\% & 46.45\% & 56.16\% & 74.8\% & 2.84\% & 10.5\% \\
			\textbf{MNIST}     & 90.91\% & 42.67\% & 31.90\% &34.5\% & 0.20\% & 0.00\% \\
			\textbf{VQA v2.0}  & 50.54\% & 3.91\% & 19.15\% & 14.2\% & 2.80\% & 0.60\% \\
			\textbf{CIFAR-10}  & 98.45\% & 43.55\% & 16.52\% & 98.9\% & 1.78\% & 0.00\% \\ 
			\hline
		\end{tabular}
	}
	\end{minipage}
	\begin{minipage}{0.285\linewidth}
		\caption{Detector performance.}
	\label{tab:defense-clustering}
	\setlength{\tabcolsep}{1.5pt}
	\resizebox{\linewidth}{!}{
		\begin{tabular}{l|c|c|c|c}
			\hline
          & \multicolumn{2}{c}{\textbf{95\% Threshold}} &\multicolumn{2}{|c}{\textbf{99\% Threshold}} \\\cline{2-3}\cline{4-5}
          & \textbf{Avg. Acc} & \textbf{Avg. F1} & \textbf{Avg. Acc} & \textbf{Avg. F1} \\
			\hline
			\textbf{NUS-WIDE}  & 74.25\% & 62.24\% & 81.76\% & 77.18\% \\
			\textbf{Credit}    & 72.41\% & 59.96\% & 77.14\% & 67.22\% \\
			\textbf{Vehicle}   & 85.09\% & 79.13\% & 89.47\% & 85.92\% \\
			\textbf{MNIST}     & 64.16\% & 52.99\% & 76.03\% & 63.78\% \\
			\textbf{VQA v2.0}  & NA      & NA & 56.52\% & 40.24\% \\
			\textbf{CIFAR-10}  & 60.40\% & 32.67\% & 67.53\% & 53.79\% \\ 
			\hline
		\end{tabular}
	}
		\end{minipage}
	\vspace*{-5pt}
\end{table*}

\noindent \textbf{Stealth of ADIs.}~We compare ADIs with normal inputs following
the same procedure of reporting \F~\ref{fig:bounded}; the results are
in Appendix \ref{sec:fuzzing-visualize}. We find that ADIs are highly similar to regular inputs. We
also present ADIs found by fuzzing and their saliency maps
in Appendix \ref{sec:fuzzing-visualize}. Findings (i.e., ADIs greatly reduce the contribution of
normal inputs) are consistent with \F~\ref{fig:observeMnist}.

\subsection{Comparing ADIs and Standard AEs}
\label{subsec:eval-compare-uae}

Following \S~\ref{sec:adi} which compares ADIs and AEs from the conceptual level,
we now empirically compare them. We first generate AEs in the VFL setting using
a classic AE generation algorithm, FGSM~\cite{goodfellow2014explaining}. We then
measure the number of AEs, when being used as inputs of \ma, that can dominate over
$95\%$ normal inputs from \mb. Similar to \T~\ref{tab:mutation}, we also report
the ADP. As a comparison, besides AEs, we set up the same evaluation using
randomly-generated inputs and using ADIs we generated in
\S~\ref{subsec:whitebox-attack}. The random inputs are uniformly sampled from
the feature space of \ma's normal inputs. The results are shown in
\T~\ref{tab:ae-compare}.

As expected, the ADPs in the AE evaluation are much lower than that of ADIs, and
are comparable to the randomly-generated inputs. Similar observations are made
for the attack success rate evaluations as well. 
For instance, the average ADPs across different datasets are 82.39\% for ADIs, 42.29\% for AEs, and 42.70\% for randomly-generated inputs.
Similarly, the average attack success rates across different datasets are 58.79\% for ADIs, 3.44\% for AEs, and 10.63\% for randomly-generated inputs.

%
The results further demonstrate the effectiveness of ADIs. As explained in
\S~\ref{subsec:motivation-adi}, each AE aims to manipulate one specific
input of \mb, and it is agnostic to other inputs from \mb. Also, AE synthesis
does not explicitly consider diminishing the contribution of \mb. In contrast,
ADI synthesis aims to find an input that can dominate the model output and
control the majority of \mb's inputs. Moreover, ADIs explicitly minimize the
contribution of the benign participants.
In sum, findings in \T~\ref{tab:ae-compare} empirically illustrates the
distinction between AEs and ADIs; we conclude that AEs and randomly-sampled
inputs are much less effective in dominating the outputs of the VFL system
compared with ADIs.

\subsection{Mitigating ADIs with Two Defense Schemes}
\label{subsec:defense}
In this section, we explore the effectiveness of two common defense methods
against AEs: clustering-based detector and adversarial training.
Clustering-based detector is commonly used to detect the out-of-distribution
data, and it has been widely-used in mitigating conventional
AEs~\cite{bai2021clustering, tian2021analysis}. We use ADIs generated in
\S~\ref{subsec:whitebox-attack} (using gradient-based whitebox synthesis) and
collect the local model outputs when processing these ADIs. We then randomly
select the same number of normal inputs from \ma's test dataset and collect the
corresponding local model outputs. We further run the K-means clustering
algorithm~\cite{macqueen1967classification} to cluster the outputs into $K$ clusters, and we mark ADI clusters and
normal clusters according to the proportion of ADI outputs and normal data
outputs in each cluster. 
In practice, we found that $K=5$ is a good option for our datasets, because on
average, we have a few hundreds data samples for each clustering task. Users may
also generate more ADIs using our fuzz testing-based ADI discover algorithm.

We repeat the clustering procedure for ten times and report the average
detection accuracy and the average F1 scores in
\T~\ref{tab:defense-clustering}. The F1 score is defined as $F1 = \frac{2 *
\textit{precision} * \textit{recall}}{\textit{precision} + \textit{recall}} \in [0, 1]$. A higher F1 score indicates better performance.
The results show that our clustering-based detection method is reasonably
effective but cannot reach very high accuracy. 
We achieve over $60\%$ accuracy and over $50\%$ F1 score for most
datasets. Due to the complexity of the task and the size of the feature space,
detectors in the VQA v2.0 and CIFAR-10 evaluations are less effective. Also,
detectors under 99\% threshold are generally more accurate than the 95\%
case. Overall, we find that to generate ADIs under 99\% threshold, more
mutations are generally needed, and therefore, the ADIs often become
more distinguishable under 99\% threshold.

Overall, we deem the detection as effective, and it is a promising and
demanding direction to improve detection accuracy further. We also emphasize
that the clustering algorithm uses local model outputs instead of the raw
inputs. Thus, the results are not contradictory to the stealthiness evaluation
of ADIs. 
Moreover, the benign participant needs to access the intermediate outputs of the
adversarial participants, and need to know in advance whether the outputs are
generated by ADIs or normal inputs to tune the detector. Thus, deploying
the mitigation is challenging and under development, especially when there is no
trusted coordinator in VFL systems.

\begin{table}[!t]
	\captionsetup{skip=2pt}
	\centering
	\scriptsize
  \caption{Adversarial training results.}
	\label{tab:adv-training}
	\setlength{\tabcolsep}{1.5pt}
		\begin{tabularx}{0.85\linewidth}{ l|c|c
			| >{\centering\arraybackslash}X
			| >{\centering\arraybackslash}X}
			\hline
          & \textbf{Accuracy on} & \textbf{Accuracy on} & \multicolumn{2}{c}{\textbf{Attack Success Rate}} \\\cline{4-5}
          & \textbf{Test Dataset} & \textbf{AE Dataset} & \textbf{95\%} & \textbf{99\%} \\
			\hline
			\textbf{NUS-WIDE} & 73.13\% & 64.06\% & 34.9\% & 16.4\% \\
			\textbf{Credit}   & 0.7347 (auc-roc) & 77.40\% & 96.0\% & 40.8\% \\
			\textbf{Vehicle}  & 84.00\% & 80.00\% & 89.6\% & 87.7\% \\
			\textbf{MNIST}    & 97.93\% & 90.46\% & 33.4\% & 13.8\% \\
			\textbf{CIFAR-10} & 85.97\% & 66.97\% & 97.9\% & 27.8\% \\ 
			\hline
		\end{tabularx}
\end{table}

In addition, we also evaluate the gradient-based ADI synthesis under the
adversarial training defense that is designed to mitigate standard AEs. We use
adversarial training techniques from~\cite{tsipras2018robustness} to train
robust models, and then generate ADIs on the robust models using gradient-based
ADI synthesis. The setting is the same as \S~\ref{subsec:whitebox-attack}. The
results presented in \T~\ref{tab:adv-training} show that the accuracy of the
trained robust models on the test datasets are close to the non-robust models in
\T~\ref{tab:original-data-ats}. And the accuracy on the AE datasets are high,
i.e., all of the robust models achieve around or higher than $65\%$
accuracy. To compare, the non-robust models' accuracy on the AE datasets is
lower than $30\%$. Also, for the VQA task, it is non-trivial to perform
adversarial training on the large models like BERT~\cite{devlinetal2019bert}.
Thus, we omit the results for VQA v2.0 in \T~\ref{tab:adv-training}. Given that
said, we view our findings as convincing, which illustrate that adversarial
training is not effective to mitigate ADIs.

Moreover, the ADI attack success rates are comparable to that of the non-robust
models, whose results are in \T~\ref{tab:mutation} (the ``whitebox'' setting).
We therefore conclude that the adversarial training methods that designed for
mitigating AEs are not effective in mitigating ADIs. As we have discussed in
\S~\ref{sec:adi} and \S~\ref{subsec:eval-compare-uae}, the objectives of ADIs
and AEs are distinct, and evaluations here show that the AE defense methods
cannot be generalized to mitigate ADIs.


\subsection{Attack without Accessing \sd}
\label{subsec:weaker-threat-model}

This section assesses gradient-based ADI synthesis when \ma\ cannot
access the dataset \sd\ of \mb. \ma\ only knows the range of \mb's input data.
Aligned with settings in previous sections, we randomly generate 20 \mb's
inputs. The generated inputs are uniformly distributed in \mb's feature ranges.

We conduct the experiments on the blackbox setting with bounded mutation
strategy, and the settings are the same as \S~\ref{subsec:whitebox-attack}. The
results in \T~\ref{tab:blind} show that the ADI synthesis algorithm can
effectively generate ADIs. The attack success rates and the ADPs are reasonably
high compared to the results when \ma\ can access \sd. For instance, we achieve
15.5\% attack success rate under the 95\% dominating threshold on MNIST and the
ADP is 78.37\%. Recall the results, when \ma\ can access \sd, are 33.9\% and
85.82\%, respectively (shown in \T~\ref{tab:mutation} and
\T~\ref{tab:mutation-dominate-number}). The observations are similar on other
datasets. 
In sum, evaluations show that non-trivial amount of ADIs can be generated, even
though \ma\ does not have access to \mb's data. The results further emphasize
the feasibility of performing ADI attacks on real-world VFL systems.

\begin{table}[!t]
	\captionsetup{skip=2pt}
	\centering
	\scriptsize
  \caption{Blackbox bounded mutation without \sd.}
	\label{tab:blind}
	\setlength{\tabcolsep}{1.5pt}
		\begin{tabularx}{0.75\linewidth}{l|>{\centering\arraybackslash}X|>{\centering\arraybackslash}X|c|c|c}
			\hline
          & \multicolumn{2}{c}{\textbf{Attack Success Rate}} & \multicolumn{3}{|c}{\textbf{ADP}} \\
		  \cline{2-3}\cline{4-6}
          & \textbf{95\%} & \textbf{99\%} & \textbf{95\%} & \textbf{99\%} & \textbf{All}\\
			\hline
			\textbf{NUS-WIDE}  & 14.9\% & 11.3\% & 99.14\% & 99.90\% & 66.50\%\\
			\textbf{Credit}    & 54.0\% & 41.2\% & 99.20\% & 99.81\% & 93.88\% \\
			\textbf{Vehicle}   & 57.2\% & 56.2\% & 99.94\% & 99.98\% & 80.81\%  \\
			\textbf{MNIST}     & 15.5\% & 1.45\% & 96.81\% & 99.01\% & 78.37\% \\
			\textbf{VQA v2.0}  & NA & 21.4\% & NA & 100\% & 53.95\% \\
			\textbf{CIFAR-10}  & 69.3\% & 52.2\% & 99.18\% & 99.70\% & 80.89\% \\ 
			\hline
		\end{tabularx}
\end{table}

\section{Discussion}
\label{sec:discussion}
\noindent \textbf{ADI Mitigation with Fine-Tuning.}~We have
evaluated the clustering-based ADI detector and adversarial training techniques
in \S~\ref{subsec:defense}. However, they cannot achieve high accuracy to form a
practical solution to mitigate ADIs.

Careful readers might ask if the fine-tuning techniques using the
generated ADIs as training data can improve the robustness of the model to
defense against ADIs. At this step, we launch a tentative experiment to
fine-tune the model with the generated ADIs using bounded mutation under the
95\% threshold on MNIST and CIFAR-10. The accuracy of the MNIST model increases by 0.04\%, and the accuracy of the CIFAR-10 model decreases by 0.62\%. And we
successfully mitigate 64.2\% and 94.0\% of MNIST and CIFAR-10 ADIs. That is, after fine-tuning, only
35.8\% and 6.0\% of the previously found ADIs still achieve a dominating rate higher than
95\%, and we do not sacrifice much of the model's performance on the test dataset. Given
that said, we believe it is hard to completely eliminate future generation of
ADIs on the fine-tuned model, as long as the participants are making nontrivial
contributions to the model prediction. Holistically, fine-tuning fixes
some known ADIs on hand by boosting the contribution of benign parties (to
“un-dominate” adversarial participants who use ADIs). But this sheds light on a
concern, such that when those benign parties are exploited and become
``adversarial'' in the future, they are powerful enough to easily dominate other
parties with ADIs. In sum, with findings in \S~\ref{subsec:defense} and
explorations here, we see it as demanding (and technically challenging) to
propose specific ADI detection and mitigation techniques with high accuracy; we
leave it as one future work. Furthermore, we anticipate to leverage
fuzzing-based ADI discovery algorithm to continously gather ADIs and fine-tune
VFL systems. We foresee a stage when the fuzzing process can hardly find
sufficient ADIs, indicating that the continuously-tuned VFL systems have
acquired a high level of robustness.
We provide further discussions on attacking tree-based models and
selection of \sd\ in Appendix \ref{subsec:append-discuss}.
\section{Conclusion}

This paper exploits VFL using ADIs. ADIs control the joint inference and
diminish benign clients' contribution. We first prove that ADIs exist in common
VFL. We then propose both gradient-based ADI synthesis and fuzz testing for
developers to perform in-house vulnerability assessment. We assess the impact of
various settings on ADI generation. Our study exposes novel VFL attack vectors,
promoting early detection of unknown threats and more secure VFL.
\section*{Acknowledgment}

We thank anonymous reviewers for their valuable feedback. 
The HKUST authors were supported in part by the research fund provided 
by HSBC.


\bibliographystyle{plain}
\bibliography{main.bib}


\appendices

\renewcommand{\thesectiondis}[2]{\Alph{section}:}
\section{Proof of Corollary~\ref{cor:vfl-var-bound}}
\label{subsec:append-proof}

In this section, we present the detailed proof on the existence of ADIs in HeteroLR and SplitNN.

\noindent \textbf{ADI in HeteroLR.}~Participant \ma\ holds the data $X_{\mathcal{A}} \in D_{\mathcal{A}}$, whereas
participant \mb\ holds the data $X_{\mathcal{B}} \in D_{\mathcal{B}}$ following
any distribution whose density function is $p(X_{\mathcal{B}})$.
\ma\ is controlled by an adversary, whereas \mb\ behaves normally.
$\theta_{\mathcal{A}}$ and $\theta_{\mathcal{B}}$ are their corresponding
coefficients. The HeteroLR $f$ takes $X_{\mathcal{A}}$ and $X_{\mathcal{B}}$ as inputs, and its output is:

\small
\setlength{\belowdisplayskip}{2pt} \setlength{\belowdisplayshortskip}{2pt}
\setlength{\abovedisplayskip}{-3pt} \setlength{\abovedisplayshortskip}{-3pt}
\begin{equation*}
    \begin{aligned}
        f(X_{\mathcal{A}}, X_{\mathcal{B}}) = g(\theta_{\mathcal{A}}^T \, X_{\mathcal{A}} + \theta_{\mathcal{B}}^T \, X_{\mathcal{B}}) \; ,
    \end{aligned}
\end{equation*}
\normalsize

\noindent where $X_{\mathcal{A}} \in \mathbb{R}^{d_1}$, $\theta_{\mathcal{A}} \in \mathbb{R}^{d_1}$,  $X_{\mathcal{B}} \in \mathbb{R}^{d_2}$, $\theta_{\mathcal{B}} \in \mathbb{R}^{d_2}$. $d_1, d_2$ denote
dimensions of the features in \ma\ and \mb, $g(t) = \frac{1}{1+e^{-t}}$ is the Sigmoid function. We have the following corollary:

\begin{corollary}[Variance of HeteroLR]
With fixed input $X_{\mathcal{A}}^*$ and varying input $X_{\mathcal{B}} \in D_{\mathcal{B}}$, the output variance of HeteroLR is:

\small
\begin{equation*}
    \begin{aligned}
        \mathbb{V}_{X_{\mathcal{B}} \in D_{\mathcal{B}}} &(f(X_{\mathcal{A}}^*, X_{\mathcal{B}})) 
        = \sum_{k=1}^{\mathcal{K}}  \, - \frac{\pi_k}{\sqrt{2\pi}} \frac{1}{\sqrt{\sigma_k'^2 + \sigma_2^2}} \exp\{ -\frac{1}{2} \frac{\mu_k'^2}{\sigma_k'^2 + \sigma_2^2} \}\\
        & \; + \sum_{k=1}^{\mathcal{K}} \pi_k \, \Phi(\frac{\mu_k'}{\sqrt{\sigma_1^{2} + \sigma_k'^2}}) (1 -  \sum_{k=1}^{\mathcal{K}} \pi_k \, \Phi(\frac{\mu_k'}{\sqrt{\sigma_1^{2} + \sigma_k'^2}})) \; ,
    \end{aligned}
\end{equation*}
\normalsize
\noindent where the density function $p(X_{\mathcal{B}})$ is approximated by Gaussian Mixture Model: $p(X_{\mathcal{B}}) \approx \sum_{k=1}^{\mathcal{K}} \pi_k \, \mathcal{N}(X_{\mathcal{B}} | \boldsymbol{\mu_k}, \boldsymbol{\Sigma_k})$, $ \sum_{k=1}^{\mathcal{K}} \pi_k = 1$, $ \pi_k > 0 $, $\mathcal{K}$ is a finite number, $\boldsymbol{\mu_k} \in \mathbb{R}^{d_2}$, $\boldsymbol{\Sigma_k} \in \mathbb{R}^{d_2 \times d_2}$, $\mu_k' = \theta_{\mathcal{A}}^T \, X_{\mathcal{A}}^* + \theta_{\mathcal{B}}^T \, \boldsymbol{\mu_k}$, $\sigma_k'^2 = \theta_{\mathcal{B}}^T \, \boldsymbol{\Sigma_k} \, \theta_{\mathcal{B}}$, $\Phi$ is the cumulative distribution function (CDF) of the standard normal distribution, $\sigma_1 = 1.699$, and $\sigma_2 = 1.630$.

\label{cor:adi-in-VFLR}

\end{corollary}

\begin{proof}
Let $ S = \theta_{\mathcal{A}}^T
\, X_{\mathcal{A}}^* + \theta_{\mathcal{B}}^T \, X_{\mathcal{B}} $. Thus, $f(X_{\mathcal{A}}^*, X_{\mathcal{B}}) = g(S)$.
Aligned with the conventions~\cite{lee2017deep, makhzani2015adversarial, hjelm2018learning, belghazi2018mutual, fei2006one, zhu1996region} in machine learning, we use Gaussian mixture model~\cite{mclachlan1988mixture} and Expectation Maximization~\cite{dempster1977maximum} to approximate the density function $p(X_{\mathcal{B}})$. Note that GMMs can approximate any smooth distributions~\cite{goodfellow2016deep}.

\small
\setlength{\belowdisplayskip}{2pt} \setlength{\belowdisplayshortskip}{2pt}
\setlength{\abovedisplayskip}{-5pt} \setlength{\abovedisplayshortskip}{-5pt}
\begin{equation*}
    \begin{aligned}
        p(X_{\mathcal{B}}) \approx \sum_{k=1}^{\mathcal{K}} \pi_k \, \mathcal{N}(X_{\mathcal{B}} | \boldsymbol{\mu_k}, \boldsymbol{\Sigma_k}) \; ,
    \end{aligned}
\end{equation*}
\normalsize
\noindent where $ \sum_{k=1}^{\mathcal{K}} \pi_k = 1$, $ \pi_k > 0 $, $\mathcal{K}$ is a finite number, $\boldsymbol{\mu_k} \in \mathbb{R}^{d_2}$ and $\boldsymbol{\Sigma_k} \in \mathbb{R}^{d_2 \times d_2}$.
If $p(X_{\mathcal{B}})$ is discrete and not smooth (e.g. $X_{\mathcal{B}}$ is categorical data), we can approximate it by taking $\boldsymbol{\Sigma_k} \rightarrow \boldsymbol{0}$ and $\mathcal{K}$ becomes the total number of distinct data in $X_{\mathcal{B}}$, the Gaussian distribution density would become a Dirac delta function~\cite{dirac1981principles} and the following calculations still hold. 

Further, we can obtain the
density function of $S$ as: $ p(S) = \sum_{k=1}^{\mathcal{K}} \pi_k \, \mathcal{N}(S | \theta_{\mathcal{A}}^T \, X_{\mathcal{A}}^* + \theta_{\mathcal{B}}^T \,
\boldsymbol{\mu_k}, \theta_{\mathcal{B}}^T \, \boldsymbol{\Sigma_k} \, \theta_{\mathcal{B}}) $. Taking $\mu_k'
= \theta_{\mathcal{A}}^T \, X_{\mathcal{A}}^* + \theta_{\mathcal{B}}^T \, \boldsymbol{\mu_k}$ and $\sigma_k'^2 =
\theta_{\mathcal{B}}^T \, \boldsymbol{\Sigma_k} \, \theta_{\mathcal{B}}$, we thus calculate the expected value of
the output:

\small
\setlength{\belowdisplayskip}{2pt} \setlength{\belowdisplayshortskip}{2pt}
\setlength{\abovedisplayskip}{-5pt} \setlength{\abovedisplayshortskip}{-5pt}
\begin{equation}
    \begin{aligned}
        \mathbb{E}&_{X_{\mathcal{B}} \in D_{\mathcal{B}}} (g(S)) 
        = \int g(t) \, \sum_{k=1}^{\mathcal{K}} \pi_k \, \mathcal{N}(t | \mu_k', \sigma_k'^2) \, dt \\
        &\approx \int \sum_{k=1}^{\mathcal{K}} \pi_k \, \Phi(\frac{t}{\sigma_1}) \, \mathcal{N}(t| \mu_k', \sigma_k'^2) \, dt
        = \sum_{k=1}^{\mathcal{K}} \pi_k \, \Phi(\frac{\mu_k'}{\sqrt{\sigma_1^{2} + \sigma_k'^2}})
        \label{formula:sig-expectation}
    \end{aligned}
\end{equation}
\normalsize
\noindent Here, the $g(t)$ is approximated by the cumulative distribution function of standard Gaussian distribution and parameter $\sigma_1$ as $\Phi(\frac{t}{\sigma_1})$, where $\sigma_1 = 1.699$ to minimize the $L^2$ error. 
The variance of the output is calculated as follows:

\small
\setlength{\belowdisplayskip}{2pt} \setlength{\belowdisplayshortskip}{2pt}
\setlength{\abovedisplayskip}{-5pt} \setlength{\abovedisplayshortskip}{-5pt}
\begin{equation}
    \begin{aligned}
        \mathbb{V}_{X_{\mathcal{B}} \in D_{\mathcal{B}}} (g(S)) 
        &= \left \langle g(S)^2 \right \rangle - \left \langle g(S) \right \rangle^2 \\
        &= \left \langle g(S) - g(S)(1-g(S)) \right \rangle - \left \langle g(S) \right \rangle^2 \\
        &= \left \langle g(S) \right \rangle (1 - \left \langle g(S) \right \rangle) - \left \langle g'(S) \right \rangle \; ,
        \label{formula:sig-variance}
    \end{aligned}
\end{equation}
\normalsize
where $\left \langle \cdot \right \rangle$ donates $\mathbb{E}_{X_{\mathcal{B}} \in D_{\mathcal{B}}} ( \cdot )$. Further, we approximate the derivation of Sigmoid function $g'(S)$ using the pdf of Gaussian distribution. 
When the $L^2$ error is minimized, we get the approximated function of $g'(S)$ with $\sigma_2 = 1.630$:

\small
\setlength{\belowdisplayskip}{2pt} \setlength{\belowdisplayshortskip}{2pt}
\setlength{\abovedisplayskip}{-5pt} \setlength{\abovedisplayshortskip}{-5pt}
\begin{equation*}
    \begin{aligned}
        g'(S) \approx \frac{1}{\sqrt{2\pi}\sigma_2} \exp\{-\frac{S^2}{2\sigma_2^2}\}
    \end{aligned}
\end{equation*}
\normalsize

Further, we can calculate the expected value of the derivation of Sigmoid function:

\small
\setlength{\belowdisplayskip}{2pt} \setlength{\belowdisplayshortskip}{2pt}
\setlength{\abovedisplayskip}{-5pt} \setlength{\abovedisplayshortskip}{-5pt}
\begin{equation}
    \begin{aligned}
        \label{formula:derivation-sigmoid-expectation}
        \mathbb{E}_{X_{\mathcal{B}} \in D_{\mathcal{B}}} (g'(S)) &= \int \sum_{k=1}^{\mathcal{K}} \, \frac{\pi_k}{2\pi\sigma_k'\sigma_2} \exp\{ -\frac{t^2}{2\sigma_2^2} - \frac{(t - \mu_k')^2}{2\sigma_k'^2} \} dt \\
        &= \sum_{k=1}^{\mathcal{K}} \, \frac{\pi_k}{\sqrt{2\pi}} \frac{1}{\sqrt{\sigma_k'^2 + \sigma_2^2}} \exp\{ -\frac{1}{2} \frac{\mu_k'^2}{\sigma_k'^2 + \sigma_2^2} \}
    \end{aligned}
\end{equation}
\normalsize

According to \E~\ref{formula:sig-expectation}, \ref{formula:sig-variance} and \ref{formula:derivation-sigmoid-expectation}:

\small
\setlength{\belowdisplayskip}{2pt} \setlength{\belowdisplayshortskip}{2pt}
\setlength{\abovedisplayskip}{-5pt} \setlength{\abovedisplayshortskip}{-5pt}
\begin{equation*}
    \begin{aligned}
        \mathbb{V}_{X_{\mathcal{B}} \in D_{\mathcal{B}}} & (g(S)) = \,\sum_{k=1}^{\mathcal{K}}  \, - \frac{\pi_k}{\sqrt{2\pi}} \frac{1}{\sqrt{\sigma_k'^2 + \sigma_2^2}} \exp\{ -\frac{1}{2} \frac{\mu_k'^2}{\sigma_k'^2 + \sigma_2^2} \}\\
        & + \sum_{k=1}^{\mathcal{K}} \pi_k \, \Phi(\frac{\mu_k'}{\sqrt{\sigma_1^{2} + \sigma_k'^2}}) (1 -  \sum_{k=1}^{\mathcal{K}} \pi_k \, \Phi(\frac{\mu_k'}{\sqrt{\sigma_1^{2} + \sigma_k'^2}}))
    \end{aligned}\qedhere
\end{equation*}
\normalsize
\end{proof}

\noindent With unbounded $X_{\mathcal{A}}^*$ and $||\theta_{\mathcal{A}}||_1 > 0$, the
range of $\theta_{\mathcal{A}}^T \, X_{\mathcal{A}}^*$ is $(-\infty, +\infty)$.
Thus, $\Lim{\theta_{\mathcal{A}}^T \, X_{\mathcal{A}}^* \rightarrow -\infty} \, \mu_k' = -\infty$ and  
$\Lim{\mu_k' \rightarrow -\infty} \mathbb{V}_{X_{\mathcal{B}} \in D_{\mathcal{B}}} (f(X_{\mathcal{A}}^*, X_{\mathcal{B}})) = 0^+$. For any $\epsilon > 0$, there exists $X_{\mathcal{A}}^*$ satisfying
$\mathbb{V}_{X_{\mathcal{B}} \in D_{\mathcal{B}}} (f(X_{\mathcal{A}}^*, X_{\mathcal{B}})) \leq \epsilon$. According to Def.~\ref{def:ADI}, $X_{\mathcal{A}}^*$ is an ADI for HeteroLR.

\noindent \textbf{ADI in SplitNN.}~~Similarly, in SplitNN, $X_{\mathcal{A}} \in D_{\mathcal{A}}$ and $X_{\mathcal{B}} \in D_{\mathcal{B}}$ are the corresponding local outputs from participants \ma\ and \mb, and
the coordinator model is a single-layer fully connected network, with ReLU as its activation function. SplitNN $f$ takes $X_{\mathcal{A}}$ and $X_{\mathcal{B}}$ as inputs, and its output is:

\small
\setlength{\belowdisplayskip}{2pt} \setlength{\belowdisplayshortskip}{2pt}
\setlength{\abovedisplayskip}{-3pt} \setlength{\abovedisplayshortskip}{-3pt}
\begin{equation*}
    \begin{aligned}
        f(X_{\mathcal{A}}, X_{\mathcal{B}}) = ReLU(w^T \, [X_{\mathcal{A}} \, || \, X_{\mathcal{B}}])
        = ReLU(w_{\mathcal{A}}^T \, X_{\mathcal{A}} + w_{\mathcal{B}}^T \, X_{\mathcal{B}}),
    \end{aligned}
\end{equation*}
\normalsize

\noindent where $||$ is concatenation and $w$ denotes the parameter of the coordinator
model. $X_{\mathcal{A}} \in \mathbb{R}^{d_1}$, $w_{\mathcal{A}} \in
\mathbb{R}^{d_1}$, $X_{\mathcal{B}} \in \mathbb{R}^{d_2}$ and $w_{\mathcal{B}}
\in \mathbb{R}^{d_2}$. $d_1, d_2$ are dimensions of the outputs in \ma\ and \mb. 
$X_{\mathcal{B}}$ follows any distribution whose density function is $p(X_{\mathcal{B}})$.
We have the following corollary.

\vspace{-0.5em}
\begin{corollary}[Variance of SplitNN]
With fixed input $X_{\mathcal{A}}^*$ and varying input $X_{\mathcal{B}} \in D_{\mathcal{B}}$, the output variance of SplitNN is:

\small
\begin{equation*}
    \begin{aligned}
        \mathbb{V}&_{X_{\mathcal{B}} \in D_{\mathcal{B}}} (f(X_{\mathcal{A}}^*,X_{\mathcal{B}}))\\
        & = (\sum_{k=1}^{\mathcal{K}} \pi_k \, \Phi(\frac{\mu_k'}{\sigma_k'})) \, 
        (\sum_{k=1}^{\mathcal{K}} \pi_k^2 \, \sigma_k'^2(1 - \frac{\mu_k'}{\sigma_k'} \frac{\phi(\frac{\mu_k'}{\sigma_k'})}{\Phi(\frac{\mu_k'}{\sigma_k'})} - (\frac{\phi(\frac{\mu_k'}{\sigma_k'})}{\Phi(\frac{\mu_k'}{\sigma_k'})})^2) \\
        & \; + (\sum_{k=1}^{\mathcal{K}} \pi_k \, (\mu_k' + \sigma_k' \frac{\phi(\frac{\mu_k'}{\sigma'_k})}{\Phi(\frac{\mu_k'}{\sigma_k'})}))^2
        (1 - \sum_{k=1}^{\mathcal{K}} \pi_k \, \Phi(\frac{\mu_k'}{\sigma_k'}))) \; ,
    \end{aligned}
\end{equation*}
\normalsize
\noindent where $p(X_{\mathcal{B}})$ is approximated by Gaussian Mixture Model: $p(X_{\mathcal{B}}) \approx \sum_{k=1}^{\mathcal{K}} \pi_k \, \mathcal{N}(X_{\mathcal{B}} | \boldsymbol{\mu_k}, \boldsymbol{\Sigma_k})$, $ \sum_{k=1}^{\mathcal{K}} \pi_k = 1$, $ \pi_k > 0 $, $\mathcal{K}$ is a finite number, $\mu_k' = w_{\mathcal{A}}^T \, X_{\mathcal{A}}^* + w_{\mathcal{B}}^T \, \boldsymbol{\mu_k}$, $\sigma_k'^2 = w_{\mathcal{B}}^T \, \boldsymbol{\Sigma_k} \, w_\mathcal{B}$, and $\Phi$ is the CDF of the standard normal distribution.

\label{cor:adi-in-VFNN}

\end{corollary}

\begin{proof}
    Similar to the proof of Cor.~\ref{cor:adi-in-VFLR}, for any distribution of $X_{\mathcal{B}}$, we use Gaussian mixture model and Expectation Maximization to approximate its density function $p(X_{\mathcal{B}})$ as:
    
    \small
    \setlength{\belowdisplayskip}{2pt} \setlength{\belowdisplayshortskip}{2pt}
    \setlength{\abovedisplayskip}{-5pt} \setlength{\abovedisplayshortskip}{-5pt}
    \begin{equation*}
        \begin{aligned}
            p(X_{\mathcal{B}}) \approx \sum_{k=1}^{\mathcal{K}} \pi_k \, \mathcal{N}(X_{\mathcal{B}} | \boldsymbol{\mu_k}, \boldsymbol{\Sigma_k}) \; ,
        \end{aligned}
    \end{equation*}
    \normalsize
    
    \noindent where $ \sum_{k=1}^{\mathcal{K}} \pi_k = 1$, $ \pi_k > 0 $, $\mathcal{K}$ is a finite number, $\boldsymbol{\mu_k} \in \mathbb{R}^{d_2}$ and $\boldsymbol{\Sigma_k} \in \mathbb{R}^{d_2 \times d_2}$. 
    Similarly, if $p(X_{\mathcal{B}})$ is highly discrete, we can approximate it by taking $\boldsymbol{\Sigma_k} \rightarrow \boldsymbol{0}$. 
    
    For fixed $X_{\mathcal{A}}^*$, let $Y = w_{\mathcal{A}}^T
    \, X_{\mathcal{A}}^* + w_{\mathcal{B}}^T \, X_{\mathcal{B}}$, $\mu_k' = w_{\mathcal{A}}^T
    \, X_{\mathcal{A}}^* + w_{\mathcal{B}}^T \, \boldsymbol{\mu_k}$, and $\sigma_k'^2 =
    w_{\mathcal{B}}^T \, \boldsymbol{\Sigma_k} \, w_\mathcal{B}$. Then, $p(Y) =
    \sum_{k=1}^{\mathcal{K}} \pi_k \, \mathcal{N} (Y | \mu_k', \sigma_k'^2)$. We have: 
    
    \small
    \setlength{\belowdisplayskip}{2pt} \setlength{\belowdisplayshortskip}{2pt}
    \setlength{\abovedisplayskip}{-5pt} \setlength{\abovedisplayshortskip}{-5pt}
    \begin{equation}
        \begin{aligned}
            \mathbb{E}_{X_{\mathcal{B}} \in D_{\mathcal{B}}} (Y | Y > 0) = \sum_{k=1}^{\mathcal{K}} \pi_k \, (\mu_k' + \sigma_k' \frac{\phi(\frac{\mu_k'}{\sigma'_k})}{\Phi(\frac{\mu_k'}{\sigma_k'})})
        \end{aligned}
        \label{formula:e-cond}
    \end{equation}
    
    \setlength{\belowdisplayskip}{2pt} \setlength{\belowdisplayshortskip}{2pt}
    \setlength{\abovedisplayskip}{-5pt} \setlength{\abovedisplayshortskip}{-5pt}
    \begin{equation}
        \begin{aligned}
            \mathcal{P} (Y > 0) = \sum_{k=1}^{\mathcal{K}} \pi_k \, \Phi(\frac{\mu_k'}{\sigma_k'})
        \end{aligned}
        \label{formula:p-cond}
    \end{equation}
    \normalsize
    And we get the expected central coordinator output:
    
    \small
    \setlength{\belowdisplayskip}{2pt} \setlength{\belowdisplayshortskip}{2pt}
    \setlength{\abovedisplayskip}{-5pt} \setlength{\abovedisplayshortskip}{-5pt}
    \begin{equation}
        \begin{aligned}
            \mathbb{E}_{X_{\mathcal{B}} \in D_{\mathcal{B}}} \, &(f(X_{\mathcal{A}}^*, X_{\mathcal{B}})) = \mathbb{E}_{X_{\mathcal{B}} \in D_{\mathcal{B}}} (Y | Y > 0) \times \mathcal{P} (Y > 0) \\
            =& (\sum_{k=1}^{\mathcal{K}} \pi_k \, (\mu_k' + \sigma_k' \frac{\phi(\frac{\mu_k'}{\sigma'_k})}{\Phi(\frac{\mu_k'}{\sigma_k'})}))
               \, (\sum_{k=1}^{\mathcal{K}} \pi_k \, \Phi(\frac{\mu_k'}{\sigma_k'}))
        \end{aligned}
        \label{formula:e-relu}
    \end{equation}
    \normalsize
    The variance of the truncated mixture Gaussian distribution:
    
    \small
    \setlength{\belowdisplayskip}{2pt} \setlength{\belowdisplayshortskip}{2pt}
    \setlength{\abovedisplayskip}{-5pt} \setlength{\abovedisplayshortskip}{-5pt}
    \begin{equation}
        \begin{aligned}
            \mathbb{V}_{X_{\mathcal{B}} \in D_{\mathcal{B}}} & (Y | Y > 0) 
            = \sum_{k=1}^{\mathcal{K}} \pi_k^2 \, \sigma_k'^2(1 - \frac{\mu_k'}{\sigma_k'} \frac{\phi(\frac{\mu_k'}{\sigma_k'})}{\Phi(\frac{\mu_k'}{\sigma_k'})} - (\frac{\phi(\frac{\mu_k'}{\sigma_k'})}{\Phi(\frac{\mu_k'}{\sigma_k'})})^2)
        \end{aligned}
        \label{formula:v-Y}
    \end{equation}
    \normalsize
    
    According to \E~\ref{formula:e-cond}, \ref{formula:p-cond}, \ref{formula:e-relu} and \ref{formula:v-Y}, we have:
    
    \small
    \setlength{\belowdisplayskip}{2pt} \setlength{\belowdisplayshortskip}{2pt}
    \setlength{\abovedisplayskip}{-5pt} \setlength{\abovedisplayshortskip}{-5pt}
    \begin{equation*}
        \begin{aligned}
            \mathbb{E}_{X_{\mathcal{B}} \in D_{\mathcal{B}}} (Y^2 | Y > 0) 
            = \, \mathbb{V}_{X_{\mathcal{B}} \in D_{\mathcal{B}}} (Y | Y > 0) + (\mathbb{E}_{X_{\mathcal{B}} \in D_{\mathcal{B}}} (Y | Y > 0))^2
        \end{aligned}
        \label{formula:e-squareY}
    \end{equation*}
    \normalsize
    
    \noindent Therefore, the variance of the coordinator model's output is:
    
    \small
    \setlength{\belowdisplayskip}{2pt} \setlength{\belowdisplayshortskip}{2pt}
    \setlength{\abovedisplayskip}{-5pt} \setlength{\abovedisplayshortskip}{-5pt}
    \begin{equation*}
        \begin{aligned}
            \mathbb{V}_{X_{\mathcal{B}} \in D_{\mathcal{B}}} &(f(X_{\mathcal{A}}^*,X_{\mathcal{B}})) = \,
            \mathbb{E}_{X_{\mathcal{B}} \in D_{\mathcal{B}}} (Y^2 | Y > 0) \times \mathcal{P} (Y > 0) 
            \\ & \qquad \qquad \qquad \qquad \, -(\mathbb{E}_{X_{\mathcal{B}} \in D_{\mathcal{B}}} \, (f(X_{\mathcal{A}}^*, X_{\mathcal{B}})))^2\\
            =& (\sum_{k=1}^{\mathcal{K}} \pi_k \, \Phi(\frac{\mu_k'}{\sigma_k'})) \, 
            (\sum_{k=1}^{\mathcal{K}} \pi_k^2 \, \sigma_k'^2(1 - \frac{\mu_k'}{\sigma_k'} \frac{\phi(\frac{\mu_k'}{\sigma_k'})}{\Phi(\frac{\mu_k'}{\sigma_k'})} - (\frac{\phi(\frac{\mu_k'}{\sigma_k'})}{\Phi(\frac{\mu_k'}{\sigma_k'})})^2) \\
            & + (\sum_{k=1}^{\mathcal{K}} \pi_k \, (\mu_k' + \sigma_k' \frac{\phi(\frac{\mu_k'}{\sigma'_k})}{\Phi(\frac{\mu_k'}{\sigma_k'})}))^2
            (1 - \sum_{k=1}^{\mathcal{K}} \pi_k \, \Phi(\frac{\mu_k'}{\sigma_k'})))
        \end{aligned}\qedhere
        \label{formula:v-relu}
    \end{equation*}
    \normalsize
        
\end{proof}

\noindent According to Cor.~\ref{cor:adi-in-VFNN}, with unbounded $X_{\mathcal{A}}^*$ and $||w_{\mathcal{A}}||_1 > 0$, 
the range of $w_{\mathcal{A}}^T \, X_{\mathcal{A}}^*$ is $(-\infty, +\infty)$. 
Thus, $\Lim{ w_{\mathcal{A}}^T \, X_{\mathcal{A}}^* \rightarrow -\infty} \, \mu_k' = -\infty$ and 
$\Lim{\mu_k' \rightarrow -\infty} \mathbb{V}_{X_{\mathcal{B}} \in D_{\mathcal{B}}}
(f(X_{\mathcal{A}}^*, X_{\mathcal{B}})) = 0^+$. 
For any $\epsilon > 0$, there must exists $X_{\mathcal{A}}^*$ satisfying $\mathbb{V}_{X_{\mathcal{B}} \in D_{\mathcal{B}}}
(f(X_{\mathcal{A}}^*,X_{\mathcal{B}})) \leq \epsilon$. 
Thus, according to Def.~\ref{def:ADI}, $X_{\mathcal{A}}^*$ is an ADI for SplitNN.
\section{Proof on the Existence of ADIs Using Bounded Mutation}
\label{sec:adi-bounded}

In line with the proof given in \S~\ref{sec:adi} by arbitrarily mutating inputs,
this section presents the following proof on the existence of ADIs using bounded
mutation. 

\begin{proof}
Let $X_{\mathcal{A}}^*$ be bounded by a space
$\mathcal{S}_{\mathcal{A}}$ in $d_{\mathcal{A}}$ dimension, $X_{\mathcal{A}}^*
\in \mathcal{S}_{\mathcal{A}}$. For HeteroLR, $\mu_k'$s are linear combinations of
features in $X_{\mathcal{A}}^*$; therefore, they are also bounded by two limited
real numbers: $\mu_k' \in [\mu_{min}' \, , \, \mu_{max}']$. Further, the ratio
$\frac{\mu_k'}{\sqrt{\sigma_1^{2} + \sigma_k'^2}} $ is bounded by $[r_{min} \, ,
  \, r_{max}]$ and the variance is also bounded:

\small
\setlength{\belowdisplayskip}{3pt} \setlength{\belowdisplayshortskip}{3pt}
\setlength{\abovedisplayskip}{0pt} \setlength{\abovedisplayshortskip}{0pt}
\begin{equation*}
    \begin{aligned}
        \mathbb{V}_{X_{\mathcal{B}} \in D_{\mathcal{B}}} & (g(S)) = \,\sum_{k=1}^{\mathcal{K}}  \, - \frac{\pi_k}{\sqrt{2\pi}} \frac{1}{\sqrt{\sigma_k'^2 + \sigma_2^2}} \exp\{ -\frac{1}{2} \frac{\mu_k'^2}{\sigma_k'^2 + \sigma_2^2} \}\\
        \qquad & + \, \sum_{k=1}^{\mathcal{K}} \pi_k \, \Phi(\frac{\mu_k'}{\sqrt{\sigma_1^{2} + \sigma_k'^2}}) (1 -  \sum_{k=1}^{\mathcal{K}} \pi_k \, \Phi(\frac{\mu_k'}{\sqrt{\sigma_1^{2} + \sigma_k'^2}}))\\
        < \, & \sum_{k=1}^{\mathcal{K}} \, \pi_k \,\Phi(\frac{\mu_k'}{\sqrt{\sigma_1^{2} + \sigma_k'^2}}) - 
        \frac{1}{\mathcal{K}} (\sum_{k=1}^{\mathcal{K}} \, \pi_k \,\Phi(\frac{\mu_k'}{\sqrt{\sigma_1^{2} + \sigma_k'^2}}))^2 \\
        < \, & \Phi(r_{max}) - \frac{1}{\mathcal{K}} \Phi(r_{max})^2
    \end{aligned}
\end{equation*}
\normalsize

If $\, r_{max}$ satisfies $\, \Phi(r_{max}) \, \le \, \frac{\mathcal{K} + \sqrt{\mathcal{K}^2 - 4 \epsilon \mathcal{K}}}{2}$ 
, the variance must be bounded by
$\epsilon$, $\mathbb{V}_{X_{\mathcal{B}} \in D_{\mathcal{B}}} (g(S)) \,
\le \, \epsilon$, which means the ADI $X_{\mathcal{A}}^*$ 
generated by bounded mutation exists.

Similarly, for SplitNN, $\mu_k'$s are also bounded, $\mu_k' \in [\mu_{min}'\, , \, \mu_{max}']$. Similarly, $\sigma_k' \in [\sigma_{min}'\, , \, \sigma_{max}']$ and $\frac{\mu_k'}{\sigma_k'} \in [r_{min}\, , \, r_{max}]$. 
When $\mu_k' \leq 0$, the variance is bounded by:

\small
\setlength{\belowdisplayskip}{3pt} \setlength{\belowdisplayshortskip}{3pt}
\setlength{\abovedisplayskip}{0pt} \setlength{\abovedisplayshortskip}{0pt}
\begin{equation*}
    \begin{aligned}
        \mathbb{V}_{X_{\mathcal{B}} \in D_{\mathcal{B}}}& (f(X_{\mathcal{A}}^*,X_{\mathcal{B}})) = \,
        \mathbb{E}_{X_{\mathcal{B}} \in D_{\mathcal{B}}} (Y^2 | Y > 0) \times \mathcal{P} (Y > 0)\\ 
        & \qquad \qquad \qquad \qquad \qquad  - \, (\mathbb{E}_{X_{\mathcal{B}} \in D_{\mathcal{B}}} \, (f(X_{\mathcal{A}}^*, X_{\mathcal{B}})))^2\\
         = \, & (\mathbb{E}_{X_{\mathcal{B}} \in D_{\mathcal{B}}} (Y | Y > 0))^2 \times \mathcal{P} (Y > 0) (1 - \mathcal{P} (Y > 0))\\
        & \qquad \qquad \qquad \qquad \qquad  + \, \mathbb{V}_{X_{\mathcal{B}} \in D_{\mathcal{B}}} (Y | Y > 0) \times \mathcal{P} (Y > 0)\\
        \leq \, & \mathbb{V}_{X_{\mathcal{B}} \in D_{\mathcal{B}}} (Y | Y > 0) \times \Phi(r_{max}) + \, \frac{1}{4} (\mathbb{E}_{X_{\mathcal{B}} \in D_{\mathcal{B}}} (Y | Y > 0))^2\\
        < \, & \Phi(r_{max}) + \frac{1}{4} \Phi(r_{max}) \mu_{min}'^2 + \frac{1}{4} \mu_{max}'^2 
        + \, \frac{1}{4} \sigma_{max}'^2 \frac{\phi(r_{max})^2}{\Phi(r_{min})^2}
    \end{aligned}
\end{equation*}
\normalsize

If the boundaries satisfy:
\small
\setlength{\belowdisplayskip}{3pt} \setlength{\belowdisplayshortskip}{3pt}
\setlength{\abovedisplayskip}{0pt} \setlength{\abovedisplayshortskip}{0pt}
\begin{equation*}
  \begin{aligned}
 \Phi(r_{max}) + \frac{1}{4} \Phi(r_{max}) \mu_{min}'^2 + \frac{1}{4} \mu_{max}'^2
+ \, \frac{1}{4} \sigma_{max}'^2 \frac{\phi(r_{max})^2}{\Phi(r_{min})^2} \leq \, \epsilon \; ,
\end{aligned}
\end{equation*}
\normalsize
\noindent the variance must be bounded by $\epsilon$: $\mathbb{V}_{X_{\mathcal{B}} \in
  D_{\mathcal{B}}} (f(x_{\mathcal{A}}^*,x_{\mathcal{B}})) \, \leq \,\epsilon$. Thus, the ADI $X_{\mathcal{A}}^*$ generated by bounded mutation exists. \qedhere
\end{proof}
\section{Further Discussion}
\label{subsec:append-discuss}

\noindent \textbf{Attack on Tree-based Models.}~This paper examines
parameterized VFL systems. In tree-based VFL
systems~\cite{wu2020privacy,cheng2021secureboost}, each of the parties will hold several nodes of the tree.
The structure of the
malicious participant's tree is explicitly revealed, allowing them to easily
modify their inputs to reach the target node of their tree and control the output at
their will. However, if some inputs from the other participants never reach the malicious participant's node,
they cannot dominate the outputs of those inputs. We deem it an interesting future
work to study ADIs in tree-based VFL.

\noindent \textbf{Selection of \sd.}~In evaluation, we select \sd\ from the
test dataset of \mb\ randomly. According to our analysis in
\S~\ref{subsec:uae} and our evaluation, we don't have strict requirements for
\sd; randomly selecting \sd\ is sufficient to synthesize ADIs. However, if
\sd\ is highly biased (e.g., the labels of samples in \sd\ are the same), then
the success rate would be lower. In practice, attackers can manually check the
distribution of \sd\ to avoid a highly biased situation.
We deem it an interesting topic to further study generating ADIs using biased \sd.


\section{Exploring The Size of $\mdutchcal{S}$}
\label{subsec:eval-subset-size}

Synthesizing ADIs requires preparing a sample dataset \sd\ that follows the
distribution of \mb's standard inputs. \ssd\ is 20 for the above experiments,
meaning that we \textit{randomly} select 20 sample inputs from the test dataset
of \mb\ to form \sd. This section explores how \ssd\ influences ADI synthesis.

We use MNIST and CIFAR-10 over two participants \ma\ and \mb. In particular, we randomly
select ten $X_\mathcal{A}^*$ as the inputs of \ma. These ten $X_\mathcal{A}^*$
will be classified into labels 0--9 and we confirm that they behave normally and
do not dominate the joint inference. We further launch ADI synthesis with
bounded mutation to generate an ADI from each $X_\mathcal{A}^*$. Model
configuration and the feature partition ratio over \ma\ and \mb\ are aligned
with evaluation in \S~\ref{subsec:whitebox-attack}, and the dominating threshold
is set as 95\%.

Recall to synthesize an ADI from $X_\mathcal{A}^*$, we iterate each data
$X_{\mathcal{B}}^i$ in \sd\ and compute a perturbation vector $V_i$ (see
\A~\ref{alg:attacking}). Each $V_i$ points toward the direction of the
adversary-specified classification region $R_{l_{target}}$. Let $h$ be the size
of \sd, to quantify the correlation between $h$ perturbation vectors, we first
define the matrix: $N = \left[ \frac{V_1}{||V_1||_2} \dots \frac{V_h}{||V_h||_2} \right]$.

\begin{figure}[!t]
	\captionsetup{skip=5pt}
	\captionsetup[sub]{skip=1pt}
	\vspace{-5pt}
    \centering
	\begin{subfigure}{.43\linewidth}
		\centering
		\resizebox{1.0\linewidth}{!}{\begin{tikzpicture}
    \begin{axis}[
      grid=major,
      ymax=20,
      ytick align=outside, ytick pos=left,
      xtick align=outside, xtick pos=left,
      xlabel={\huge Singular Index},
      ylabel={\huge Singular Value},
      legend pos=outer north east,
      legend style={draw=none}]

\addplot+[
black, mark options={scale=0.001},
smooth, 
error bars/.cd, 
    y fixed,
    y dir=both, 
    y explicit
] table [x=x, y=y] {fig/SVD-value-random.txt};
\addlegendentry{\Large Random}

\addplot+[
  blue, mark options={scale=0.001},
  smooth, 
  error bars/.cd, 
      y fixed,
      y dir=both, 
      y explicit
  ] table [x=x, y=y] {fig/SVD-value-0-h10000.txt};
  \addlegendentry{\Large Label 0}

\addplot+[
  green, mark options={scale=0.001},
  smooth, 
  error bars/.cd, 
      y fixed,
      y dir=both, 
      y explicit
  ] table [x=x, y=y] {fig/SVD-value-1-h10000.txt};
  \addlegendentry{\Large Label 1}

\addplot+[
  orange, mark options={scale=0.001},
  smooth, 
  error bars/.cd, 
      y fixed,
      y dir=both, 
      y explicit
  ] table [x=x, y=y] {fig/SVD-value-2-h10000.txt};
  \addlegendentry{\Large Label 2}

\addplot+[
  purple, mark options={scale=0.001},
  smooth, 
  error bars/.cd, 
      y fixed,
      y dir=both, 
      y explicit
  ] table [x=x, y=y] {fig/SVD-value-3-h10000.txt};
  \addlegendentry{\Large Label 3}

\addplot+[
  teal, mark options={scale=0.001},
  smooth, 
  error bars/.cd, 
      y fixed,
      y dir=both, 
      y explicit
  ] table [x=x, y=y] {fig/SVD-value-4-h10000.txt};
  \addlegendentry{\Large Label 4}

\addplot+[
  olive, mark options={scale=0.001},
  smooth, 
  error bars/.cd, 
      y fixed,
      y dir=both, 
      y explicit
  ] table [x=x, y=y] {fig/SVD-value-5-h10000.txt};
  \addlegendentry{\Large Label 5}

\addplot+[
darkgray, mark options={scale=0.001},
smooth, 
error bars/.cd, 
    y fixed,
    y dir=both, 
    y explicit
] table [x=x, y=y] {fig/SVD-value-6-h10000.txt};
\addlegendentry{\Large Label 6}

\addplot+[
magenta, mark options={scale=0.001},
smooth, 
error bars/.cd, 
    y fixed,
    y dir=both, 
    y explicit
] table [x=x, y=y] {fig/SVD-value-7-h10000.txt};
\addlegendentry{\Large Label 7}

\addplot+[
  red, mark options={scale=0.001},
  smooth, 
  error bars/.cd, 
      y fixed,
      y dir=both, 
      y explicit
  ] table [x=x, y=y] {fig/SVD-value-8-h10000.txt};
  \addlegendentry{\Large Label 8}

\addplot+[
  cyan, mark options={scale=0.001},
  smooth, 
  error bars/.cd, 
      y fixed,
      y dir=both, 
      y explicit
  ] table [x=x, y=y] {fig/SVD-value-9-h10000.txt};
  \addlegendentry{\Large Label 9}

\end{axis}
\end{tikzpicture}}
		\caption{\small Singular values.}
		\label{fig:SVD}
	\end{subfigure}%
	\begin{subfigure}{.43\linewidth}
		\centering
		\resizebox{1.0\linewidth}{!}{\begin{tikzpicture}
    \begin{axis}[
      grid=major,
      ymin=0, ymax=105,
      ytick align=outside, ytick pos=left,
      xtick align=outside, xtick pos=left,
      xlabel={\huge \#Singular Vectors},
      ylabel={\huge Dominating Rate},
      legend pos= outer north east,
      legend cell align=left,
      legend style={draw=none}]

\addplot+[
  blue, mark options={scale=0.75},
  smooth, 
  error bars/.cd, 
      y fixed,
      y dir=both, 
      y explicit
  ] table [x=x, y=y, col sep=comma] {fig/SVD-recons-0-h10000.txt};
  \addlegendentry{\Large Label 0}

\addplot+[
  green, mark options={scale=0.75},
  smooth, 
  error bars/.cd, 
      y fixed,
      y dir=both, 
      y explicit
  ] table [x=x, y=y, col sep=comma] {fig/SVD-recons-1-h10000.txt};
  \addlegendentry{\Large Label 1}

\addplot+[
  orange, mark options={scale=0.75},
  smooth, 
  error bars/.cd, 
      y fixed,
      y dir=both, 
      y explicit
  ] table [x=x, y=y, col sep=comma] {fig/SVD-recons-2-h10000.txt};
  \addlegendentry{\Large Label 2}

\addplot+[
  purple, mark options={scale=0.75},
  smooth, 
  error bars/.cd, 
      y fixed,
      y dir=both, 
      y explicit
  ] table [x=x, y=y, col sep=comma] {fig/SVD-recons-3-h10000.txt};
  \addlegendentry{\Large Label 3}

\addplot+[
  teal, mark options={scale=0.75},
  smooth, 
  error bars/.cd, 
      y fixed,
      y dir=both, 
      y explicit
  ] table [x=x, y=y, col sep=comma] {fig/SVD-recons-4-h10000.txt};
  \addlegendentry{\Large Label 4}

\addplot+[
  olive, mark options={scale=0.75},
  smooth, 
  error bars/.cd, 
      y fixed,
      y dir=both, 
      y explicit
  ] table [x=x, y=y, col sep=comma] {fig/SVD-recons-5-h10000.txt};
  \addlegendentry{\Large Label 5}

\addplot+[
  darkgray, mark options={scale=0.75},
  smooth, 
  error bars/.cd, 
      y fixed,
      y dir=both, 
      y explicit
  ] table [x=x, y=y, col sep=comma] {fig/SVD-recons-6-h10000.txt};
  \addlegendentry{\Large Label 6}

\addplot+[
  magenta, mark options={scale=0.75},
  smooth, 
  error bars/.cd, 
      y fixed,
      y dir=both, 
      y explicit
  ] table [x=x, y=y, col sep=comma] {fig/SVD-recons-7-h10000.txt};
  \addlegendentry{\Large Label 7}

\addplot+[
  red, mark options={scale=0.75},
  smooth, 
  error bars/.cd, 
      y fixed,
      y dir=both, 
      y explicit
  ] table [x=x, y=y, col sep=comma] {fig/SVD-recons-8-h10000.txt};
  \addlegendentry{\Large Label 8}

\addplot+[
  cyan, mark options={scale=0.75},
  smooth, 
  error bars/.cd, 
      y fixed,
      y dir=both, 
      y explicit
  ] table [x=x, y=y, col sep=comma] {fig/SVD-recons-9-h10000.txt};
  \addlegendentry{\Large Label 9}

\end{axis}
\end{tikzpicture}}
		\caption{\small Reconstructed results.}
		\label{fig:reconstruct}
	\end{subfigure}
	\caption{Singular values and reconstructed results using different numbers of singular vectors on MNIST.}
	\label{fig:singular}
	\vspace{-5pt}
\end{figure}
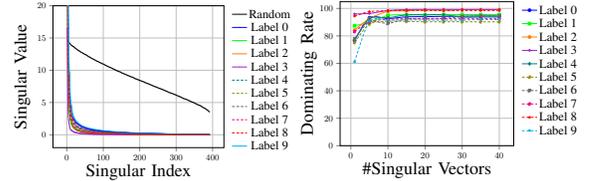

\begin{figure}[!t]
	\captionsetup{skip=5pt}
	\captionsetup[sub]{skip=1pt}
	\vspace{-5pt}
    \centering
	\begin{subfigure}{.43\linewidth}
		\centering
		\resizebox{1.0\linewidth}{!}{\begin{tikzpicture}
    \begin{axis}[
      grid=major,
      ymax=15,
      ytick align=outside, ytick pos=left,
      xtick align=outside, xtick pos=left,
      xlabel={\huge Singular Index},
      ylabel={\huge Singular Value},
      legend pos=outer north east,
      legend style={draw=none}]

\addplot+[
black, mark options={scale=0.001},
smooth, 
error bars/.cd, 
    y fixed,
    y dir=both, 
    y explicit
] table [x=x, y=y] {fig/SVD-value-random.txt};
\addlegendentry{\Large Random}

\addplot+[
  blue, mark options={scale=0.001},
  smooth, 
  error bars/.cd, 
      y fixed,
      y dir=both, 
      y explicit
  ] table [x=x, y=y] {fig/SVD-value-0-cifar-h10000.txt};
  \addlegendentry{\Large Label 0}

\addplot+[
  green, mark options={scale=0.001},
  smooth, 
  error bars/.cd, 
      y fixed,
      y dir=both, 
      y explicit
  ] table [x=x, y=y] {fig/SVD-value-1-cifar-h10000.txt};
  \addlegendentry{\Large Label 1}

\addplot+[
  orange, mark options={scale=0.001},
  smooth, 
  error bars/.cd, 
      y fixed,
      y dir=both, 
      y explicit
  ] table [x=x, y=y] {fig/SVD-value-2-cifar-h10000.txt};
  \addlegendentry{\Large Label 2}

\addplot+[
  purple, mark options={scale=0.001},
  smooth, 
  error bars/.cd, 
      y fixed,
      y dir=both, 
      y explicit
  ] table [x=x, y=y] {fig/SVD-value-3-cifar-h10000.txt};
  \addlegendentry{\Large Label 3}

\addplot+[
  teal, mark options={scale=0.001},
  smooth, 
  error bars/.cd, 
      y fixed,
      y dir=both, 
      y explicit
  ] table [x=x, y=y] {fig/SVD-value-4-cifar-h10000.txt};
  \addlegendentry{\Large Label 4}

\addplot+[
  olive, mark options={scale=0.001},
  smooth, 
  error bars/.cd, 
      y fixed,
      y dir=both, 
      y explicit
  ] table [x=x, y=y] {fig/SVD-value-5-cifar-h10000.txt};
  \addlegendentry{\Large Label 5}

\addplot+[
darkgray, mark options={scale=0.001},
smooth, 
error bars/.cd, 
    y fixed,
    y dir=both, 
    y explicit
] table [x=x, y=y] {fig/SVD-value-6-cifar-h10000.txt};
\addlegendentry{\Large Label 6}

\addplot+[
magenta, mark options={scale=0.001},
smooth, 
error bars/.cd, 
    y fixed,
    y dir=both, 
    y explicit
] table [x=x, y=y] {fig/SVD-value-7-cifar-h10000.txt};
\addlegendentry{\Large Label 7}

\addplot+[
  red, mark options={scale=0.001},
  smooth, 
  error bars/.cd, 
      y fixed,
      y dir=both, 
      y explicit
  ] table [x=x, y=y] {fig/SVD-value-8-cifar-h10000.txt};
  \addlegendentry{\Large Label 8}

\addplot+[
  cyan, mark options={scale=0.001},
  smooth, 
  error bars/.cd, 
      y fixed,
      y dir=both, 
      y explicit
  ] table [x=x, y=y] {fig/SVD-value-9-cifar-h10000.txt};
  \addlegendentry{\Large Label 9}

\end{axis}
\end{tikzpicture}}
		\caption{\small Singular values.}
		\label{fig:SVD-cifar}
	\end{subfigure}%
	\begin{subfigure}{.43\linewidth}
		\centering
		\resizebox{1.0\linewidth}{!}{\begin{tikzpicture}
    \begin{axis}[
      grid=major,
      ymin=0, ymax=105,
      ytick align=outside, ytick pos=left,
      xtick align=outside, xtick pos=left,
      xlabel={\huge \#Singular Vectors},
      ylabel={\huge Dominating Rate},
      legend pos= outer north east,
      legend cell align=left,
      legend style={draw=none}]

\addplot+[
  blue, mark options={scale=0.75},
  smooth, 
  error bars/.cd, 
      y fixed,
      y dir=both, 
      y explicit
  ] table [x=x, y=y, col sep=comma] {fig/SVD-recons-0-cifar-h10000.txt};
  \addlegendentry{\Large Label 0}

\addplot+[
  green, mark options={scale=0.75},
  error bars/.cd, 
      y fixed,
      y dir=both, 
      y explicit
  ] table [x=x, y=y, col sep=comma] {fig/SVD-recons-1-cifar-h10000.txt};
  \addlegendentry{\Large Label 1}

\addplot+[
  orange, mark options={scale=0.75},
  smooth, 
  error bars/.cd, 
      y fixed,
      y dir=both, 
      y explicit
  ] table [x=x, y=y, col sep=comma] {fig/SVD-recons-2-cifar-h10000.txt};
  \addlegendentry{\Large Label 2}

\addplot+[
  purple, mark options={scale=0.75},
  smooth, 
  error bars/.cd, 
      y fixed,
      y dir=both, 
      y explicit
  ] table [x=x, y=y, col sep=comma] {fig/SVD-recons-3-cifar-h10000.txt};
  \addlegendentry{\Large Label 3}

\addplot+[
  teal, mark options={scale=0.75},
  error bars/.cd, 
      y fixed,
      y dir=both, 
      y explicit
  ] table [x=x, y=y, col sep=comma] {fig/SVD-recons-4-cifar-h10000.txt};
  \addlegendentry{\Large Label 4}

\addplot+[
  olive, mark options={scale=0.75},
  smooth, 
  error bars/.cd, 
      y fixed,
      y dir=both, 
      y explicit
  ] table [x=x, y=y, col sep=comma] {fig/SVD-recons-5-cifar-h10000.txt};
  \addlegendentry{\Large Label 5}

\addplot+[
  darkgray, mark options={scale=0.75},
  smooth, 
  error bars/.cd, 
      y fixed,
      y dir=both, 
      y explicit
  ] table [x=x, y=y, col sep=comma] {fig/SVD-recons-6-cifar-h10000.txt};
  \addlegendentry{\Large Label 6}

\addplot+[
  magenta, mark options={scale=0.75},
  smooth, 
  error bars/.cd, 
      y fixed,
      y dir=both, 
      y explicit
  ] table [x=x, y=y, col sep=comma] {fig/SVD-recons-7-cifar-h10000.txt};
  \addlegendentry{\Large Label 7}

\addplot+[
  red, mark options={scale=0.75},
  smooth, 
  error bars/.cd, 
      y fixed,
      y dir=both, 
      y explicit
  ] table [x=x, y=y, col sep=comma] {fig/SVD-recons-8-cifar-h10000.txt};
  \addlegendentry{\Large Label 8}

\addplot+[
  cyan, mark options={scale=0.75},
  smooth, 
  error bars/.cd, 
      y fixed,
      y dir=both, 
      y explicit
  ] table [x=x, y=y, col sep=comma] {fig/SVD-recons-9-cifar-h10000.txt};
  \addlegendentry{\Large Label 9}

\end{axis}
\end{tikzpicture}}
		\caption{\small Reconstructed results.}
		\label{fig:reconstruct-cifar}
	\end{subfigure}
	\caption{Singular values and reconstructed results using different numbers of singular vectors on CIFAR-10.}
	\label{fig:singular-cifar}
	\vspace{-5pt}
\end{figure}
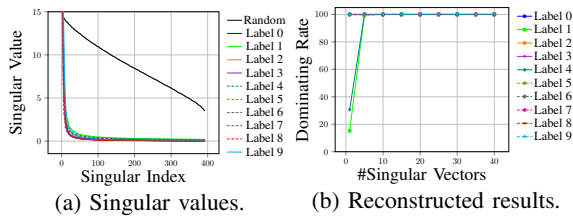

To analyze the correlation of perturbation vectors, we use entire test datasets to
form matrix $N$. 
\F~\ref{fig:SVD} and \F~\ref{fig:SVD-cifar}
report the singular values. Compared with the matrix uniformly sampled at random
from the unit sphere, singular values of $N$ decay faster. This indicates the
existence of primary correlations and redundancies in mutations launched during
ADI synthesis using different inputs of \mb. This also implies there exists a
low dimensional subspace that sufficiently captures the correlations among
different inputs of \mb.

We thus hypothesize that ADI exists because of a low-dimensional subspace that
captures correlations between different classification regions. To test our
hypothesis, we reconstruct the mutation using different numbers of singular
vectors in decreasing order of their singular values and test the ADI dominating
rates by perturbing $X_{\mathcal{A}}^*$. \F~\ref{fig:reconstruct} and \F~\ref{fig:reconstruct-cifar} report the
results: mutations reconstructed using only five singular vectors
already reach high dominating rates (over 80\% for MNIST and close to 100\% for CIFAR-10), and ten singular vectors with
the highest singular values obtain close to 90\% dominating rates for MNIST and close to 100\% for CIFAR-10. 
The results are consistent on the datasets we evaluate.
We interpret that the results support our hypothesis. This hypothesis also explains
that most perturbations are heading to a similar direction. Therefore, we do not
need a large \ssd\ to successfully generate ADIs.

\begin{figure}
	\vspace{-5pt}
	\captionsetup{skip=5pt}
	\captionsetup[sub]{skip=1pt}
    \centering
	\begin{subfigure}{.40\linewidth}
		\centering
		\resizebox{1.0\linewidth}{!}{\begin{tikzpicture}
    \begin{axis}[
      grid=major,
      ymin=0,
      xlabel={\huge \ssd},
      ylabel={\huge Attack success rate},]
    \addplot+[
    red, mark options={scale=0.75},
    smooth, 
    error bars/.cd, 
        y fixed,
        y dir=both, 
        y explicit
    ] table [x=x, y=y, col sep=comma] {fig/subsetsize-whitebox.txt}; \label{size-whitebox}
    \end{axis}

\begin{axis}[
    ylabel near ticks, 
    ylabel style={rotate=180},
    yticklabel pos=right,
    ymin=0,
    axis x line=none,
    ylabel={\huge Process time},
    legend pos= south east]
    \addlegendimage{/pgfplots/refstyle=size-whitebox}\addlegendentry{\Large bounded mutation}

    \addplot+[
        blue, mark options={scale=0.75},
        smooth, 
        error bars/.cd, 
            y fixed,
            y dir=both, 
            y explicit
        ] table [x=x, y=y, col sep=comma] {fig/subsetsize-whitebox-time.txt}; \label{size-time}
        \addlegendimage{/pgfplots/refstyle=size-time}\addlegendentry{\Large hours / 1k data samples}
\end{axis}

\end{tikzpicture}}
		\caption{\small ADI synthesis.}
		\label{fig:subset-whitebox}
	\end{subfigure}%
	\quad
	\begin{subfigure}{.35\linewidth}
		\centering
		\resizebox{1.0\linewidth}{!}{\begin{tikzpicture}
    \begin{axis}[
      grid=major,
      ymin=0,
      ytick align=outside, ytick pos=left,
      xtick align=outside, xtick pos=left,
      xlabel={\huge \ssd},
      ylabel={\huge ADI numbers},
      legend pos=north west,
      legend style={draw=none}]
\addplot+[
blue, mark options={scale=0.75},
smooth, 
error bars/.cd, 
    y fixed,
    y dir=both, 
    y explicit
] table [x=x, y=y, col sep=comma] {fig/subsetsize-fuzz.txt};
\end{axis}
\end{tikzpicture}}
		\caption{\small Fuzz testing.}
		\label{fig:subsetsize-fuzz}
	\end{subfigure}
	\caption{Performance under different \ssd\ on MNIST.}
	\label{fig:alg-performance-subsets}
	\vspace{-5pt}
\end{figure}
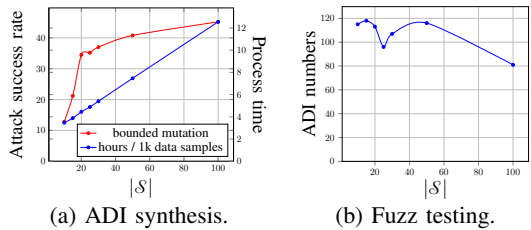

\begin{figure}
        \captionsetup{skip=5pt}
        \captionsetup[sub]{skip=1pt}
        \centering
        \begin{subfigure}{.40\linewidth}
            \centering
            \resizebox{1.0\linewidth}{!}{\begin{tikzpicture}
    \begin{axis}[
      grid=major,
      ymin=0,
      xlabel={\huge \ssd},
      ylabel={\huge Attack success rate},]
    \addplot+[
    red, mark options={scale=0.75},
    smooth, 
    error bars/.cd, 
        y fixed,
        y dir=both, 
        y explicit
    ] table [x=x, y=y, col sep=comma] {fig/subsetsize-whitebox-append.txt}; \label{size-whitebox-append}
    \end{axis}

\begin{axis}[
    ylabel near ticks, 
    ylabel style={rotate=180},
    yticklabel pos=right,
    ymin=0,
    axis x line=none,
    ylabel={\huge Process time},
    legend pos= south east]
    \addlegendimage{/pgfplots/refstyle=size-whitebox}\addlegendentry{\Large bounded mutation}

    \addplot+[
        blue, mark options={scale=0.75},
        smooth, 
        error bars/.cd, 
            y fixed,
            y dir=both, 
            y explicit
        ] table [x=x, y=y, col sep=comma] {fig/subsetsize-whitebox-time-append.txt}; \label{size-time-append}
        \addlegendimage{/pgfplots/refstyle=size-time}\addlegendentry{\Large hours / 1k data samples}
\end{axis}

\end{tikzpicture}}
            \caption{\small ADI synthesis.}
            \label{fig:subset-whitebox-append}
        \end{subfigure}%
        \quad
        \begin{subfigure}{.35\linewidth}
            \centering
            \resizebox{1.0\linewidth}{!}{\begin{tikzpicture}
    \begin{axis}[
      grid=major,
      ymin=0,
      ytick align=outside, ytick pos=left,
      xtick align=outside, xtick pos=left,
      xlabel={\huge \ssd},
      ylabel={\huge ADI numbers},
      legend pos=north west,
      legend style={draw=none}]
\addplot+[
blue, mark options={scale=0.75},
smooth, 
error bars/.cd, 
    y fixed,
    y dir=both, 
    y explicit
] table [x=x, y=y, col sep=comma] {fig/subsetsize-fuzz-append.txt};
\end{axis}
\end{tikzpicture}}
            \caption{\small Fuzz testing.}
            \label{fig:subsetsize-fuzz-append}
        \end{subfigure}
        \caption{Performance under different \ssd\ on CIFAR-10.}
        \label{fig:alg-performance-subsets-append}
    \end{figure}
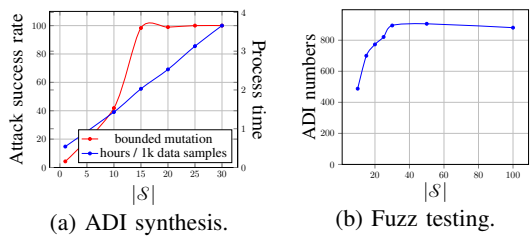

We further evaluate how \ssd\ influences ADI generation, whose results
are in \F~\ref{fig:alg-performance-subsets} and
\F~\ref{fig:alg-performance-subsets-append}. The ADI synthesis follows
\S~\ref{subsec:whitebox-attack}. Fuzzing is launched for 12 hours for different
\ssd\ and we record the number of discovered ADIs.
As expected, \F~\ref{fig:subset-whitebox} and
\F~\ref{fig:subset-whitebox-append} show that larger \ssd\ leads to increased
success rates in synthesizing ADIs. However, as seen in the
\textcolor{blue}{blue} lines of \F~\ref{fig:subset-whitebox} and
\F~\ref{fig:subset-whitebox-append}, ADI synthesis takes substantially longer
time as \ssd\ increases (we report the total time taken by perturbing 1,000
inputs of \ma). Recall given a normal input of \ma, we iterate every element in
\sd\ to compute the perturbation vector, which explains the linear growth of
processing time in \F~\ref{fig:subset-whitebox} and
\F~\ref{fig:subset-whitebox-append}.
We have similar observations in \F~\ref{fig:subsetsize-fuzz} and
\F~\ref{fig:subsetsize-fuzz-append}. Overall, while larger \ssd\ may increase
the likelihood of generating ADIs (as in \F~\ref{fig:subset-whitebox} and
\F~\ref{fig:subset-whitebox-append}), it may also limit the fuzzing throughput.
It is seen that the ``sweet spot'' is around 15--20 in
\F~\ref{fig:subsetsize-fuzz} for the medium complex MNIST dataset. For the
complex dataset CIFAR-10, we find that \ssd\ between $20$ and $50$ are good
for our setting. The range is a bit larger than that of MNIST, given that
CIFAR-10 has a larger feature space. Overall, fuzzing can be boosted by taking a
slightly larger \ssd, such that the mutation vector can capture contributions of
\mb\ for different input data more accurately. Consistent with MNIST, the number
of found ADIs for CIFAR-10 is starting to decrease when \ssd\ is greater than
$50$, since an overly large \ssd\ limits the fuzzing throughput.

Determining \ssd\ should take
the VFL task specification, feature dimensions, and protocol complexity into
account. Moreover, we emphasize that synthesizing ADIs does not require a large
\ssd: as in \F~\ref{fig:SVD} and \F~\ref{fig:SVD-cifar}, perturbations applied
toward different data samples are highly correlated.

\bigskip
\section{Rewards Estimation}
\label{sec:reward}

Several schemes are proposed to analyze the fairness and rewarding in federated
learning and collaborative machine learning~\cite{yang2017designing,
  yu2020fairness, song2019profit, sim2020collaborative, shrikumar2017learning}.
Despite the difference in implementations, they primarily assess the importance
of a participant's contribution in a joint prediction. The VFL frameworks we
evaluated do not offer an ``out-of-the-box'' rewarding estimation module.
Therefore, to estimate rewards for each participant in a joint inference, we
take a general approach to measuring the $L_1$ norm of the saliency map (line 3
in \A~\ref{alg:attacking}) derived from each participant's input data.

\T~\ref{tab:contribution-Estimation} reports the normalized estimated rewards
for different datasets. To generate ADIs, we use bounded mutation at the 95\%
threshold. It's easy to see that given normal inputs used by both ends, each
participant is estimated to receive approximately the same reward (except VFVQA;
see discussion below). For instance, given one unit of reward offered for each
joint inference over NUS-WIDE, \ma\ and \mb\ anticipate receiving 0.492 and
0.508 units of rewards, respectively. In contrast, when \ma\ uses ADIs, \ma\ can
receive 0.859 units of rewards for at least 95\% of joint inferences. This is
consistent with our discussion in \S~\ref{subsec:threat-model}.

For VQA v2.0, we observe that the contribution of \mb\ drops when
\ma\ uses ADIs. However, we clarify that results in
\T~\ref{tab:contribution-Estimation} indeed \textit{underestimate} the dominance
of ADIs.
Recall for this VFVQA task, \mb\ raises natural-language questions toward images
possessed by \ma. The gradients of natural-language question inputs in VFVQA
measure the \textit{local} sensitivity of the VFVQA system. However, different
from other tasks, the perturbed questions will not form reasonable text if they
are only perturbed by gradients measuring local sensitivity. This explains
why the contribution of \mb\ in \T~\ref{tab:contribution-Estimation}, even for
normal inputs, is larger than \ma.

While studies on assessing attention to natural-language questions or images
exist~\cite{hermann2015teaching, lu2016hierarchical}, comparing their
contribution is an open problem. We follow~\cite{Goyal2017CVPR} to approximate
and compare their contributions. The approach proposed by~\cite{Goyal2017CVPR}
is conceptually consistent with our evaluation in \S~\ref{sec:evaluation}, where
the questions or images are replaced by other data samples in the dataset to see
how these replacements will influence the joint inference. We report that when
using this method to compare the contributions of \ma\ and \mb , \ma\ and \mb\
will get 0.511 and 0.489 units of rewards on average when both participants use
normal inputs. In contrast, when \ma\ uses ADIs, \ma\ will hog all the rewards.


\begin{table}[!t]
	\centering
	\scriptsize
	\captionsetup{skip=2pt}
  \caption{Contribution Estimation.}
	\label{tab:contribution-Estimation}
	\setlength{\tabcolsep}{2.0pt}
	\resizebox{0.7\linewidth}{!}{
		\begin{tabular}{l|c|c|c}
			\hline
         & & \multicolumn{1}{c|}{\textbf{\ma}} & \multicolumn{1}{c}{\textbf{\mb}} \\
			\hline
			\multirow{2}{*}{\textbf{NUS-WIDE}} & {Normal Inputs Used by Both Sides}  & 0.492 & 0.508 \\
			& {ADIs in \ma , Normal Inputs in \mb}    & 0.859 & 0.141 \\
			\hline
			\multirow{2}{*}{\textbf{Credit}} & {Normal Inputs Used by Both Sides}  & 0.457 & 0.543 \\
			& {ADIs in \ma , Normal Inputs in \mb}    & 0.856 & 0.144 \\
			\hline
			\multirow{2}{*}{\textbf{Vehicle}} & {Normal Inputs Used by Both Sides}  & 0.504 & 0.496 \\
			& {ADIs in \ma , Normal Inputs in \mb}    & 0.784 & 0.216 \\
			\hline
			\multirow{2}{*}{\textbf{MNIST}} & {Normal Inputs Used by Both Sides}  & 0.429 & 0.571 \\
			& {ADIs in \ma , Normal Inputs in \mb}    & 0.833 & 0.167 \\
			\hline
			\multirow{2}{*}{\textbf{VQA v2.0}} & {Normal Inputs Used by Both Sides}  & 0.279 & 0.721 \\
			& {ADIs in \ma , Normal Inputs in \mb}    & 0.352 & 0.648 \\
			\hline
			\multirow{2}{*}{\textbf{CIFAR-10}} & {Normal Inputs Used by Both Sides}  & 0.425 & 0.575 \\
			& ADIs in \ma , Normal Inputs in \mb    & 0.748 & 0.252 \\
			\hline
		\end{tabular}
	}
\end{table}
\section{Explore Feature Partition Ratios}
\label{subsec:eval-feature-allocation}

\begin{table}[t]
	\captionsetup{skip=2pt}
	\centering
	\scriptsize
  \caption{Dominating rates of the standard dataset under different feature
    partition ratios.}
	\label{tab:alpha-dominating-rate}
	\setlength{\tabcolsep}{1.5pt}
	\resizebox{0.90\linewidth}{!}{
		\begin{tabular}{c|c|c|c|c}
			\hline
         \multirow{2}{*}{\textbf{Dataset}} & \textbf{Feature}        & \textbf{Dominating} & \textbf{Dominating} & \textbf{Model Accuracy} \\
         & \textbf{Partition Ratio}   & \textbf{Rate on \ma}     & \textbf{Rate on \mb}     & \textbf{on Test Datasets} \\
			\hline
			\multirow{6}{*}{\textbf{MNIST}} & \textbf{0.40}  & 0.00\%  & 94.30\%  &  97.55\% \\
			& \textbf{0.65}  & 0.00\%  & 20.49\%  &  97.81\% \\
			& \textbf{1.00}  & 0.87\%  & 0.43\%   &  97.78\% \\
			& \textbf{1.33}  & 5.79\%  & 0.00\%   &  98.01\% \\
			& \textbf{1.80}  & 50.40\% & 0.00\%   &  98.20\% \\
			& \textbf{2.11}  & 72.40\% & 0.00\%   &  97.62\% \\
			\hline
			\multirow{6}{*}{\textbf{CIFAR-10}} & \textbf{0.40}  & 0.00\%  & 47.6\%  &  86.56\% \\
			& \textbf{0.65}  & 0.30\%  & 31.0\%  &  86.34\% \\
			& \textbf{1.00}  & 2.63\%  & 7.20\%  &  86.55\% \\
			& \textbf{1.33}  & 13.9\%  & 2.10\%  &  86.22\% \\
			& \textbf{1.80}  & 31.3\%  & 0.10\%  &  86.21\% \\
			& \textbf{2.11}  & 36.2\%  & 0.10\%  &  86.38\% \\
			\hline
		\end{tabular}
	}
	\vspace{-5pt}
\end{table}

Recall in a typical VFL setting where each user ID $i$ has a number of features
$X$, each participant possesses a subset of features. This section studies how
the success rate of ADI generation is affected by the feature partition ratio
using MNIST and CIFAR-10. We re-run experiments in \T~\ref{tab:original-data-ats} to compare
the dominating rates of standard datasets under various feature partition ratios
(``feature'' is the number of vertical dimensions of an image).
\T~\ref{tab:alpha-dominating-rate} reports the findings: a feature partition
ratio $x$ denotes that \ma\ possesses $\sfrac{x}{(1+x)}$ percent of the features
in each data sample while \mb\ has the rest. To be fair, we initialize all
hyperparameters to the same value as \T~\ref{tab:original-data-ats}. The model
accuracy on the test dataset is sufficiently high, indicating that models
are well-trained.

\begin{figure}[!t]
	\captionsetup{skip=5pt}
	\captionsetup[sub]{skip=3pt}
	\centering
	\begin{subfigure}{.43\linewidth}
		\centering
		\resizebox{1.0\linewidth}{!}{\begin{tikzpicture}
    \begin{axis}[
      grid=major,
      ytick align=outside, ytick pos=left,
      xtick align=outside, xtick pos=left,
      xlabel={\huge Feature Parition Ratio},
      ylabel={\huge Success Rate (\%)},
      legend pos=south east,
      legend cell align={left},
      legend style={draw=none}]

\addplot+[
  red, mark options={scale=0.75},
  smooth, 
  error bars/.cd, 
      y fixed,
      y dir=both, 
      y explicit
  ] table [x=x, y=y, col sep=comma] {fig/featureimpactub.txt};
  \addlegendentry{\large MNIST random mutation}

\addplot+[
blue, mark options={scale=0.75},
smooth, 
error bars/.cd, 
    y fixed,
    y dir=both, 
    y explicit
] table [x=x, y=y, col sep=comma] {fig/featureimpactwhitebox.txt};
\addlegendentry{\large MNIST bounded mutation}

\addplot+[
  orange, mark options={scale=0.75},
  smooth, 
  error bars/.cd, 
      y fixed,
      y dir=both, 
      y explicit
  ] table [x=x, y=y, col sep=comma] {fig/featureimpactub-append.txt};
  \addlegendentry{\large CIFAR-10 random mutation}

\addplot+[
cyan, mark options={scale=0.75},
smooth, 
error bars/.cd, 
    y fixed,
    y dir=both, 
    y explicit
] table [x=x, y=y, col sep=comma] {fig/featureimpactwhitebox-append.txt};
\addlegendentry{\large CIFAR-10 bounded mutation}

\end{axis}
\end{tikzpicture}}
		\caption{\small Gradient-based synthesis.}
		\label{fig:info-whitebox}
	\end{subfigure}%
	\quad
	\begin{subfigure}{.43\linewidth}
		\centering
		\resizebox{1.0\linewidth}{!}{\begin{tikzpicture}
    \begin{axis}[
      grid=major,
      ytick align=outside, ytick pos=left,
      xtick align=outside, xtick pos=left,
      xlabel={\huge Feature Partition Ratio},
      ylabel={\huge \#Found ADIs},
      legend pos=south east,
      legend cell align={left},
      legend style={draw=none}]
\addplot+[
blue, mark options={scale=0.75},
smooth, 
error bars/.cd, 
    y fixed,
    y dir=both, 
    y explicit
] table [x=x, y=y, col sep=comma] {fig/featureimpact.txt};
\addlegendentry{\large MNIST}

\addplot+[
cyan, mark options={scale=0.75},
smooth, 
error bars/.cd, 
    y fixed,
    y dir=both, 
    y explicit
] table [x=x, y=y, col sep=comma] {fig/featureimpact-append.txt};
\addlegendentry{\large CIFAR-10}

\end{axis}
\end{tikzpicture}}
		\caption{\small Fuzz testing.}
		\label{fig:info-fuzz}
	\end{subfigure}
	\caption{Performance under different feature ratios.}
	\label{fig:info}
	\vspace{-3pt}
\end{figure}
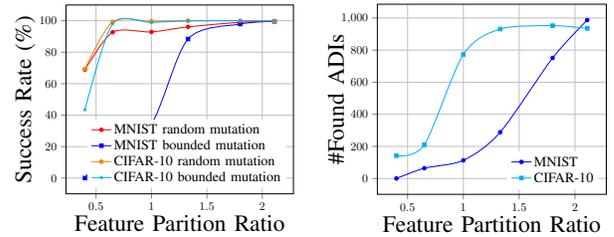

Higher ratios imply that \ma\ holds more features.
\T~\ref{tab:alpha-dominating-rate} reports that when more features are allocated
to a participant, more ADIs are likely found on its side and dominate others.
For instance, for MNIST dataset, when the ratio reaches 2.11 (i.e., \ma\ holds 19 columns while
\mb\ holds 9 columns of the image), the dominating rate is 72.40\%, indicating
that even without perturbation, 72.40\% of \ma's normal inputs can control the
inference. When the ratio falls below 0.65 (\ma\ holds less than 8 columns),
none of the original data samples in \ma\ can control the inference, whereas
data samples in \mb\ dominate \ma. When data samples are partitioned evenly
(each participant gets 14 columns), the dominating rates on both sides are lower
than 1\%.
For CIFAR-10, we observe similar results that when more features are allocated
to a participant, more ADIs are likely found on its side and dominate others.
In sum, we view this evaluation reveals a critical observation:

\begin{tcolorbox}[size=small]
In VFL, numbers of features allocated on each participant should not largely
deviate from each other; by contributing a comparable amount of ``knowledge'' to
the joint inference, the model is seen to be more robust to ADIs. 
\end{tcolorbox}
  
In contrast, unbalanced feature partitioning causes concerns of dominating
inputs and attack opportunities.
\F~\ref{fig:info-whitebox} reports the success rate of ADI synthesis using
different feature partitions. Consistent with
\T~\ref{tab:alpha-dominating-rate}, \ma\ with more features can achieve greater
success rates. The success rates are near 100\% when the feature partition ratios
surpasse 1.8 for MNIST and 1.0 for CIFAR-10.
\F~\ref{fig:info-fuzz} explores how feature partitioning influences fuzzing. The
fuzz testing, consistent with \S~\ref{subsec:greybox-fuzzing}, starts with a
corpus of 1,000 seeds. We run a 12-hour fuzzing campaign for each feature
partition ratio. We find that when assigning \ma\ with more features, fuzzing
can uncover more ADIs. For CIFAR-10, we also observe that when the
feature partition ratio is greater than $1.33$, the number of found ADIs
increases much slower. We deem that this is because we use a fixed amount of
initial seeds, i.e., $1,000$. Therefore, the number of found ADIs is likely to
converge around $1,000$.

VFL faces serious ADI issues when feature partitioning is largely unbalanced. To
understand the root cause, assuming \ma\ and \mb\ hold feature partition dimensions $d_1$
and $d_2$ as their inputs where $d_1$ is notably larger than $d_2$. 
The mutation ${V}_{d_1}$
and ${V}_{d_2}$ are performed on inputs of \ma\ and \mb, respectively. The
parameters of \ma\ and \mb\ are $\theta_{\mathcal{A}_{d_1}}$ and
$\theta_{\mathcal{B}_{d_2}}$. According to our observation, for \ma\ holding
$d_1$ and \mb\ holding $d_2$ dimension of features, their parameters are
roughly the same order of magnitude, meaning that
$\frac{||\theta_{\mathcal{A}_{d_1}}||_2}{d_1} \approx
\frac{||\theta_{\mathcal{B}_{d_2}}||_2}{d_2}$. Thus, the upper bound of
$||\theta_{\mathcal{A}_{d_1}}^T {V}_{d_1} ||_1$ is: $||\theta_{\mathcal{A}_{d_1}}^T {V}_{d_1}||_1 \leq {||{V}_{d_1}||_2 \times
  ||\theta_{\mathcal{A}_{d_1}}||_2}$. 
Similarly, $||\theta_{\mathcal{B}_{d_2}}^T {V}_{d_2}||_1 \leq {||{V}_{d_2}||_2 \times
  ||\theta_{\mathcal{B}_{d_2}}||_2}$.
With $||{V}_{d_1}||_2 = ||{V}_{d_2}||_2$, and given $d_1$ is notably larger than
$d_2$, we have $||\theta_{\mathcal{A}_{d_1}}||_2 >
||\theta_{\mathcal{B}_{d_2}}||_2$. Thus, the upper bound of
$||\theta_{\mathcal{A}_{d_1}}^T {V}_{d_1}||_1$ becomes greater than
$||\theta_{\mathcal{B}_{d_2}}^T {V}_{d_2}||_1$, meaning that the same mutation
($||{V}_{d_1}||_2 = ||{V}_{d_2}||_2$) may have a more significant effect on the
output when it is performed on inputs of \ma\ whose dimension $d_1$ is larger.

\section{Exploring Number of Participants}
\label{subsec:eval-client-numbers}

\begin{table}[!t]
	\captionsetup{skip=2pt}
	\centering
	\scriptsize
  \caption{Dominating rate of the original datasets under different numbers of participants.}
	\label{tab:multiclient-acc}
	\setlength{\tabcolsep}{1.5pt}
	\resizebox{0.8\linewidth}{!}{
		\begin{tabular}{c|c|c|c}
			\hline
			\textbf{Dataset} & \textbf{\#Participants} & \textbf{Dominating Rate} & \textbf{Model Accuracy}\\
			\hline
			\multirow{3}{*}{\textbf{MNIST}} & \textbf{2}   & 0.87\% & 97.78\% \\
			& \textbf{3}   & 0.00\% & 95.97\% \\
			& \textbf{5}   & 0.00\% & 97.84\% \\
			\hline
			\multirow{3}{*}{\textbf{CIFAR-10}} & \textbf{2} & 2.63\% & 86.55\% \\
			& \textbf{3}   & 0.00\% & 85.61\% \\
			& \textbf{5}   & 0.00\% & 85.55\% \\
			\hline
		\end{tabular}
	}
  \vspace{-3pt}
\end{table}

\begin{table}[!t]
	\captionsetup{skip=2pt}
	\centering \scriptsize
  \caption{Performance of gradient- and fuzz
  testing-based algorithms w.r.t. different numbers of participants.}
	\label{tab:multiclient-performance}
	\setlength{\tabcolsep}{1.5pt}
	\resizebox{0.93\linewidth}{!}{
		\begin{tabular}{c|c|c|c|c}
			\hline
			\textbf{Dataset} & \textbf{\#Participants} & \textbf{Gradient Random} & \textbf{Gradient Bounded} & \textbf{Fuzz Testing}\\
			\hline
			\multirow{3}{*}{\textbf{MNIST}} & \textbf{2}   & 92.9\% & 34.5\% & 113\\
			& \textbf{3}   & 56.0\% & 9.40\% & 44 \\
			& \textbf{5}   & 48.0\% & 3.20\% & 9 \\
			\hline
			\multirow{3}{*}{\textbf{CIFAR-10}} & \textbf{2}   & 99.6\% & 98.9\% & 773\\
			& \textbf{3}   & 83.7\% & 51.2\% & 169 \\
			& \textbf{5}   & 64.4\% & 49.3\% & 38 \\
			\hline
		\end{tabular}
	}
  \vspace{-5pt}
\end{table}

As introduced in \S~\ref{sec:background}, the number of participants in most
VFL system designs is two. We now benchmark the attack effectiveness regarding
different number of participants. At this step, we use the MNIST and CIFAR-10 datasets and
extend the SplitNN protocol for multiple participants (i.e., 3 and 5).

MNIST images are sparse: pixels close to the image corner/border are usually
``dark,'' conveying trivial information. To ensure that image features are
partitioned evenly among participants, we divide an image into columns based on
the average numbers of pixels greater than zero in each column. This way, to
partition images for three participants, \ma\ receives the leftmost 11 columns,
participant $\mathcal{B}_1$ gets the middle six, while participant
$\mathcal{B}_2$ gets the rightmost 11. For five participants, \ma\ gets the
leftmost eight, $\mathcal{B}_{1-3}$ equally split the next 12 columns, and
$\mathcal{B}_4$ gets the rightmost eight. For CIFAR-10, we evenly
distribute the embeddings of the VGG16 outputs to 3 and 5 participants.

We take 95\% as the threshold for this study. We report model accuracy and
dominating rates of \ma's normal inputs in \T~\ref{tab:multiclient-acc}. We
interpret \ma's normal inputs have a trivial dominating rate, which is
consistent with our findings in \T~\ref{tab:original-data-ats}.
We further report the performance of ADI synthesis and fuzzing in
\T~\ref{tab:multiclient-performance}. Because data features are evenly
partitioned among participants, the dimension of data features held by \ma\
shrinks as \#participants grows. Thus, the second and third columns of
\T~\ref{tab:multiclient-performance} show substantially lower success rates.
Similarly, fuzz testing finds fewer ADIs in 12 hours. Having more participants
means having fewer features allocated to each (malicious) participant. That is,
the findings in \T~\ref{tab:multiclient-performance} are consistent with those
in \T~\ref{tab:alpha-dominating-rate} and
Appendix \ref{subsec:eval-feature-allocation}. Theoretically, the
unbounded mutation can always generate an ADI as long as the malicious
participant can make non-trivial contributions to the VFL prediction. However,
under the bounded mutation strategy, the proportion of features held by the
malicious participant will affect the attack success rate. We deem this as less
concerned, since VFL's participant number is typically smaller than five (as
noted in \S~\ref{sec:background}). Thus, we deem that the bounded mutation will
be generally applicable for VFL under different settings.

\section{Multiple Malicious Participants}
\label{subsec:multiple-attacker}

As stated in \S~\ref{sec:background}, unlike HFL, most VFL research designs a VFL
system with two participants (maximum four). Hence, we regard only one
participant as malicious; this is a practical and stealthy setting. Considering
the case where multiple participants are malicious and use ADIs simultaneously
(they are not collaborative), it is likely that one of them will dominate the
inference results. Here, the same setting as in \S~\ref{subsec:whitebox-attack}
are used to synthesize ADIs on MNIST for \mb. Then, we let \ma\ and \mb\ both use ADIs,
and observe that all the outputs are dominated by either \ma\ or \mb\ (about 70\%
outputs dominated by \mb\ and the remaining dominated by \ma).

We also run experiments when there are normal participants and multiple attackers on MNIST and CIFAR-10. 
Specifically, we first consider three parties, where two parties (\ma\ and \mb) input ADIs and the other normal participant \mc\ inputs normal data during inference. We found that on MNIST, 98.67\% of the output results are dominated by either \ma\ or \mb. Particularly, 45.87\% of the results are dominated by \ma, while the remaining are dominated by \mb. On CIFAR-10, 96.45\% of the output results are dominated by either \ma\ or \mb, with 65.61\% of the results dominated by \ma\ and the rest by \mb.

Furthermore, we then consider five parties, where two (\ma\ and \mb) input ADIs, and the other normal participants \mc, \md, and \me\ input normal data during inference. We found that on MNIST, 93.48\% of the output results are dominated by either \ma\ or \mb. Particularly, 43.24\% of the results are dominated by \ma, and the rest are dominated by \mb. On CIFAR-10, 90.38\% of the results are dominated by either \ma\ or \mb, with 43.22\% of the results dominated by \ma\ and the rest by \mb. 

We observed that the ratio of the dominated outputs is lower compared with the 3-party setting (93.48\% vs. 98.67\% and 90.38\% vs. 96.45\%). The reason is that in the 5-party setting, the attackers control fewer features compared to the 3-party setting. Also, synthesizing ADIs assumes that the other parties will input normal data during inference. When multiple attackers simultaneously input ADIs, such assumptions do not hold anymore. Thus, it is possible that the results are not dominated by any of the attackers' ADIs. And we observe that when the attackers hold more features, their ADIs are more likely to dominate the inference. This observation is consistent with our results in Appendix \ref{subsec:eval-feature-allocation} and Appendix \ref{subsec:eval-client-numbers}.

Overall, the dominating observation could be influenced by the local
model, the feature importance, and factors that can affect the ADI generation
process, such as \ssd. See our study on feature partition ratios and \ssd\ in
Appendix \ref{subsec:eval-feature-allocation} and
Appendix \ref{subsec:eval-subset-size}.

\section{Training and Implementation Details of Learning Protocols}
\label{sec:lr-nn}

\setlength{\belowdisplayskip}{5pt} \setlength{\belowdisplayshortskip}{5pt}
\setlength{\abovedisplayskip}{-5pt} \setlength{\abovedisplayshortskip}{-5pt}

\smallskip
\noindent \textbf{HeteroLR.}~~Given $n$ training samples $X_i$ whose label is
$y_i \in \{0,1\}$ and coefficient vector $\theta$, logistic regression can be
described in a basic form as follows:

\begin{equation}
    \begin{aligned}
        \mathbb{P}(y_i=1|X_i,\theta) = h_{\theta}(X_i) = \frac{1}{1 + e^{-\theta^TX_i}}
    \end{aligned}
\end{equation}

In typical VFL settings, $X_i$ is partitioned as $[X_{\mathcal{A}}^i || X_{\mathcal{B}}^i]$, while
$\theta$ is partitioned as $[\theta_{\mathcal{A}} || \theta_{\mathcal{B}}]$. Therefore, HeteroLR in VFL
can be formulated as follows:

\begin{equation}
    \begin{aligned}
        \mathbb{P}(y_i=1|X_{\mathcal{A}}^i, X_{\mathcal{B}}^i,\theta_{\mathcal{A}}, \theta_{\mathcal{B}}) &= h_{\theta_{\mathcal{A}}, \theta_{\mathcal{B}}}(X_{\mathcal{A}}^i,X_{\mathcal{B}}^i)\\
        & = \frac{1}{1 + e^{-\theta_{\mathcal{A}}^TX_{\mathcal{A}}^i - \theta_{\mathcal{B}}^TX_{\mathcal{B}}^i}}
    \end{aligned}
\end{equation}

\noindent whose training objective is to minimize the loss function:

\begin{equation}
    \begin{aligned}
        \mathbb{L}_{\theta} = -\frac{1}{n} \sum_{i=1}^n & \; y_i \, log(h_{\theta_{\mathcal{A}}, \theta_{\mathcal{B}}}(X_{\mathcal{A}}^i,X_{\mathcal{B}}^i)) \\
        & \, + (1 - y_i) \, log(1 - h_{\theta_{\mathcal{A}}, \theta_{\mathcal{B}}}(X_{\mathcal{A}}^i,X_{\mathcal{B}}^i))
    \end{aligned}
\end{equation}

We have introduced the general procedure of HeteroLR in background section.
At the training stage, when computing gradients, Taylor approximation and
homomorphic encryption scheme can be utilized to make training more efficient.
To make a joint inference for user ID $i$, \mc\ collects local prediction
results $\theta_{\mathcal{A}}^T X_{\mathcal{A}}^i$ and $\theta_{\mathcal{B}}^T
X_{\mathcal{B}}^i$. It then computes and sends the probability $\frac{1}{1 +
  e^{-\theta_{\mathcal{A}}^TX_{\mathcal{A}}^i -
    \theta_{\mathcal{B}}^TX_{\mathcal{B}}^i}}$ back to \ma\ and \mb. Note that
all the intermediate data is encrypted.

\begin{figure}[!h]
    \centering
    \includegraphics[width=1.0\linewidth]{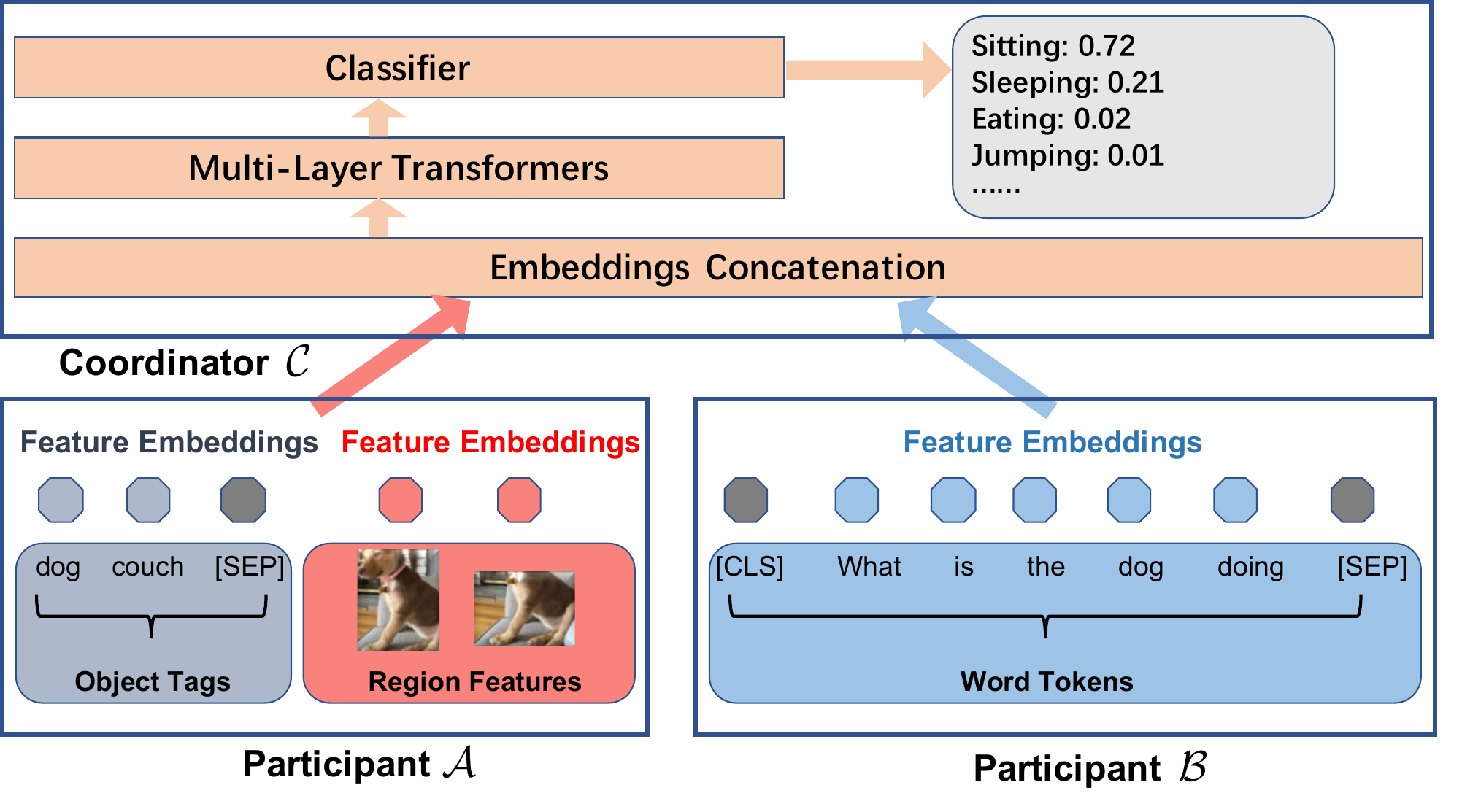}
    \caption{Architecture of VFVQA.}
    \label{fig:VFVQA}
\end{figure}

\begin{figure*}[!t]
    \centering
    \includegraphics[width=1\linewidth]{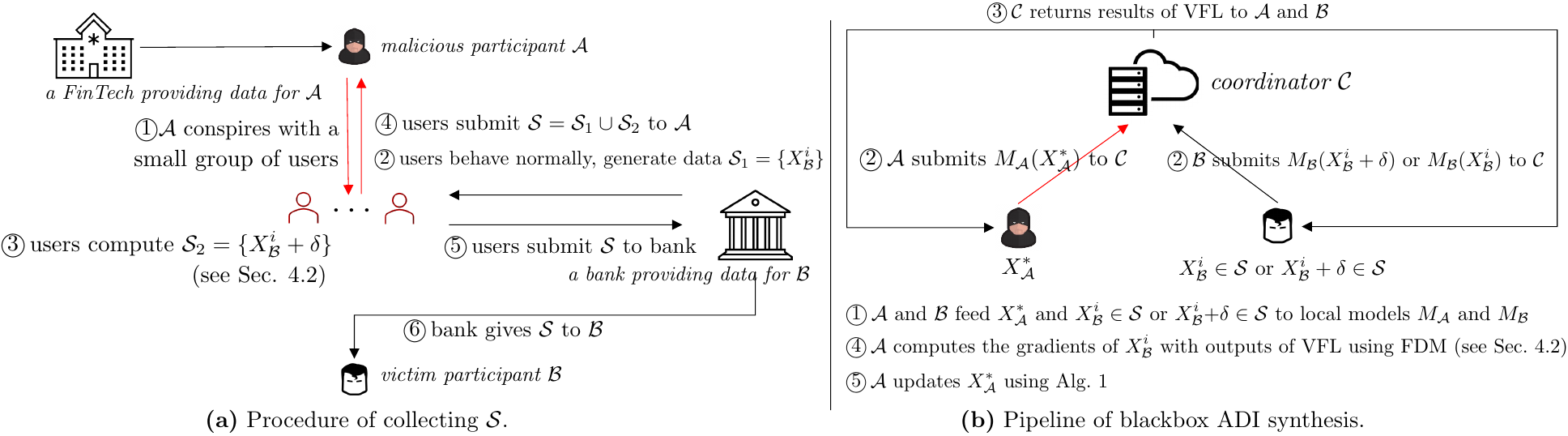}
    \caption{Pipeline of Blackbox ADI Synthesis.}
    \label{fig:adi-pipeline}
\end{figure*}

\noindent \textbf{VFVQA.}~~
Visual Question Answering~\cite{VQA}, a popular
multimodal learning task, answers open-ended natural-language questions about
images. VQA models develop a joint understanding of questions and images and
have been employed in privacy-sensitive scenarios like medical image
diagnosis~\cite{vqamed,lubna2019mobvqa}. VFL provides a practical and effective
solution to facilitate privacy-preserving VQA~\cite{liu2020federated}. In our
evaluation, we modify the state-of-the-art VQA model Oscar~\cite{li2020oscar}
and construct VFVQA as illustrated in \F~\ref{fig:VFVQA}.

As introduced in background section, VFVQA concretizes and extends the SplitNN protocol, which involves considerable
engineering efforts to link with several modern computer vision and natural
language processing models. \ma\ holds the image data and their associated
object tags. It computes the feature representation of data using
FasterRCNN~\cite{ren2016faster} and BERT~\cite{li2020oscar}, where
$X_{\mathcal{A}_1}$ and $X_{\mathcal{A}_2}$ represent the image data and the
corresponding tags, respectively. \mb\ holds the textual data (i.e.,
``questions'' in VQA) and computes the embedding results of each question using BERT.
In particular, \ma\ holds the image data and their associated object tags. It
uses FasterRCNN~\cite{ren2016faster} and BERT~\cite{li2020oscar} to process
$X_{\mathcal{A}_1}$ and $X_{\mathcal{A}_2}$, corresponding to images and tags.
\mb\ process the textual data (i.e., ``questions'' in VQA tasks) using BERT:

\begin{equation}
    \begin{aligned}
        L_{\mathcal{A}} &= [FasterRCNN(X_{\mathcal{A}_1}, w_{\mathcal{A}_1}) \, || \, BERT(X_{\mathcal{A}_2}, w_{\mathcal{A}_2})] \\
        L_{\mathcal{B}} &= BERT(X_{\mathcal{B}}, w_{\mathcal{B}})
    \end{aligned}
\end{equation}

As specified by the SplitNN protocol, local results on each participant are sent
to the coordinator \mc. Then, \mc\ predicts the answer, another natural
language sentence, to the raised question on \mb\ based on the image feature
shared by \ma\ using multi-layer transformers:

\begin{equation}
    \begin{aligned}
        L_{\mathcal{C}} = Transformer([L_{\mathcal{A}} \, || \, L_{\mathcal{B}}], w_{\mathcal{C}})
    \end{aligned}
\end{equation}

The backward propagation phase is performed consistently with SplitNN. And all
intermediate data in the above phases are encrypted.

\section{Pipeline of Blackbox ADI Synthesis}
\label{sec:workflow}

In this section, we present an end-to-end illustration of ADI synthesis in the
blackbox setting. We consider the VFL example given in \S~\ref{sec:introduciton},
where a FinTech and a bank jointly predict a user's credit score. Before
launching the attack and synthesizing ADIs, we assume that a VFL system has been
well trained, where \ma\ takes training data from the FinTech and \mb\ takes
the training data from the bank.

\F~\hyperref[fig:adi-pipeline]{\ref{fig:adi-pipeline}a} depicts the procedure of
collecting \sd, whose high-level procedure has been mentioned in
\S~\ref{subsec:threat-model}. In particular, the malicious participant
\ma\ conspires with a small group of users: these users behave normally, but are
willing to share their records in the bank with \ma\ (\textcircled{1}). These users
behave \textit{normally} and register their information in the bank. Their
provided bank records will form our target sample dataset $\mathcal{S}_1$ (\textcircled{2}).
Then, as discussed in \S~\ref{subsec:blackbox-adi}, these users further compute
another collection of bank records, by adding
permutation $\delta$ to each bank record $X_{\mathcal{B}}^i \in \mathcal{S}_1$.
These $X_{\mathcal{B}}^{i} + \delta$ form another sample dataset $\mathcal{S}_2$ (\textcircled{3}).
Given that users are conspiring with \ma, both $\mathcal{S}_1$ and $\mathcal{S}_2$ are shared with
\ma\ (\textcircled{4}). Similarly, $\mathcal{S}_2$ will need to be provided to the bank, and
therefore, both $\mathcal{S}_1$ and $\mathcal{S}_2$ are accessible to \mb\ (\textcircled{5}, \textcircled{6}).The tiny collection of data samples \sd\ is formed by $\mathcal{S}_1$ and $\mathcal{S}_2$: $\mathcal{S} = \mathcal{S}_1 \cup \mathcal{S}_2$.


With \sd\ becomes accessible to both \ma\ and \mb,
\ma\ can start to synthesize ADIs via estimated gradients (see
\S~\ref{subsec:blackbox-adi}).
\F~\hyperref[fig:adi-pipeline]{\ref{fig:adi-pipeline}b} shows the pipeline of
our blackbox ADI synthesis.

Typically, before launching a joint inference, \mc\ will coordinate and inform
\ma\ and \mb\ about the user ID. Thus, \ma\ can easily decide whether the
current input of \mb\ is from the sample set \sd\ according to the
current user ID. If so, \ma\ can start to update its ADI input
$X_{\mathcal{A}}^*$. To do so, \ma\ feeds ADI $X_{\mathcal{A}}^*$ to local model
$M_{\mathcal{A}}$ and \mb\ feeds $X_{\mathcal{B}}^i \in \mathcal{S}$ or
$X_{\mathcal{B}}^{i} + \delta \in \mathcal{S}$ to local model $M_{\mathcal{B}}$ and
get the corresponding outputs (\textcircled{1}). Then, \ma\ and \mb\ submit their
local outputs to \mc\ and receive the joint inference output of the VFL model
from \mc\ (\textcircled{2} and \textcircled{3}). Then, as we discuss in
\S~\ref{subsec:blackbox-adi}, \ma\ can compute the gradients of
$X_{\mathcal{B}}^i \in \mathcal{S}$ by FDM (\textcircled{4}). Finally, \ma\ updates
ADI $X_{\mathcal{A}}^*$ by \A~\ref{alg:attacking} without accessing the local
model of \mb\ (\textcircled{5}).

\section{Cooperating Fuzz Testing with VFL Protocols}
\label{sec:fuzzing-vfl}
\smallskip

\begin{table*}[!t]
	\centering
    \scriptsize
    \newcommand{\tabincell}[2]{\begin{tabular}{@{}#1@{}}#2\end{tabular}}
  \caption{Detail steps to cooperate fuzz testing with the VFL protocols.}
    \label{tab:fuzz-implementation}
	\resizebox{1.00\linewidth}{!}{
        \begin{tabular}{|l|c|c|c|}
            \hline
             & \textbf{Participant \ma} & \textbf{Participant \mb} & \textbf{Coordinator \mc} \\
			\hline
            \textbf{1.}  & \tabincell{c}{Generate index Seeds \textbf{S},\\ maintain a list \textbf{Q} storing the corresponding data.} & \tabincell{c}{Compute all local results of data in participant \mb , send to \mc} & \\
            \hline
            \textbf{2.}  & \tabincell{c}{a) Choose next from \textbf{S} as $index_a$.\\ b) Add noise to \textbf{Q}[$index_a$] for $\beta$ times.\\ c) Compute local results of the noised data, send to \mc.} &  & \\
            \hline
            \textbf{3.}  &   &  & \tabincell{c}{a) Randomly choose a local result of participant \mb\\ and the corresponding index is $index_b$.\\ b) Compute the outputs of all noised data and gradients, \\send to \ma\ and \mb.}\\
            \hline
            \textbf{4.}  & \tabincell{c}{Compute the saliency score of all noised data,\\ send the highest $saliency\_score_a$\\ and corresponding $index\_noise_a$ to \mc.} & \tabincell{c}{Compute the ${saliency\_score\_orig_b}$, send to \mc.} & \\
            \hline
            \textbf{5.}  &   &  & \tabincell{c}{Compute Attack Success Rate, send to \ma.} \\
            \hline
            \textbf{6.}  & \tabincell{c}{a) Maintain the $orig\_acc$.\\ b) Compute the Saliency Mask based on the gradients received.\\ c) Compute $data\_noise\_mask_a$ based on the Mask.\\ d) Compute the local result of $data\_noise\_mask_a$, send to \mc.}   &  &  \\
            \hline
            \textbf{7.}  &   &  & \tabincell{c}{Compute output, gradients and Attack Success Rate ($masked\_acc$),\\ send to \ma\ and \mb.}\\
            \hline
            \textbf{8.}  & \tabincell{c}{Compute ${saliency\_score\_masked_a}$, send to \mc.}  & \tabincell{c}{Compute ${saliency\_score\_masked_b}$, send to \mc.} &  \\
            \hline
            \textbf{9.}  &   &  & \tabincell{c}{Compute $ratio_a$ = ${saliency\_score\_masked_a}$ / ${saliency\_score\_orig_a}$ \\and $ratio_b$ = ${saliency\_score\_masked_b}$ / ${saliency\_score\_orig_b}$,\\ send to \ma.}  \\
            \hline
            \textbf{10.}  & \tabincell{c}{If $masked\_acc > orig\_acc$ and $ratio_a > ratio_b$,\\    update \textbf{Q}[$index_a$] = $data\_noise\_mask_a$.\\ If $masked\_acc > $ threshold,\\    add \textbf{Q}[$index_a$] to \ma,\\ delete $index_a$ from \textbf{S} and return to Step 2.} &  &  \\
            \hline
            \textbf{11.}  & \tabincell{c}{Return to step 3 for $\gamma$ times.} &  &  \\
            \hline
            \textbf{12.}  & \tabincell{c}{Return to step 2 for $\Gamma$ times.} &  &  \\
			\hline
		\end{tabular}
	}
\end{table*}

As reported in evaluation section, we launch fuzz testing toward two
widely-used VFL protocols (i.e. HeteroLR and SplitNN) on the basis of
real-world VFL platforms FATE and FedML. We also launch fuzz testing toward
VFVQA, whose protocol is consistent with SplitNN. A detailed procedure to cope
with fuzz testing with learning protocols is given in
\T~\ref{tab:fuzz-implementation}. In all, greybox fuzzing can be accordingly
performed on the distributed versions of learning protocols without sharing much
information about the original data.

\begin{figure*}[!h]
    \captionsetup{skip=5pt}
    \captionsetup[sub]{skip=1pt}
    \vspace{-10pt}
\centering
    \begin{subfigure}{.16\linewidth}
        \centering
        \resizebox{1\linewidth}{!}{
        \begin{tikzpicture}
            \node (img) {\includegraphics[trim={1cm 0.5cm 1cm 0.5cm},clip]{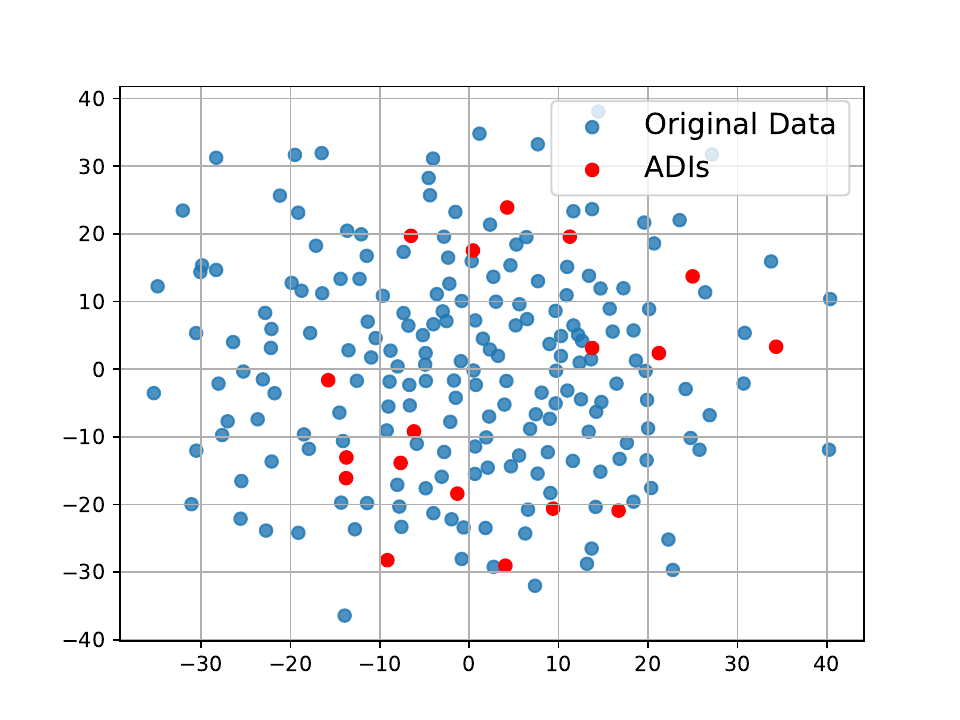}};
        \end{tikzpicture}
        }
    \caption{\small NUS-WIDE.}
    \end{subfigure}
    \begin{subfigure}{.16\linewidth}
        \centering
        \resizebox{1\linewidth}{!}{
        \begin{tikzpicture}
            \node (img) {\includegraphics[trim={1cm 0.5cm 1cm 0.5cm},clip]{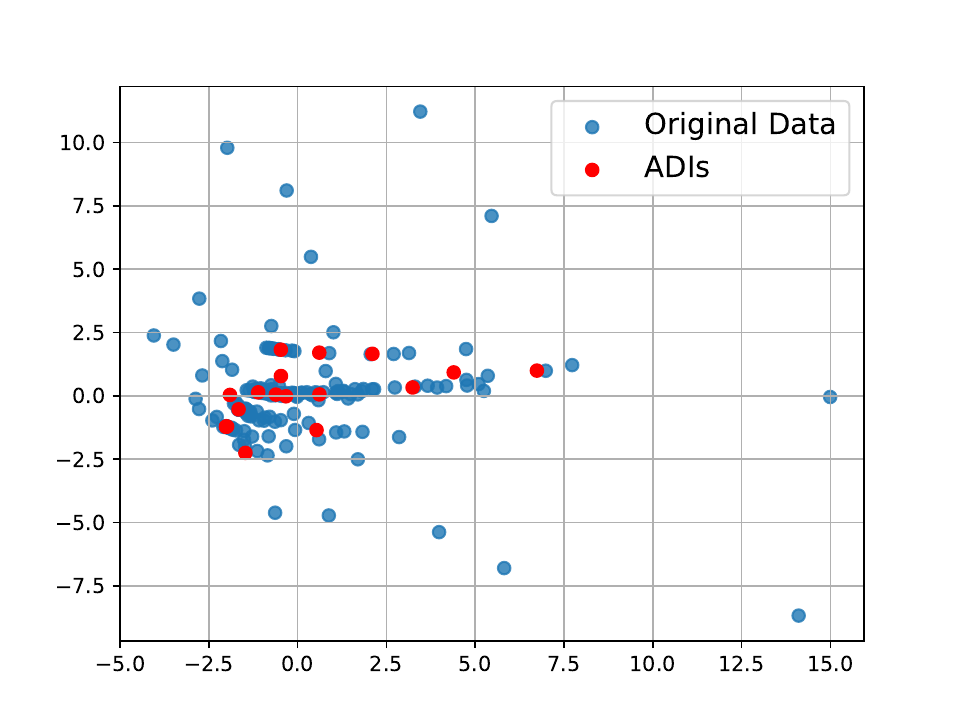}};
        \end{tikzpicture}
        }
    \caption{\small Credit.}
    \end{subfigure}
    \begin{subfigure}{.16\linewidth}
        \centering
        \resizebox{1\linewidth}{!}{
        \begin{tikzpicture}
        \node (img) {\includegraphics[trim={1cm 0.5cm 1cm 0.5cm},clip]{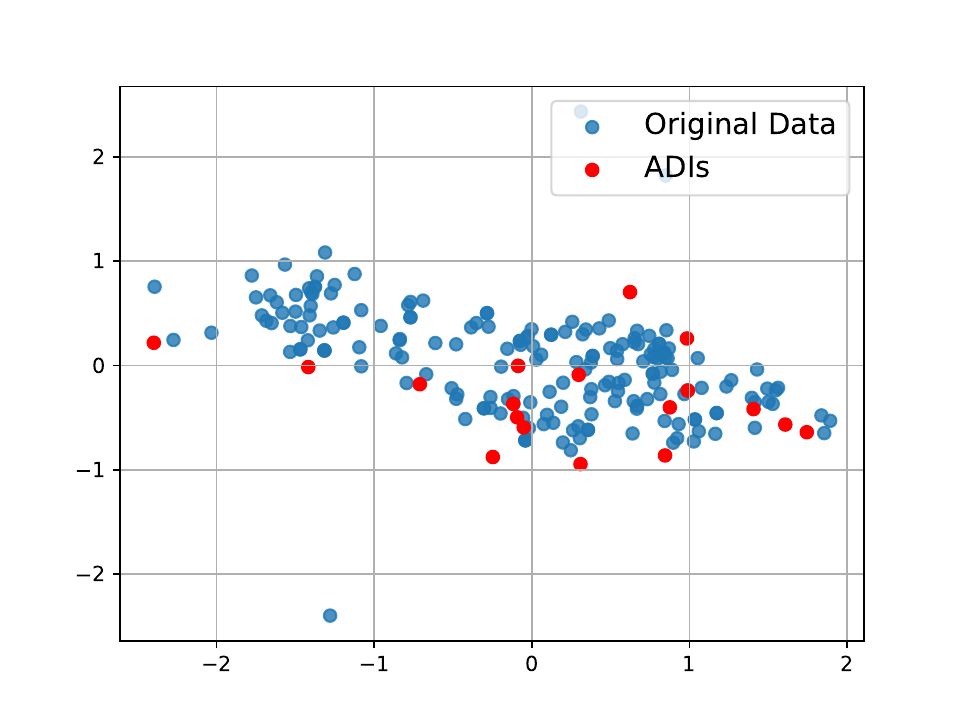}};
        \end{tikzpicture}
        }
    \caption{\small Vehicle.}
    \end{subfigure}
    \begin{subfigure}{.16\linewidth}
        \centering
        \resizebox{1\linewidth}{!}{
        \begin{tikzpicture}
            \node (img) {\includegraphics[trim={1cm 0.5cm 1cm 0.5cm},clip]{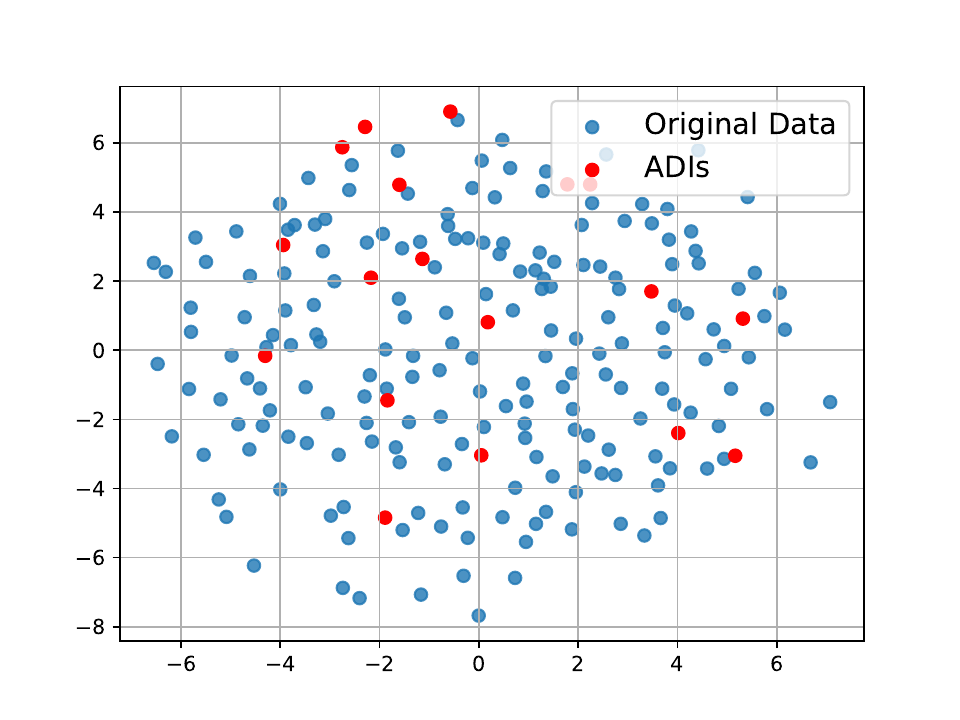}};
        \end{tikzpicture}
        }
    \caption{\small MNIST.}
    \end{subfigure}
    \begin{subfigure}{.16\linewidth}
        \centering
        \resizebox{1\linewidth}{!}{
        \begin{tikzpicture}
            \node (img) {\includegraphics[trim={1cm 0.5cm 1cm 0.5cm},clip]{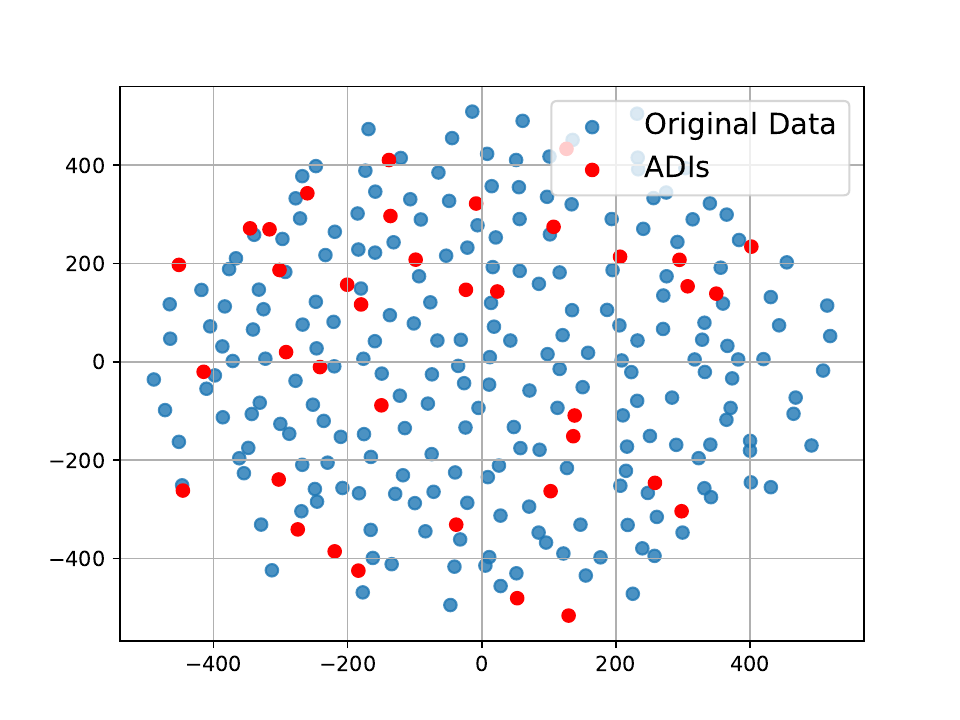}};
        \end{tikzpicture}
        }
    \caption{\small VQA v2.0.}
    \end{subfigure}
    \begin{subfigure}{.16\linewidth}
        \centering
        \resizebox{1\linewidth}{!}{
        \begin{tikzpicture}
            \node (img) {\includegraphics[trim={1cm 0.5cm 1cm 0.5cm},clip]{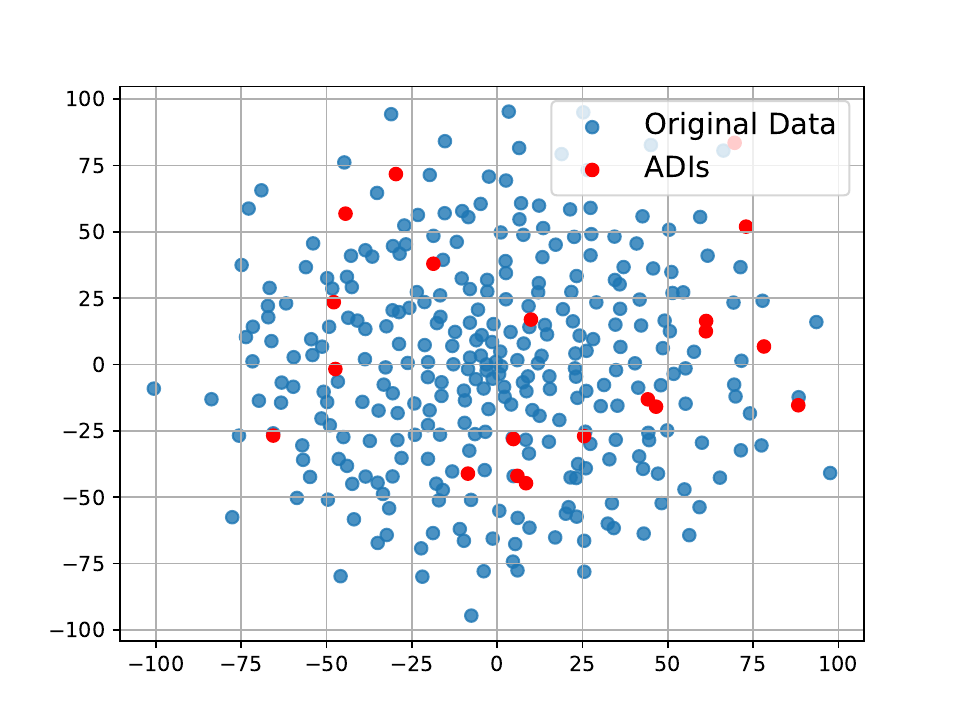}};
        \end{tikzpicture}
        }
    \caption{\small CIFAR-10.}
    \end{subfigure}

    \centering
    \caption{Normal inputs and ADIs generated by fuzz testing projected to 2D figures.}
    \label{fig:fuzz-project}
    \vspace{-3pt}
\end{figure*}

\begin{figure}[!t]
    \centering
    \includegraphics[width=0.95\linewidth]{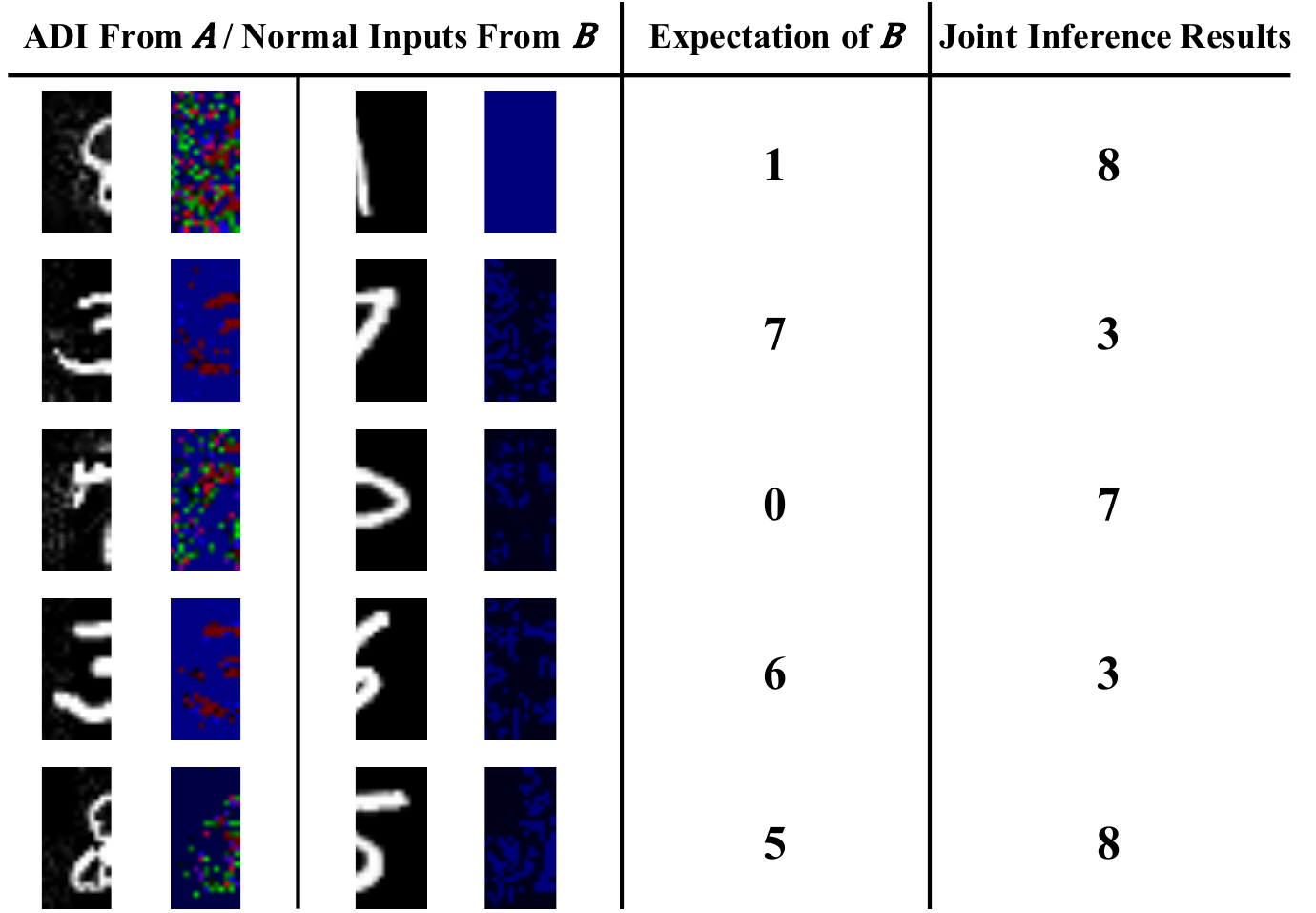}
    \caption{ADIs found in MNIST using fuzz testing.}
    \label{fig:fuzz-mnist-mask}
\end{figure}
\section{Datasets Information}
\label{sec:dataset}
We use six popular datasets in our evaluation \S~\ref{sec:evaluation}: NUS-WIDE~\cite{chua2009nus}, Credit~\cite{credit},
Vehicle~\cite{vehicle}, MNIST~\cite{lecun1998gradient}, VQA
v2.0~\cite{lin2014microsoft}, and CIFAR-10~\cite{krizhevsky2009learning}.
NUS-WIDE dataset contains 269,648 images and the associated tags. Each
sample has 634 low-level image features and 1,000 text features. Thus, it is
suitable for a feature-partition setting and is therefore commonly used to
benchmark VFL. We set up a joint classification task of ten labels via two
participants \ma\ and \mb. We use the top ten labels in NUS-WIDE having the
maximum number of data samples. \ma\ holds the image features whereas \mb\ holds
the text features.

Credit dataset is a popular benchmark dataset for VFL that comprises the
payment records of 30,000 customers of a bank, where 5,000 customers are
malicious while the rest are benign. Each data sample has 23 integer or
floating-point number features. We set up a VFL logistic regression task over
participants \ma\ and \mb\ to predict whether a payment is from a malicious
customer or not. Participant \ma\ holds 13 features and participant \mb\ holds
10.

Vehicle dataset is a multi-classification dataset containing 946 samples of
four vehicle categories. Each sample has 18 features. We set up a VFL logistic
regression task over two participants \ma\ and \mb\ to classify data samples
into the vehicle categories. Both \ma\ and \mb\ get nine features each.

MNIST is a handwritten digital image dataset with 60,000 training samples and
10,000 testing samples, each with dimensions of 28$\times$28 pixels. We set up a
VFL classification task with ten labels over participants \ma\ and \mb. Each
image is vertically partitioned into two such that \ma\ gets the left piece
(with 28$\times$14 pixels) while \mb\ gets the right piece.

VQA v2.0 is a large-scale dataset widely used to train VQA models. We set up
VFVQA over participants \ma\ and \mb\ to facilitate VQA, where \ma\ holds images
and \mb\ raises natural-language questions. There are 82,783 images and 443,757
questions in the training set, and 40,504 images and 214,354 questions in the
validation set. The images are embedded as a feature vector of size 2,054 while
the questions are embedded as a word embedding matrix with shape 128$\times$512.

CIFAR-10 is a relatively complex colour image dataset with 50,000 training samples and 10,000 testing samples, each with dimensions of $32\times 32 \times 3$ pixels.
There are 10 labels for this dataset.
Similar to MNIST, we set up a VFL classification task with 10 labels iver participants \ma\ and \mb. 
Each image is vertically partitioned into two, such that \ma\ gets the left piece and \mb\ gets the right piece.
\ma\ and \mb\ use VGG16 to get a feature embedding with 10,752 dimensions and use the embeddings as the inputs for the VFL model.
\section{Fuzz Testing Results Visualization}
\label{sec:fuzzing-visualize}

The ADIs generated by fuzz testing and original inputs of participant \ma\ are
projected to 2D figures by multidimensional scaling. As shown in
\F~\ref{fig:fuzz-project}, the ADIs are distributed close to the original data.
We interpret that the generated ADIs (marked in \textcolor{red}{red}) are
\textit{stealthy}: they can be hardly distinguished by the data distribution nor
by the distances between the data points.

We also present several MNIST cases generated by fuzz testing in
\F~\ref{fig:fuzz-mnist-mask}. Similar to \F~\ref{fig:observeMnist}, we report
the ADIs used by \ma, the normal inputs used by \mb, and also the corresponding
saliency maps. It is easy to see that ADIs on \ma\ dominate the joint inference,
and make the contribution of \mb\ negligible.



\end{document}